\documentclass[12pt,thmsa]{article}
\usepackage{amssymb}

\usepackage{sw20lart}

\input{tcilatex}
\begin{document}

\begin{center}
{\LARGE Quantum search processes in the cyclic }

{\LARGE group state space\bigskip s}

{\normalsize Xijia Miao}$^{*}$

June, 2005; Somerville, Massachusetts\bigskip

Abstract
\end{center}

The hardness to solve an unstructured quantum search problem by a standard
quantum search algorithm mainly originates from the low efficiency to
amplify the amplitude of the unknown marked state in the Hilbert space of an 
$n-$qubit pure-state quantum system by the oracle unitary operation
associated with other known unitary operations. A standard quantum search
algorithm generally can achieve only a square speedup over the best known
classical counterparts. In order to bypass this square speedup limitation it
is necessary to develop other type of quantum search algorithms. In the
paper an oracle-based quantum dynamical method has been developed to solve
the quantum search problem in the cyclic group state space of the Hilbert
space. The binary dynamical representation for a quantum state in the
Hilbert space of the $n-$qubit quantum system is generalized to the
multi-base dynamical representation for a quantum state in the cyclic group
state space. Thus, any quantum state such as the marked state and its
corresponding oracle unitary operation in the cyclic group state space may
be described completely in terms of a set of dynamical parameters that are
closely related to the symmetric property and structure of the cyclic group.
The quantum search problem then may be solved through determining the set of
dynamical parameters that describe completely the marked state instead by
directly measuring the marked state which is a necessary step in the
standard quantum search algorithm. The quantum dynamical method makes it
possible to manipulate at will the unknown marked state and its oracle
unitary operation. By a similar method used extensively in the hidden
subgroup problems, a cyclic group state space may be formed by mapping all
the group elements of a cyclic group one-to-one onto the specific states of
the Hilbert space of the $n-$qubit quantum system. It carries the symmetric
property and structure of a cyclic group. An unstructured quantum search
process in the Hilbert space may be affected greatly by the symmetric
property and structure of the cyclic group when the quantum search problem
is solved in the cyclic group state space. When the cyclic group is high
symmetric, that is, the cyclic group with order $p$ is a product group of
many cyclic subgroups and each cyclic subgroup has an order $\thicksim
O(\log p)$, the quantum search problem in the cyclic group state space could
be solved better through reducing it from the cyclic group state space with
dimension $p$ to the cyclic group state subspaces with dimension $\thicksim
O(\log p)$ of these cyclic subgroups, for any quantum search problem can be
efficiently solved in these subspaces due to their much small dimension. The
main attempt of the paper is to make use of the symmetric properties and
structures of groups to help solving the unstructured quantum search problem
in the Hilbert space. It is shown how the quantum search process could be
reduced efficiently from the cyclic group state space to its cyclic group
state subspaces with the help of the symmetric property and structure of the
cyclic group on an ideal universal quantum computer. \newline
\newline
{\large 1. Introduction}

The quantum search is tremendously valuable as it has an extensive
application in computation science. In classical computation most important
problems are either polynomial-time or NP-complete [1]. The conventional
computers based on the classical physical principles are much suitable for
solving efficiently the polynomial-time problems, but inherently they are
not enough powerful to treat efficiently all the NP-hard problems [1]. On
the other hand, it has been shown that all the NP-complete problems in the
classical computation could be solved efficiently on a quantum computer if
there existed a polynomial-time unstructured quantum search algorithm. Thus,
a great progress could be achieved in quantum computation if an efficient
quantum search algorithm could be found. In the past decade a great effort
has been devoted to attacking this extremely important problem in quantum
computation. A number of quantum search algorithms [2-13] have been proposed
and developed since the standard Grover quantum search algorithm was
suggested [2]. The famous include the standard Grover search algorithm [2],
the quantum adiabatic search algorithm [4, 5], and the
amplitude-amplification search algorithm [6]. Most of these oracle-based or
block-box-based quantum search algorithms are based on the quantum-state
tomographic method. These quantum search algorithms usually start with a
superposition of the Hilbert space of a pure-state quantum system, then
perform an iterative sequence of unitary operations which include the oracle
unitary operation and other known unitary operations to amplify the
amplitude of the marked state of the quantum search problem, and after the
unitary operation sequence measure the generated state, in which the marked
state has a high probability ($\thicksim 1)$, to output directly the
computing result, i.e., the complete information of the marked state. Since
the efficiency is low to amplify amplitude of the marked state with these
unitary operation sequences these search algorithms usually can only achieve
a square speedup over the best known classical counterparts. It has been
proven that this square speedup for all these unstructured quantum search
algorithms is optimal [3, 6, 9, 13]. More generally, it has been shown that
many oracle-based quantum algorithms (not limited to the quantum search
algorithms) based on the quantum-state tomography are subjected to
polynomial bounds in speedup [14], that is, these quantum algorithms can
only achieve a polynomial speedup over their best classical counterparts. In
order to bypass this speedup obstacle inherently for the oracle-based
quantum algorithms based on the quantum-state tomography it is necessary to
develop other types of quantum algorithms to solve the quantum search
problem and other problems [15]. Due to the fact that there is a low
efficiency to amplify the amplitude of the marked state in these quantum
search algorithms [2, 3, 6, 15], in developing new type of quantum search
algorithms a direct quantum measurement on the marked state with a high
probability ($\thicksim 1$) should be avoided becoming a necessary step so
that amplification of the amplitude of the marked state may not be the key
component in algorithm, while the quantum measurement to output the
computing results could be carried out on those states which are closely
related to the marked state and carry the complete information of the marked
state [15]. It is particularly important to be able to manipulate at will
any quantum state even the unknown marked state in the Hilbert space in
developing efficient quantum algorithms. This is an important step towards
the goal to realize that any quantum state in the Hilbert space of an $n-$%
qubit quantum system is able to be described and characterized completely in
a parameterization form [15]. Such a parameterization description for a
quantum state is different from the conventional quantum-state tomographic
method. Since any quantum state in the Hilbert space can be described and
characterized completely by a set of dynamical parameters called the
quantum-state unit-number vector [15] and there is a one-to-one
correspondence between the oracle unitary operation and the marked state in
the quantum search problem, it becomes possible to manipulate at will the
oracle unitary operation and the unknown marked state. Due to the fact that
the unknown marked state can be described completely by the set of dynamical
parameters it is possible to solve the quantum search problem by determining
the set of dynamical parameters instead directly through measuring the
marked state [15]. This gives a possibility to avoid a direct amplification
of amplitude of the marked state which is a key component in the
conventional quantum search algorithms [2-13]. This strategy to solve the
quantum search problem opens a large space to develop new quantum search
algorithms.

Generally, the unknown marked state of the quantum search problem can be any
possible state of the Hilbert space and the quantum search space for the
problem must contain the marked state in a quantum search algorithm.
Therefore, for these conventional quantum search algorithms the quantum
search spaces usually are taken as the whole Hilbert space and hence the
initial state is a superposition over the whole Hilbert space. The usual
quantum search algorithms also have showed that the low efficiency to
amplify the amplitude of the marked state is strongly dependent on the
dimensional size of the search space, that is, the efficiency generally is
inversely proportional to the square root of the dimensional size of the
search space [2, 3, 6] and it has been shown that the efficiency is optimal
[3, 6, 9, 13]. One possible scheme to increase this efficiency therefore
could be that the quantum search space is limited to a small state subspace
of the Hilbert space for a quantum search problem [16]. Generally, this
scheme will meet difficulty and is not feasible if the marked state is not
in the subspace. To make the scheme feasible one must convert the marked
state from the whole Hilbert space to the small subspace. Because there is
the rotation symmetry in spin space in the $n-$qubit quantum spin system the
whole Hilbert space of the spin system can be divided into $(n+1)$ state
subspaces according to the angular momentum theory in quantum mechanics and
it can be shown that any unknown quantum state such as the marked state can
be converted efficiently from a small subspace into a larger subspace in the
Hilbert space [16]. This fact directly leads to that in a single $n-$qubit
quantum system the quantum search problem can be reduced efficiently from
the whole Hilbert space into its largest subspace. This search space
reduction speeds up the conventional quantum search process, although this
speedup is limited and does not change essentially the computational
complexity for the quantum search problem. However, it is very important for
the fact that the symmetric properties and structures of quantum systems may
be exploited to speed up the quantum search process, for one can go a
further step to use the special group symmetric properties and structures to
help solving the quantum search problem. This idea will play an important
role to guide the construction of quantum search algorithms in the cyclic
group state space in the paper. Generally, the whole Hilbert space of the $%
n- $qubit quantum system may not have some specific group symmetric
properties and structures, but a specific state subset of the Hilbert space
could carry the symmetric property and structure of a specific group such as
a cyclic group. Then quantum computation may be affected greatly by the
symmetric property and structure of the group if it is carried out on the
state subset. In order to make use of the symmetric property and structure
of a finite group in developing new quantum algorithms one may first form
this specific state subspace in the Hilbert space of the $n-$qubit quantum
system. By mapping all the group elements of the group one-to-one onto these
specific states of the Hilbert space, which means that each group element
corresponds one-to-one to a state of the specific state subspace and hence
the mapping is isomorphic, then all these specific states form a state
subspace of the Hilbert space and evidently this subspace is an invariant or
closed state subspace under the action of the group operations. This state
subspace is called the group state space of the Hilbert space. This subspace
can be thought of as an artificially-formed state space of the Hilbert space
which carries the information of the symmetric property and structure of the
group. The dimension of the group state space is just the order of the
finite group. The similar scheme to the group state space has been
extensively used previously in the hidden subgroup problems [17]. If the
group is high symmetric, which means that the group is a product of many its
factor subgroups, then the corresponding group state space also may contain
many state subspaces which one-to-one correspond to these factor subgroups
of the group. Since the dimensional size of the group is a product of the
dimensional sizes of all these factor subgroups, the dimension of the group
state space is also a product of those of the group state subspaces of the
factor subgroups. The dimensional size (denoted as $p$ here) of the group
state space can be very large, $p\thicksim 2^{n},$ and may increase
exponentially as the qubit number, but since it is a product of the
dimensional sizes of many state subspaces of these factor subgroups the
dimensional sizes of the state subspaces of the factor subgroups can be much
small, $\thicksim O(\log p),$ and may increase only polynomially as the
qubit number. Then the quantum search problem in the whole group state space
would be solved efficiently if the problem could be efficiently reduced from
the whole group state space to the state subspaces of these subgroups. As
the main purpose, the paper intends to achieve such a search space reduction
for the quantum search problem in the cyclic group state space with the help
of the symmetric property and structure of a cyclic group. Here the
symmetric property and structure of a cyclic group are employed to help
solving the quantum search problem as a cyclic group is one of the simplest
groups and its symmetric property and structure are very simple and have
been studied in detail and thoroughly [18]. \newline
\newline
{\large 2. Quantum search model in the cyclic group state space}\newline
\newline
{\large 2.1. The binary dynamical representation and the multi-base
dynamical representation of quantum states}

In the Hilbert state space of an $n-$qubit pure-state quantum system a
quantum state can be characterized and described completely by a set of $n$
dynamical parameters $\{a_{k}^{s}=\pm 1,$ $k=1,2,...,n\},$ which has been
called the quantum-state unit-number vector in the papers [15]. This is a
parameterization description for quantum states in the $2^{n}-$dimensional
Hilbert space and different from the conventional quantum-state tomographic
method. By measuring the set of the dynamical parameters one can determine
uniquely the corresponding quantum state. This dynamical description picture
not only is able to describe completely a quantum state $|s\rangle $ in the $%
2^{n}-$dimensional Hilbert space of the pure-state quantum system but also
is used to describe the corresponding quantum state $\rho _{s}=|s\rangle
\langle s|,$ which is represented by a diagonal density operator, in the
Liouville operator space of the quantum ensemble of the quantum system. For
instance, in the Hilbert state space a conventional computational basis
state $|s\rangle $ can be described completely with the parameter vector $%
\{a_{k}^{s}\}$ by $|s\rangle =\bigotimes_{k=1}^{n}(\frac{1}{2}%
T_{k}+a_{k}^{s}S_{k})$ with $T_{k}=|0_{k}\rangle +|1_{k}\rangle $ and $S_{k}=%
\frac{1}{2}(|0_{k}\rangle -|1_{k}\rangle ),$ while in the corresponding
Liouville operator space the quantum state is described completely by the
diagonal density operator $\rho _{s}=|s\rangle \langle
s|=\bigotimes_{k=1}^{n}(\frac{1}{2}E_{k}+a_{k}^{s}I_{kz})$ which is also
determined uniquely by the vector $\{a_{k}^{s}\}$. By the parameter vector $%
\{a_{k}^{s}\}$ one can set up a one-to-one correspondence between a quantum
state $|s\rangle $ or $\rho _{s}=|s\rangle \langle s|$ and the selective
rotation unitary operation $C_{s}(\theta )=\exp (-i\theta D_{s})$ with the
Hermitian diagonal operator $D_{s}=|s\rangle \langle s|=\bigotimes_{k=1}^{n}(%
\frac{1}{2}E_{k}+a_{k}^{s}I_{kz}),$ which selectively acts on only the
quantum state $|s\rangle $ or $\rho _{s}$ and is described completely also
by the vector $\{a_{k}^{s}\}$. The diagonal operator $D_{s}$ is called the
quantum-state diagonal operator since it is a diagonal operator and also
equals the state $\rho _{s}$ formally$.$ The unitary evolution process of a
quantum system or its corresponding quantum ensemble under the action of the
selective rotation unitary operation $C_{s}(\theta )$ is described
completely by the vector parameters $\{a_{k}^{s}\}$ and in this sense the
vector parameters $\{a_{k}^{s}\}$ are also called the dynamical parameters.
The quantum-state diagonal operator $D_{s}$ makes it possible for one
manipulating at will the evolution process of an unknown quantum state in
the Hilbert space of a quantum system. This representation for the quantum
state via the dynamical parameter vector $\{a_{k}^{s}\}$ is called the
binary dynamical representation, for the vector parameters $\{a_{k}^{s}\}$
can take only two values $+1$ and $-1.$ In the quantum dynamics any
quantum-state search problem can be reduced to determining the dynamical
parameter vector $\{a_{k}^{s}\}$ of the marked state. The direct measurement
on the marked state to output the information of the marked state in the
conventional quantum search algorithms [2-13] therefore could not be
necessary in the quantum dynamical method, for there are a number of
possible methods in quantum dynamics which work either in a pure-state
quantum system or in a quantum ensemble to determine the dynamical parameter
vectors $\{a_{k}^{s}\}$ for quantum states including the marked state [15].
Therefore, the quantum search algorithms based on the quantum dynamics have
an important difference from the conventional ones [2-13] that it is not
necessary to measure directly the marked state to output the complete
information of the marked state in algorithm. The quantum dynamic method
opens a large space to develop new type of quantum search algorithms.

Besides the binary dynamical representation for a quantum state in the
Hilbert space of an $n-$qubit quantum system it is possible to use other
multi-base dynamical representations to describe completely a quantum state
of a quantum system. The multi-base dynamical representations for a quantum
state could be a better choice for the quantum search problem in the cyclic
group state space. Before the multi-base dynamical representations can be
described the group state space of a cyclic group is firstly defined in the
Hilbert space. A cyclic group is an Abelian group in which any group
elements are commutable to one another [18]. A cyclic group $G$ can be
generated by a fixed generator $g,$ that is, $G=\langle g\rangle
=\{E,g,g^{2},...,g^{n_{r}-1}\},$ here $E$ is the unity element and $n_{r}$
the order of the cyclic group. In an analogue way to the method used
extensively in the hidden subgroup problems [17], now each group element of
the cyclic group $G$ is mapped one-to-one onto the specific state of the
Hilbert state space of an $n-$qubit quantum system. Then these specific
states of the Hilbert space that correspond to all the group elements of the
cyclic group form a state subset and this state subset is the cyclic group
state space of the Hilbert space. The cyclic group state space is an
invariant state subspace under the action of any group operation (element)
of the cyclic group. Suppose further that the unity element $E$ of the group 
$G$ is mapped onto the specific state $|\varphi _{0}\rangle $ of the Hilbert
space, then the cyclic group state space $S(G)$ may be given formally by 
\[
S(G)=\{|\varphi _{0}\rangle ,g|\varphi _{0}\rangle ,g^{2}|\varphi
_{0}\rangle ,...,g^{n_{r}-1}|\varphi _{0}\rangle \}. 
\]
Here the cyclic group state space $S(G)$ is within the Hilbert space and its
dimension is just the order $n_{r}$ of the cyclic group. In quantum
computation a convenient state basis in the Hilbert space of an $n-$qubit
quantum system is the conventional computational basis. This basis set
consists of the integer states $\{|Z_{2^{n}}\rangle \}=\{|0\rangle
,|1\rangle ,|2\rangle ,...,|2^{n}-1\rangle \}$. Then the state basis set of
the cyclic group state space of the Hilbert space is the specific state
subset of the integer state set $\{|Z_{2^{n}}\rangle \}$. Now consider the
integer set $Z_{m}=\{0,1,2,...,m-1\}$. The integer set $Z_{m}$ is a Ring $%
(Z_{m}=Z/mZ)$ under the modular arithmetic operation ($\func{mod}m$) in
number theory [19] and also an additive cyclic group under the modular $(m)$
additive operation [18, 20]. As the multiplicative operation is often used
in quantum computation, the integer set $Z_{p-1}=\{0,1,...,p-2\}$ can be
mapped by the modular exponentiation: $z\rightarrow g^{z}\func{mod}p$ to the
positive integer set $Z_{p}^{+}=\{g^{z}\func{mod}p\}=\{1,2,...,p-1\},$ where
the integer $z\in Z_{p-1}$ and $p$ is a known prime and $g$ a known
primitive root ($\func{mod}p)$. The integer set $Z_{p}^{+}$ forms a
multiplicative cyclic group under modular multiplication operation [18, 20].
Both the additive cyclic group $Z_{p-1}$ and the multiplicative cyclic group 
$Z_{p}^{+}$ have an order $p-1.$ Both the cyclic groups have a one-to-one
correspondence. In fact, all the same order cyclic groups are isomorphic to
one another [18]. Hereafter $p$ is denoted as a known prime, $g$ as a
primitive root or a generator of a cyclic group, and $C_{m}$ the
multiplicative cyclic group such as $Z_{m}^{+}$ with order $m.$ If any of
the two cyclic groups is mapped onto the Hilbert space, one obtains their
corresponding cyclic group state spaces. For the additive cyclic group $%
Z_{p-1}$ the mapping between the group elements and the corresponding states
in the Hilbert space may be given conveniently by $s\rightarrow |s\rangle $
for the group element $s\in Z_{p-1}=\{0,1,...,p-2\}$ ($|\varphi _{0}\rangle
=|0\rangle $), and for the multiplicative cyclic group $C_{p-1}$ the mapping
may be conveniently written as $f(s)=g^{s}\func{mod}p\rightarrow |g^{s}\func{%
mod}p\rangle $ ($|\varphi _{0}\rangle =|1\rangle $) for the group element $%
g^{s}\equiv g^{s}\func{mod}p\in Z_{p}^{+}=\{1,2,...,p-1\}.$ Therefore, any
quantum state of the cyclic group state space $S(C_{p-1})=\{|g^{s}\func{mod}%
p\rangle \}=\{|1\rangle ,|2\rangle ,...,|p-1\rangle \}$ of the
multiplicative cyclic group $(C_{p-1})$ can be expressed generally as 
\[
|\varphi _{s}\rangle =|g^{s}\func{mod}p\rangle ,\text{ }s\in Z_{p-1}, 
\]
where the state $|\varphi _{s}\rangle $ is also a conventional computational
base. Since $\varphi _{s}>0$ for any $s\in Z_{p-1}$ the state $|0\rangle $
is not included in the cyclic group state space $S(C_{p-1})$. There is a
one-to-one correspondence between the modular exponential function $%
f(s)=g^{s}\func{mod}p$ of the integer set $Z_{p}^{+}$ and the index $s$ of
the integer set $Z_{p-1}.$ The index $s$ is really the discrete logarithmic
function of the function $f(s)$, that is, $s=\log _{g}f(s)$ with $g$ a
logarithmic base. In other words, the index $s$ is the inversion function of
the modular exponential function $f(s)=g^{s}\func{mod}p$, i.e., $%
s=f(s)^{-1}. $ In classical computation the modular exponential function is
not hard to be computed, but the discrete logarithmic function usually is
hard and this forms the basis of the classical public secure key
cryptography based on the discrete logarithms [21]. However, the Shor
discrete logarithmic quantum algorithm shows that the discrete logarithmic
function now is not hard yet to be computed in quantum computation [22].

Actually, the additive cyclic group state space $S(Z_{p-1})$ is just the
state subset $\{|Z_{p-1}\rangle \}$ consisting of the first $p-2$
conventional computational bases of the Hilbert space $\{|Z_{2^{n}}\rangle
\} $. Apparently it can not see any difference between the additive cyclic
group state space $S(Z_{p-1})$ and the state subset $\{|Z_{p-1}\rangle \}$
of the Hilbert space if one does not care about the symmetric properties and
structures of the two state subsets in the quantum search problem. However,
their difference could be great if their symmetric properties and structures
are taken into consideration. For the multiplicative cyclic group state
space $S(C_{p-1})$ whose state bases are the modular exponentiation states $%
\{|g^{s}\func{mod}p\rangle \}$, one may easily imagine that there exists
difference between the two state subsets $S(C_{p-1})$ and $\{|Z_{p-1}\rangle
\}$. However, only from the symmetric property and structure of the cyclic
group state space can one understand deeply that the difference could lead
to a completely different result in quantum computation. According to the
fundamental theorem of arithmetic (see the Theorem 2 in Ref. [19]) the order 
$p-1$ of the multiplicative cyclic group $C_{p-1}$ which is also the
dimension of the cyclic group state space $S(C_{p-1})$ can be expressed as a
product of distinct primes, $p-1=p_{1}^{a_{1}}p_{2}^{a_{2}}...p_{r}^{a_{r}},$
where $p_{1},p_{2},...,p_{r}$ are distinct primes and the exponents $%
a_{1},a_{2},...,a_{r}>0$. Then correspondingly the cyclic group $C_{p-1}$
can be decomposed as a product of its factor cyclic subgroups (see Chapter
One and Two in Ref. [18]) : 
\begin{equation}
C_{p-1}=C_{p_{1}^{a_{1}}}\times C_{p_{2}^{a_{2}}}\times ...\times
C_{p_{r}^{a_{r}}},  \label{1}
\end{equation}
where the factor cyclic subgroup $C_{p_{k}^{a_{k}}}$ has an order $%
p_{k}^{a_{k}}$ for $k=1,2,...,r.$ Thus, the order $p-1$ of the cyclic group $%
C_{p-1}$ is a product of the orders $\{p_{k}^{a_{k}}\}$ of the factor cyclic
subgroups $\{C_{p_{k}^{a_{k}}}\}$. This shows that though the order $p-1$ of
the cyclic group $C_{p-1}$ can be a large number (even $p\thicksim 2^{n})$,
the orders $\{p_{k}^{a_{k}}\}$ of the factor cyclic subgroups $%
\{C_{p_{k}^{a_{k}}}\}$ may take much small numbers $\thicksim O(\log p)$.
Just like the cyclic group state space $S(C_{p-1})$ the cyclic group state
subspace $S(C_{p_{k}^{a_{k}}})$ ($k=1,2,...,r$) also can be formed by
mapping all the elements of the cyclic subgroup $C_{p_{k}^{a_{k}}}$ onto the
specific states of the Hilbert space, and it is a state subspace of the
cyclic group state space $S(C_{p-1})$ and also of the Hilbert space. Since
the dimensional size of a cyclic group state space is just the order of the
cyclic group, the cyclic group state subspaces $\{S(C_{p_{k}^{a_{k}}})\}$
may also have much small dimensional sizes $\thicksim O(\log p)$ even if the
whole cyclic group state space $S(C_{p-1})$ has a large dimension $(p$ $%
\thicksim 2^{n}).$ It is well known that a problem could be difficult to be
solved in a large dimension, but generally it may be fast solved in a small
dimension even in classical computation. Since the quantum search speed for
a search problem is generally inversely proportional to the square root of
the dimensional size of the problem [2, 6], then the quantum search problem
could be efficiently solved even in the whole cyclic group state space $%
S(C_{p-1})$\ if it could be efficiently reduced from the whole cyclic group
state space to the state subspaces $\{S(C_{p_{k}^{a_{k}}})\}$ of the factor
cyclic subgroups $\{C_{p_{k}^{a_{k}}}\}$. Thus, the main purpose in the
paper is how to achieve the reduction for the quantum search problem from
the cyclic group state space $S(C_{p-1})$\ to the cyclic group state
subspaces $\{S(C_{p_{k}^{a_{k}}})\}.$

A quantum state $|s\rangle $ of the additive cyclic group state space $%
S(Z_{p-1})$ may be described completely by the dynamical parameter vector $%
\{a_{k}^{s}\}$. Since $g^{s}\func{mod}p$ is an integer of the positive
integer set $Z_{p}^{+}$ any quantum state $|g^{s}\func{mod}p\rangle $ of the
cyclic group state space $S(C_{p-1})$ is a usual computational basis and
also can be described completely by the dynamical parameter vector $%
\{a_{k}^{t}\},$ where the parameter vector $\{a_{k}^{t}\}$ may not be equal
to the vector $\{a_{k}^{s}\}$ and the two vectors $\{a_{k}^{s}\}$ and $%
\{a_{k}^{t}\}$ are related by the one-to-one correspondence $%
s\leftrightarrow g^{s}\func{mod}p$. However, a better method to describe
completely a quantum state $|g^{s}\func{mod}p\rangle $ of the multiplicative
cyclic group state space $S(C_{p-1})$ could be to use the multi-base
dynamical representation in quantum computation. Notice that the cyclic
group $C_{p-1}$ is a product of the cyclic subgroups $\{C_{p_{k}^{a_{k}}}\}$%
, each of which has an order $p_{k}^{a_{k}}$. Suppose that the cyclic
subgroup $C_{p_{k}^{a_{k}}}$ is generated by a generator $g_{k},$ that is, $%
C_{p_{k}^{a_{k}}}=\langle g_{k}\rangle .$ Then any element of the cyclic
subgroup $C_{p_{k}^{a_{k}}}$ can be generally written as $g_{k}^{l_{k}}$ for
the index $l_{k}=0,1,...,p_{k}^{a_{k}}-1.$ Corresponding to the product
decomposition (1) for the cyclic group $C_{p-1}$ each group element $g^{s}$
of the cyclic group $C_{p-1}$ is also a product of the group elements $%
\{g_{k}^{s_{k}}\}$ of the factor cyclic subgroups $\{C_{p_{k}^{a_{k}}}\},$ 
\newline
\begin{equation}
g^{s}=g_{1}^{n_{1}s_{1}}\times g_{2}^{n_{2}s_{2}}\times ...\times
g_{r}^{n_{r}s_{r}},  \label{2}
\end{equation}
where the generator $g_{k}$ of the subgroup $C_{p_{k}^{a_{k}}}$ can written
as $g_{k}=g^{M_{k}}\func{mod}p$ [18] for $k=1,2,...,r$ and the positive
integers $M_{k}$ and $n_{k}$ will be determined later. The product
decomposition (2) for a group element of the cyclic group $C_{p-1}$ is
really written according to the Chinese remainder theorem (see the Theorem
121 in Ref. [19]). Actually, there exists a one-to-one correspondence
between the index $s$ of the group element $g^{s}$ of the cyclic group $%
C_{p-1}$ and the index vector $\{s_{k}\}$ of the group elements $%
\{g_{k}^{n_{k}s_{k}}\}$ of the factor cyclic subgroups $\{C_{p_{k}^{a_{k}}}%
\}.$ Note that the order $p-1$ of the cyclic group $C_{p-1}$ is decomposed
as a product of the distinct prime factors $\{p_{k}^{a_{k}}\}:$ $%
p-1=p_{1}^{a_{1}}p_{2}^{a_{2}}...p_{r}^{a_{r}}$. For convenience, here
denote that $m_{k}=p_{k}^{a_{k}}$ for $k=1,2,...,r$, and $%
p-1=m=m_{1}m_{2}...m_{r}.$ Evidently, the integers $\{m_{k}\}$ are coprime
to one another in pair, that is, the highest common divisor for any pair of
the integers $m_{i}$ and $m_{j}$ equals one: $(m_{i},m_{j})=1$ for $1\leq
i<j\leq r.$ Since the index $s=s\func{mod}(p-1)$ and if the index $s_{k}$ is
written as $s_{k}=s\func{mod}m_{k}$ for $k=1,2,...,r,$ then the index $s$
can be uniquely expressed as a linear combination ($\func{mod}(p-1)$) of the
indices $\{s_{k}\}$ according to the Chinese remainder theorem [19], 
\begin{equation}
s=(n_{1}M_{1}s_{1}+n_{2}M_{2}s_{2}+...+n_{r}M_{r}s_{r})\func{mod}(p-1),
\label{3}
\end{equation}
where $p-1=m_{k}M_{k}$ for $k=1,2,...,r.$ Note that $(m_{k},M_{k})=1.$ The
integer $n_{k}$ is just the multiplicative inverse to the integer $M_{k}$ ($%
\func{mod}m_{k}$) that satisfies $n_{k}M_{k}=1\func{mod}m_{k}.$ Using the
known integers $m_{k}$ and $M_{k}$ one can efficiently calculate the integer 
$n_{k}$ by the Euclidean algorithm [20]. Because the integer $M_{k}$
satisfies $M_{k}=(p-1)/m_{k},$ that is, $M_{k}$ is a divisor of the order $%
p-1$, it follows from the Theorem 1.4.3 in Ref. [18] that the generator $%
g_{k}$ of the cyclic subgroup $C_{p_{k}^{a_{k}}}$ is just $g_{k}=g^{M_{k}}%
\func{mod}p$ and the order of the subgroup $C_{p_{k}^{a_{k}}}$ is $%
p_{k}^{a_{k}}$ and hence the cyclic subgroup is written as $%
C_{p_{k}^{a_{k}}}=\langle g^{M_{k}}\func{mod}p\rangle .$ Then the state
subspace of the cyclic subgroup $C_{p_{k}^{a_{k}}}=\langle g^{M_{k}}\func{mod%
}p\rangle $ is given by 
\[
S(C_{p_{k}^{a_{k}}})=\{|(g^{M_{k}})^{l_{k}}\func{mod}p\rangle
,l_{k}=0,1,...,p_{k}^{a_{k}}-1\}. 
\]
The dimension of the state subspace $S(C_{p_{k}^{a_{k}}})$ is just the order 
$p_{k}^{a_{k}}$ of the cyclic subgroup $C_{p_{k}^{a_{k}}}.$ The cyclic group
state subspace $S(C_{p_{k}^{a_{k}}})$ is an invariant subspace under the
action of any group operation $g_{k}^{l_{k}}$ of the cyclic subgroup $%
C_{p_{k}^{a_{k}}}.$

Given a set of the indices $\{s_{k}\}$ for the group elements $%
\{(g^{M_{k}})^{n_{k}s_{k}}\func{mod}p\}$ of the factor cyclic subgroups $%
\{C_{p_{k}^{a_{k}}}\}$ with the generators $\{g^{M_{k}}\func{mod}p\}$, here
the integers $\{n_{k}\}$ are known, then according to the equations (2) one
can compose a unique group element $g^{s}\func{mod}p$ for the cyclic group $%
C_{p-1}$ with the index $s$ determined by the equation (3). In turn, if one
is given any element $g^{s}\func{mod}p$ of the cyclic group $C_{p-1}$ with
the index $s,$ then according to the equation (2) the element can be
decomposed uniquely as a product of the elements $\{(g^{M_{k}})^{n_{k}s_{k}}%
\func{mod}p\}$ of the cyclic subgroups $\{C_{p_{k}^{a_{k}}}\}$ and the
indices $\{s_{k}\}$ are given by $s_{k}=s\func{mod}m_{k}.$ Therefore, there
is a one-to-one correspondence between the index $s$ of the group element $%
g^{s}\func{mod}p$ of the cyclic group $C_{p-1}$ and the index set $\{s_{k}\}$
of the group elements $\{(g^{M_{k}})^{n_{k}s_{k}}\func{mod}p\}$ of the
factor cyclic subgroups $\{C_{p_{k}^{a_{k}}}\}$. This one-to-one
correspondence shows that one can also use the set of indices $\{s_{k}\}$ to
describe completely the index state $|s\rangle $ and the cyclic group state $%
|g^{s}\func{mod}p\rangle $ as well in addition to the dynamical parameter
vector $\{a_{k}^{s}\}$. Furthermore, because $s_{k}=s\func{mod}p_{k}^{a_{k}}$
the index $s_{k}$ can be expanded in the field $GF(p_{k}^{a_{k}})$ [19, 20,
21], 
\begin{equation}
s_{k}=s\func{mod}p_{k}^{a_{k}}=\stackrel{a_{k}-1}{\stackunder{l=0}{\sum }}%
h_{kl}^{s}p_{k}^{l},  \label{4}
\end{equation}
where the coefficients $\{h_{kl}^{s}\}$ satisfy $0\leq h_{kl}^{s}<p_{k}$ for 
$l=0,1,...,a_{k}-1$ and $k=1,2,...,r.$ This expansion could be thought of as
the $p_{k}-$base expansion for the index $s_{k}$ similar to the conventional
binary expansion for a number. Clearly, given the prime $p_{k}$ the index $%
s_{k}=s\func{mod}p_{k}^{a_{k}}$ is determined uniquely by the coefficients $%
h_{kl}^{s}$ for $l=0,1,...,a_{k}-1.$ Therefore, it is needed $r$ indices $%
\{s_{k}\}$ or $\sum_{k=1}^{r}a_{k}$ coefficients $\{h_{kl}^{s}\}$ to
describe completely the index state $|s\rangle $ or the cyclic group state $%
|g^{s}\func{mod}p\rangle ,$ while in the binary representation it need only $%
n$ parameters $\{a_{k}^{s}\}$ for the complete description for the index
state $|s\rangle $ in the Hilbert space of an $n-$qubit quantum system. It
seems to be that the multi-base representation $\{h_{kl}^{s}=0,1,...,p_{k}-1%
\}$ or the index vector $\{s_{k}\}$ for the index state $|s\rangle $ or the
cyclic group state $|g^{s}\func{mod}p\rangle $ is more complicated than the
binary representation $\{a_{k}^{s}=+1,-1\}$ in the Hilbert space. However,
the importance is that the multi-base representation $\{h_{kl}^{s}\}$ or the
index vector $\{s_{k}\}$ is related to the symmetric property and structure
of the cyclic group $C_{p-1}$, while this symmetric property and structure
could have a great impact on the quantum computation that is carried out in
the cyclic group state space. Thus, it could be better in the quantum search
problem in the cyclic group state space that the binary dynamical
representation $\{a_{k}^{s}\}$ is replaced with the index vector $\{s_{k}\}$
or the multi-base dynamical representation $\{h_{kl}^{s}\}$ to represent
completely the quantum states and to describe the quantum dynamics of the
quantum search process. Now in the cyclic group state space the quantum
search process to find the marked state is just to determine completely the
index vector $\{s_{k}\}$ or the parameter vector $\{h_{kl}^{s}\}$, which is
the same process as the previous one that searching for the marked state is
just to determine the dynamical parameter vector $\{a_{k}^{s}\}$ [15]. 
\newline
\newline
{\large 2.2. The oracle unitary operation acting on the cyclic group state
space}

The quantum search process in the cyclic group state space may be carried
out either in the additive cyclic group state space $S(Z_{p-1})$ or in the
multiplicative cyclic group state space $S(C_{p-1})$. Correspondingly the
marked state of the search problem can be represented either by the index
state $|s\rangle $ of the additive cyclic group state space $S(Z_{p-1})$ or
the modular exponentiation state $|g^{s}\func{mod}p\rangle $ of the
multiplicative cyclic group state space $S(C_{p-1})$. The index state $%
|s\rangle $ and the modular exponentiation state $|g^{s}\func{mod}p\rangle $
can be efficiently converted into each other by a unitary transformation
which is given in next section. It might be more convenient that the quantum
search process is carried out in the multiplicative cyclic group state space 
$S(C_{p-1})$ as the multiplicative unitary operations usually are easily
constructed and used. Suppose that the quantum search process is used to
solve a specific problem such as an NP problem which has only one solution
and the possible solution to the problem is within the integer set $%
Z_{p}^{+}=\{g^{k}\func{mod}p\}=\{1,2,...,p-1\},$ here assume that the
possible solution can be represented with the integer index variable $x\in
Z_{p}^{+}$. In the quantum search problem there is a block box or an oracle
to compute a function $f(x)$ for the variable $x=g^{k}\func{mod}p\in
Z_{p}^{+}.$ If the variable $x=g^{s}\func{mod}p$ is the solution to the
problem, then the function $f(x)=f(g^{s}\func{mod}p)=1$; otherwise, $f(x)=0$%
. The quantum computational process to compute the function $f(x)$ in the
block box can be represented by a unitary operation. This basic unitary
operation is called the oracle unitary operation. It is the unique unitary
operation that can access directly the unknown marked state $|g^{s}\func{mod}%
p\rangle $ in the quantum search problem, where the marked state corresponds
to the unique solution $x=g^{s}\func{mod}p$ to the problem. Generally, if
the marked state is defined as $|g^{s}\func{mod}p\rangle ,$ then the
corresponding oracle unitary operation $U_{o}=U_{os}(\theta )$ in the cyclic
group state space $S(C_{p-1})$ can be defined by 
\begin{eqnarray*}
U_{os}(\theta )|g^{x}\func{mod}p\rangle |a\rangle &=&\exp [-i\theta f(g^{x}%
\func{mod}p)]|g^{x}\func{mod}p\rangle |a\rangle \\
&=&\left\{ 
\begin{array}{c}
\exp (-i\theta )|g^{s}\func{mod}p\rangle |a\rangle ,\text{ if }x=s \\ 
|g^{x}\func{mod}p\rangle |a\rangle ,\text{ if }x\neq s
\end{array}
\right.
\end{eqnarray*}
where the auxiliary state $|a\rangle =\frac{1}{\sqrt{2}}(|0\rangle
-|1\rangle ).$ According to this definition the oracle unitary operation $%
U_{os}(\theta )$ is really equivalent to the selective rotation operation in
the cyclic group state space $S(C_{p-1}):$%
\[
U_{os}(\theta )=\exp [-i\theta D_{s}(g)]. 
\]
Here, the quantum-state diagonal operator $D_{s}(g)$ [15] which is applied
to the cyclic group state space $S(C_{p-1})$ can be generally expressed in
terms of the cyclic group state, 
\[
D_{s}(g)=|g^{s}\func{mod}p\rangle \langle g^{s}\func{mod}p|. 
\]
Note that the diagonal operator $D_{s}(g)$ is different from the
conventional one $D_{s}=|s\rangle \langle s|$ in the Hilbert space of an $n-$%
qubit quantum system. Actually, the diagonal operator $D_{s}(g)$ can also be
expressed in terms of the dynamical parameter vector $\{b_{k}^{s}\}$, 
\[
D_{s}(g)=\stackrel{n}{\stackunder{k=1}{\bigotimes }}(\frac{1}{2}%
E_{k}+b_{k}^{s}I_{kz}), 
\]
but here the vector $\{b_{k}^{s}=\pm 1\}$ corresponds to the state $|g^{s}%
\func{mod}p\rangle ,$ while the conventional vector $\{a_{k}^{s}=\pm 1\}$ is
assigned to the index state $|s\rangle $ and the state $D_{s}=|s\rangle
\langle s|.$ Through the quantum-state diagonal operator $D_{s}(g)$ one can
set up one-to-one correspondence between the oracle unitary operation $%
U_{os}(\theta )$ and the cyclic group state $|g^{s}\func{mod}p\rangle .$
This correspondence makes it possible to calculate explicitly the time
evolution of a quantum system under the action of the oracle unitary
operation $U_{os}(\theta )$ [15]$,$ and it may also provide a convenience
for manipulating at will the time evolution of a quantum system by the
oracle unitary operation. If the auxiliary state is taken as $|a\rangle
=|0\rangle ,$ then the oracle unitary operation $U_{o}$ is simply defined as 
\begin{eqnarray*}
U_{os}|g^{x}\func{mod}p\rangle |0\rangle &=&|g^{x}\func{mod}p\rangle |f(g^{x}%
\func{mod}p)\rangle \\
&=&\left\{ 
\begin{array}{c}
|g^{s}\func{mod}p\rangle |1\rangle ,\text{ if }x=s \\ 
|g^{x}\func{mod}p\rangle |0\rangle ,\text{ if }x\neq s
\end{array}
\right. .
\end{eqnarray*}
The quantum search problem in the cyclic group state space is how to find
the marked state $|g^{s}\func{mod}p\rangle ,$ given the oracle unitary
operation $U_{os}.$ This is different from the quantum discrete logarithmic
problem which states that given a positive integer $\varphi _{s}$ such that $%
\varphi _{s}=g^{s}\func{mod}p,$ how to compute the index $s,$ while the
quantum search problem is really equivalent to that given the oracle unitary
operation $U_{os},$ how to determine the index $s.$

The conventional quantum search process usually is carried out in the $%
2^{n}- $dimensional Hilbert space of a single $n-$qubit quantum system. If
besides the given work register used for searching task there are also other
auxiliary registers, then the oracle unitary operation $U_{o}$ could be
thought of as a non-selective oracle unitary operation with respect to any
states of those auxiliary registers, since in addition to the auxiliary
state $|a\rangle ,$ which loads the functional values $f(g^{x}\func{mod}p)$,
the oracle unitary operation $U_{o}$ can only apply to the work register. If
there are any other auxiliary registers the oracle unitary operation $U_{o}$
will not make any effect on any states of all these auxiliary registers. If
the quantum search process is carried out in such a multi-register quantum
system that contains a work register and several auxiliary registers in
addition to the auxiliary state $|a\rangle $ and each register could consist
of a single $n-$qubit quantum subsystem, then in order that the search space
still has the same dimensional size as before all the states of the
auxiliary registers should be set to a given state, e.g., the state $|%
\mathbf{R0}\rangle =|00...0\rangle .$ This search space is really a small
subspace of the whole Hilbert space of the multi-register quantum system.
Corresponding to this search subspace the subspace-selective oracle unitary
operation $\overline{U}_{o}$ should be defined by 
\[
\overline{U}_{os}(\theta )|\Psi \rangle |g^{x}\func{mod}p\rangle |a\rangle
=\left\{ 
\begin{array}{c}
\exp (-i\theta )|\mathbf{R0}\rangle |g^{s}\func{mod}p\rangle |a\rangle ,%
\text{ } \\ 
\qquad \qquad \qquad \qquad \text{if }x=s\text{ and }|\Psi \rangle =|\mathbf{%
R0}\rangle ; \\ 
|\mathbf{R0}\rangle |g^{x}\func{mod}p\rangle |a\rangle ,\text{ } \\ 
\qquad \qquad \qquad \quad \quad \text{if }x\neq s\text{ and }|\Psi \rangle
=|\mathbf{R0}\rangle ; \\ 
|\Psi \rangle |g^{x}\func{mod}p\rangle |a\rangle ,\text{ if }|\Psi \rangle
\neq |\mathbf{R0}\rangle ;
\end{array}
\right. 
\]
where the states $|\Psi \rangle $ and $|\mathbf{R0}\rangle $ belong to the
auxiliary registers and the oracle unitary operation works in the cyclic
group state space $S(C_{p-1})$ of the work register. Here $|\mathbf{R0}%
\rangle $ also denotes the auxiliary register library with the specific
state $|00...0\rangle $ (see next sections). The subspace-selective oracle
unitary operation $\overline{U}_{os}(\theta )$ acts on selectively the state 
$|\mathbf{R0}\rangle $ but does not have any effect on any other state $%
|\Psi \rangle $ of the auxiliary registers. It is really equivalent to the
selective rotation operation in the Hilbert space of the multi-register
quantum system where the work register is in the cyclic group state space $%
S(C_{p-1})$, 
\[
\overline{U}_{os}(\theta )=\exp [-i\theta \overline{D}_{s}(g)] 
\]
with the diagonal operator $\overline{D}_{s}(g)=|\mathbf{R0}\rangle |g^{s}%
\func{mod}p\rangle \langle g^{s}\func{mod}p|\langle \mathbf{R0}|.$ Why using
many auxiliary registers here? This is mainly because the conventional
mathematical-logic gates usually need to use a large space to perform their
reversible operations, while these mathematical-logic gates have been used
extensively in constructing the quantum search algorithms in the Hilbert
space [16] and also in the cyclic group state space (see next sections).
However, it must be careful as there could be a potential risk that the
auxiliary registers could enlarge greatly the search space for the quantum
search problem and as a result the quantum search process could become
degraded.

The effect of the oracle unitary operation $U_{o},$ which acts on only the
marked state, on the evolution process of an $n-$qubit quantum system is so
small that it is hard to be detected quantum mechanically when the qubit
number $n$ is large [2, 3, 6, 15]. This results in that the quantum search
problem generally is hard to be solved in a large Hilbert space. Any
superposition of the Hilbert space with dimension $N=2^{n}$ could be
converted partly into the marked state under the action of the oracle
unitary operation associated with other known quantum operations, but each
time for the action this conversion efficiency of the marked state is
proportional to $1/\sqrt{N}$ [2, 3, 6]. In order to achieve an observable
amplitude for the marked state a standard quantum search algorithm needs to
call $\thicksim \sqrt{N}$ times the oracle unitary operation and thus, the
quantum search time to find the marked state with a high probability ($%
\thicksim 1$) is proportional to the square root $(\sqrt{N})$ of the
dimensional size $(N)$ of the search space which here is the whole Hilbert
space. This search time therefore increases exponentially as the qubit
number $n.$ This low amplitude-amplification efficiency results in that a
standard quantum search algorithm usually can have only a square speedup
over the best known classical counterparts, and it has been also shown that
this square speedup is optimal and hence can not be further improved
essentially [3, 6, 9, 13]. A number of quantum search algorithms [2-13] have
been proposed to achieve this optimal efficiency (with respect to the
dimensional variable $N$) which include the standard Grover search algorithm
[2], the amplitude-amplification search algorithm [6], and the quantum
adiabatic search algorithm [4, 5]. All these search algorithms are based on
the quantum-state tomography. A direct quantum measurement on the marked
state is necessary to output the information of the marked state in these
search algorithms and hence it is required in algorithm that the amplitude
of the marked state be first amplified by a suitable unitary sequence that
contains $\thicksim \sqrt{N}$ oracle unitary operations so that the
probability for the marked state is high enough $(\thicksim 1)$ for
observation. In recent years a great effort has been made to develop other
type of quantum search algorithms [15, 16] in order to break through the
square speedup limitation. These quantum search algorithms are based on the
quantum dynamical principles. In these quantum-dynamical search algorithms a
direct measurement on the marked state may not be necessary so that a direct
amplification for the amplitude of the marked state could be avoided,
instead the quantum measurement to output the computing results could be
carried out on some other states that are closely related to the marked
state and carry the complete information of the marked state and the
computing results are further used to obtain the complete information of the
marked state [15]. The basis behind the quantum-dynamical search algorithms
is that $(i)$ any quantum state such as the unknown marked state in the
Hilbert space can be described completely in a parameterization form by a
set of dynamical parameters and the quantum searching for the marked state
therefore is reduced to determining the set of the dynamical parameters; $%
(ii)$ by the set of the dynamical parameter the oracle unitary operation and
the unknown marked state is set up a one-to-one correspondence, and it
becomes possible to manipulate at will the evolution process of a quantum
system under the oracle unitary operation in the quantum search process.
This quantum search method may avoid a direct quantum measurement on the
marked state in the quantum search problem and hence it could not be
necessary to achieve an enough high probability for the marked state to be
observable. Regarding the fact that there is a low efficiency to amplify the
amplitude of the marked state by the oracle unitary operation and this
efficiency is closely related to the dimensional size of the search space of
the search problem, that is, the larger the search space, the lower the
efficiency, one simple scheme to increase the efficiency is that the search
space of the search problem is limited to a small subspace of the Hilbert
space [16]. This scheme is feasible only if the marked state is in the
subspace. Hence to make the scheme feasible one may convert the marked state
from the whole Hilbert space to the subspace. It is well known that the
structure of the Hilbert space of a quantum system generally is closely
related to the symmetric property and structure of the quantum system.
Because there is a rotation symmetry in spin space in the $n-$qubit quantum
spin system according to the angular momentum theory in quantum mechanics,
the Hilbert space of the $n-$qubit spin system can be divided into $(n+1)$
state subspaces. Then it can be shown that the quantum search problem in the
Hilbert space of the $n-$qubit spin system can be efficiently reduced from
the whole Hilbert space to the largest subspace of the $(n+1)$ state
subspaces [16]. The conventional quantum search process therefore may be
sped up, although this speedup is limited and does not change essentially
the computational complexity for the search problem. However, the importance
for the fact that the symmetric properties and structures of quantum systems
may be employed to speed up the quantum computational process is that one
may further use the symmetric property and structure of a group to help
solving the quantum search problem. This is just the main purpose of the
paper that the symmetric property and structure of a cyclic group are
employed to help solving the quantum search problem. \newline
\newline
{\large 2.3. The structural quantum search in the cyclic group state space}

The conventional unstructured and structural search problems are referred to
the problems themselves [7, 8]. The structure for a quantum search problem
in a cyclic group state space has a different sense from the conventional
one. It is referred to the symmetric structure of the cyclic group used to
help solving the unstructured quantum search problem in the Hilbert space of
the $n-$qubit quantum system. A cyclic group is one of the simplest groups.
It is an Abelian group and any two elements of a cyclic group are commutable
to one another. Its property and structure have been studied in detail and
extensively [18, 20]. As shown in equation (1), a cyclic group can be
decomposed as a direct product of its factor cyclic subgroups because every
Abelian group can be decomposed as a direct product of cyclic groups [18].
The cyclic groups of prime order are the only Abelian simple groups. They
have not any nontrivial and proper subgroup. A cyclic group of non-prime
order must have a nontrivial and proper subgroup at least. Here, that a
cyclic group is highly symmetric means that the cyclic group has many factor
cyclic subgroups. Thus, a highly symmetric cyclic group can be expressed as
a direct product of its factor cyclic subgroups, as shown in equation (1).
The quantum search problem in the cyclic group state space is either
unstructured or structural only dependent on the symmetric structure of the
cyclic group no matter what the quantum search problem itself is
unstructured or structural in the Hilbert space of the $n-$qubit quantum
system. If the quantum search is performed in a cyclic group state space
whose cyclic group has a prime order, then it is said to be an unstructured
quantum search. However, generally a quantum search is carried out in the
cyclic group state space of a highly symmetric cyclic group so that the
symmetric property and structure of the cyclic group can be employed to help
solving the quantum search problem. Therefore, the quantum search proposed
in the paper generally is structural in the cyclic group state space.
Generally, the Hilbert space of the $n-$qubit quantum system does not have
some specific group symmetric properties and structures, but a specific and
artificial state subset of the Hilbert space which could be formed by
mapping all the group elements of a specific group such as a cyclic group
onto the Hilbert space may have the symmetric property and structure of the
group. Then quantum computation which is carried out on the state subset may
be affected greatly by the group symmetric property and structure.
Consequently, though the quantum search problem in the Hilbert space of the $%
n-$qubit quantum system is unstructured, it is affected inevitably by the
symmetric property and structure of the group if it can be reduced to and
therefore is solved in the group state space. The effect of the group
symmetric properties and structures could lead to a significant speedup for
some quantum computation processes. How the cyclic group symmetric property
and structure influence on the speedup of the unstructured quantum search
process in the whole Hilbert space is important research project that comes
to be investigated in detail in the paper and in the future work. \newline
\newline
{\large 3. The efficient state transformation between the additive and
multiplicative cyclic group state spaces}

In the quantum factoring problem and the discrete logarithmic problem a
number of reversible mathematical-logic operations such as the modular
addition, modular multiplication, and modular exponentiation unitary
operations have been used extensively [22, 23, 24, 25]. In the reversible
computation one basic principle to construct a reversible mathematical-logic
operation is that all the input states are also kept together with the
output states after the logic operation [26, 27]. A mathematical-logic
operation usually needs to use many auxiliary registers so that the
operation process can be made reversible. Both the classical irreversible
computation and the reversible computation usually are equivalent in
computational complexity in time and space [27, 28]. Therefore, the
classical irreversible computation generally can be efficiently simulated by
the reversible one. The reversible logic operations can be performed in a
quantum system as well, but they usually consume much more qubits in space
than the conventional unitary operators quantum mechanically in the quantum
system. They could influence on the unitary evolution process of a quantum
system in a different manner from the conventional unitary operators quantum
mechanically. This is because a reversible logic operation usually acts on
only some specific states of the quantum system, while the conventional
unitary operators quantum mechanically usually act on any states of the
quantum system. It must be careful to use a reversible logic operation to
manipulate the quantum dynamical process of a quantum system. A reversible
mathematical-logic operation could be thought of as a selective unitary
operation of a quantum system because there are usually a number of quantum
states of the quantum system independent of the action of the logic
operation. Quantum physically there are a number of unitary evolution
pathways in a multi-qubit quantum system, but under the reversible
mathematical-logic operations there are only few unitary evolution pathways
to be allowed in the quantum system. This just shows that the mathematical
principles can make constraints on the unitary evolution process of a
quantum system. On the other hand, quantum computation is a physical process
or exactly a unitary evolution process quantum physically, as pointed out by
Deutsch [29, 38]. Therefore, the quantum computational process for a given
problem obeys not only the quantum physical laws but also is compatible with
the used mathematical principles and its computational complexity not only
is dependent on the quantum dynamical process but also on the used
mathematical principles. One large advantage to use the reversible
mathematical-logic operations in solving some mathematical problems is that
one could easily trace the unitary evolution pathways for some quantum
states in the Hilbert space of the quantum system under the action of the
logic operations.

The discrete logarithmic problem is an important problem in classical public
secure key cryptography [21]. It can be stated that given an integer $%
b=a^{s}>0$, how to calculate the discrete logarithmic function $s=(\log
_{a}b)\func{mod}p$, which is also called the index of the discrete
logarithm, where the positive integer $a$ is a given logarithmic base and $p$
a known prime. In the classical computation it is hard to calculate the
logarithmic function of a large integer $b$. This is the basis for the
classical public key cryptographic systems based on the discrete logarithm
[21]. It has been shown [22, 25, 30] that the discrete logarithmic problem
can be solved in polynomial time in quantum computation. Shor first gave an
efficient quantum algorithm to calculate the index of the discrete logarithm
[22]. Later this quantum algorithm was improved in a determination form
[30a] with the help of the amplitude amplification method [6]. Here, with
the help of these quantum algorithms [6, 22, 25, 30a] an efficient unitary
sequence is constructed to generate the index state $|s\rangle $ of the
discrete logarithm from the modular exponentiation state $|g^{s}\func{mod}%
p\rangle $. By this efficient unitary sequence any quantum state $|g^{s}%
\func{mod}p\rangle $ of the multiplicative cyclic group state space $%
S(C_{p-1})$ can be efficiently converted into the corresponding index state $%
|s\rangle $ of the additive cyclic group state space $S(Z_{p-1})$. In
constructing this efficient unitary sequence many efficient
mathematical-logic operations have been employed extensively, such as the
modular exponentiation operation, the modular multiplication operation, and
the quantum Fourier transform and so on, and some mathematical knowledge of
number theory are also used necessarily. Because the index $s$ and the
modular exponential function $f(s)=g^{s}\func{mod}p$ have a one-to-one
correspondence, there exists a unitary operator $U_{\log }(g)$ such that $%
U_{\log }(g)|g^{s}\func{mod}p\rangle =|s\rangle $ and $U_{\log
}^{+}(g)|s\rangle =|g^{s}\func{mod}p\rangle ,$ here $g$ is the logarithmic
base and also a primitive root ($\func{mod}p$) or a generator of the
multiplicative cyclic group $C_{p-1}$. Note that here there is not any extra
auxiliary register to be used by the unitary operator $U_{\log }(g)$.
Generally, such a discrete logarithmic unitary operator $U_{\log }(g)$ is
hard to be constructed. However, with the help of the reversible
computational techniques [22, 23, 24, 26, 27] an alternative construction to
the discrete logarithmic unitary operator $U_{\log }(g)$ could be achieved
conveniently by using many extra auxiliary registers. The construction can
be divided into two steps [23, 24, 27]. One step is to construct by using
two registers the modular exponentiation unitary operation: $V_{f}|s\rangle
|0\rangle =|s\rangle |g^{s}\func{mod}p\rangle .$ It is well known that the
modular exponentiation unitary operation can be efficiently built up in the
reversible computation [21, 22, 23, 24]. Another is to construct the unitary
operation of inversion function of the modular exponential function: $%
V_{f^{-1}}|g^{s}\func{mod}p\rangle |0\rangle =|g^{s}\func{mod}p\rangle
|s\rangle ,$ here also by using two registers$.$ Then the discrete
logarithmic unitary operation $U_{\log }(g)$ may be expressed equivalently
by $U_{\log }(g)=V_{f}^{+}SV_{f^{-1}},$ where the SWAP unitary operation $S$
is defined by $S|s\rangle |g^{s}\func{mod}p\rangle =|g^{s}\func{mod}p\rangle
|s\rangle .$ This is due to the fact that there holds $U_{\log }(g)|g^{s}%
\func{mod}p\rangle |0\rangle =V_{f}^{+}SV_{f^{-1}}|g^{s}\func{mod}p\rangle
|0\rangle =|s\rangle |0\rangle ,$ which further indicates that by omitting
the auxiliary register with the state $|0\rangle $ the unitary operation
sequence $V_{f}^{+}SV_{f^{-1}}$ is really the discrete logarithmic unitary
operation $U_{\log }(g).$ In effect the unitary operation sequence $%
V_{f}^{+}SV_{f^{-1}}$ is equivalent to the unitary operator $U_{\log }(g)$
of the discrete logarithm, but it must be careful when the unitary operation
sequence is performed in a quantum system since the unitary sequence
requires the auxiliary registers of the quantum system be in the specific
state $|0\rangle $ before and after the operation, while the quantum system
may be in any state. Though the modular exponentiation unitary operation $%
V_{f}$ can be built up efficiently [22, 23, 24], it is generally hard to
build up the unitary operation $V_{f^{-1}}$ of the inversion function of the
modular exponential function. It is this unitary operation $V_{f^{-1}}$ that
makes it hard to construct the discrete logarithmic unitary operation $%
U_{\log }(g)$. A functional unitary operation may exist but its
inversion-functional unitary operation could or could not, which usually is
dependent on the mathematical property of the function. Obviously, some
mathematical functions have their own unique inversion functions but some do
not have in some given functional or variable value ranges. If the functions
do not have their own unique inversion functions in some given value ranges,
then the unitary operations for the inversion functions usually could not
exist uniquely in these value ranges although the functions may have their
own unitary operations. When both a function and its inversion function
exist in a given value range, they usually have their own unitary
operations, respectively, and sometime their unitary operations are the same
up to the conjugate relation. But in general a functional and its
inversion-functional unitary operations can be different completely. As the
modular exponential function $f(s)=g^{s}\func{mod}p$ and its index variable $%
s$ have a one-to-one correspondence, this makes the modular exponential
function $f(s)$ and its inversion function, i.e., the discrete logarithmic
function or the index variable $s$, have their own unitary operations.
Because the unitary operation $V_{f}$ can be built up efficiently the
discrete logarithmic unitary sequence $U_{\log }(g)=V_{f}^{+}SV_{f^{-1}}$ is
mainly dependent on the unitary operation $V_{f^{-1}}$ in computational
complexity. Below it is devoted to the construction for the efficient
unitary operation $V_{f^{-1}}$ of inversion function of the modular
exponential function. Before building up the unitary operation $V_{f^{-1}}$
several conventional reversible mathematical-logic operations are introduced.

$(i)$ The modular addition operation $ADD_{L}(\alpha ,\beta )$. The modular
addition operation is defined as 
\[
ADD_{L}(1,2)|x\rangle |y\rangle =|x\rangle |x+y\func{mod}L\rangle ,\text{ }%
x,y\in Z_{L}. 
\]
Here the integer set $Z_{L}=\{0,1,...,L-1\}.$ The indices $\alpha $ and $%
\beta $ denote the registers that are acted on by the modular addition
operation $ADD_{L}(\alpha ,\beta )$. The modular addition operation $%
ADD_{L}(1,2)$ is performed by adding the integer $x$ of the first register
to the second register and taking modulus $L$. It can be implemented in
polynomial time $\thicksim O(\log L)$ [22]. The modular addition operation $%
ADD_{L}(1,2)$ is a reversible operation since the integer $y$ can be derived
uniquely from $x$ and $(x+y)\func{mod}L$ if $0\leq x,y\leq L-1.$ As a
specific modular addition unitary operation the COPY unitary operation $%
COPY(\alpha ,\beta )$ is defined as 
\[
COPY(1,2)|x\rangle |0\rangle =|x\rangle |x\rangle ,\text{ }x\in Z_{L}. 
\]
The inverse COPY unitary operation $[COPY(\alpha ,\beta )]^{+}$ is really
the subtraction unitary operation: $[COPY(1,2)]^{+}|x\rangle |x\rangle
=|x\rangle |0\rangle ,$ $x\in Z_{L}.$

$(ii)$ The modular multiplication unitary operation $M_{L}(\alpha ,\beta
,\gamma )$ is defined as 
\[
M_{L}(1,2,3)|x\rangle |y\rangle |0\rangle =|x\rangle |y\rangle |xy\func{mod}%
L\rangle ,\text{ }x,y\in Z_{N} 
\]
where $x,y$ are integer variables, the integer $L$ is modulus and the
integer $N$ may be different from $L$. The indices $\alpha $ and $\beta $
denote the two registers whose integer variables $x$ and $y$ are multiplied
to one another and the index $\gamma $ marks the third register that loads
the multiplication operation result. As an example, the modular
multiplication unitary operation is applied on the cyclic group state: 
\begin{eqnarray*}
&&M_{p}(1,2,3)|g^{x}\func{mod}p\rangle |g^{y}\func{mod}p\rangle |0\rangle \\
&=&|g^{x}\func{mod}p\rangle |g^{y}\func{mod}p\rangle |g^{x+y}\func{mod}%
p\rangle ,\text{ }x,y\in Z_{p-1}.
\end{eqnarray*}

$(iii)$ The modular exponentiation unitary operation. First consider the
modular multiplication operation $U_{a,N}(\alpha )$ which is defined as 
\[
U_{a,N}(1)|x\rangle =|xa\func{mod}N\rangle ,\text{ }(a,N)=1,\text{ }x\in
Z_{N}. 
\]
This operation is unitary only when the integer $a$ is coprime to the
integer $N$ [17, 31]$,$ i.e., $(a,N)=1$. This unitary operation need not any
additional auxiliary register in principle, But when the unitary operation
is constructed by the mathematical-logic operations it still needs many
extra auxiliary registers. The index $\alpha $ denotes the register acted on
by the operation $U_{a,N}(\alpha )$. Generally, the modular exponentiation
operation may be taken as $[U_{a,N}(\alpha )]^{l}$ for any positive integer $%
l.$ The conditional modular exponentiation operation $U_{a,L}^{c}(\alpha
,\beta )$ may be defined with the help of the modular multiplication
operation $U_{a,L}(\beta ):$%
\begin{eqnarray*}
U_{a,L}^{c}(1,2)|x\rangle |y\rangle &=&|x\rangle [U_{a,L}(2)]^{x}|y\rangle \\
&=&|x\rangle |ya^{x}\func{mod}L\rangle ,\text{ }x\in Z_{N},\text{ }y\in
Z_{L}.
\end{eqnarray*}
This conditional modular exponentiation operation is unitary only if the
integer $a$ is coprime to the integer $L$. However, using one more auxiliary
register a general conditional modular exponentiation operation, which is
unitary even if the integer $a$ is not coprime to the integer $L,$ may be
constructed by 
\[
U_{a,L}^{c}(1,2,3)|x\rangle |y\rangle |0\rangle =|x\rangle |y\rangle |ya^{x}%
\func{mod}L\rangle ,x,y\in Z_{N}. 
\]
In particular, the two-variable conditional modular exponentiation operation 
$U_{f}=U_{a,b,L}^{c}(1,2,3)$ have been used extensively in the discrete
logarithmic problem [22, 25, 30]: $U_{a,b,L}^{c}(1,2,3)|x\rangle |y\rangle
|0\rangle =|x\rangle |y\rangle |b^{x}a^{y}\func{mod}L\rangle ,$ $x,y\in
Z_{N},$ where $a$ and $b$ are constant integers and usually $N\geq L$. These
modular multiplication and modular exponentiation unitary operations may be
built up efficiently by the basic reversible logic operations [21-27] and
generally can be efficiently implemented in polynomial time $\thicksim
O(\log ^{2}N)$ and $\thicksim O(\log ^{3}N),$ respectively [22]. The qubit
number used to implement these modular exponentiation operations generally
is $\thicksim O(\log N)$ [21, 22, 23, 24].

Besides these conventional mathematical-logic unitary operations introduced
above mathematically or quantum physically many important unitary operators,
unitary operations, elementary propagators, or quantum gates also can be
employed in construction of a unitary sequence. A large advantage for the
type of unitary operations is that the unitary operations usually are
non-selective unitary operators and hence need not any auxiliary qubits. But
the artificial conditional unitary operations, which also can be thought of
as the selective unitary operations, may need few auxiliary qubits to help
achieving the specific conditional operations.

$(iv)$ The $SWAP$ unitary operation and other elementary quantum gates [32].
The $SWAP(\alpha ,\beta )$ unitary operation is defined as 
\[
SWAP(1,2)|x\rangle |y\rangle =|y\rangle |x\rangle ,x,y\in Z_{N}. 
\]

$(v)$ The quantum Fourier transforms in the Hilbert space. The conventional
quantum Fourier transform [22, 33, 34] usually is defined in the regular
Hilbert space $\{|Z_{N}\rangle \},$%
\begin{equation}
|l\rangle \stackrel{Q_{NFT}}{\rightarrow }\frac{1}{\sqrt{N}}\stackrel{N-1}{%
\stackunder{k=0}{\sum }}\exp [i2\pi kl/N]|k\rangle ,\text{ }k,l\in Z_{N}.
\label{5}
\end{equation}
For the integer $N=2^{n}$ the quantum circuit $Q_{NFT}$ for the quantum
Fourier transform is very simple and consists of $\thicksim O(n^{2})$ basic
quantum gates. Note that there is not any auxiliary qubit in construction of
the quantum circuit $Q_{2^{n}FT}$. For the case that the integer $N$ is not
a power of two the quantum circuit $Q_{NFT}$ also can be constructed with $%
\thicksim O(\log ^{2}N)$ basic quantum gates or even less [30, 31, 34, 35],
but many auxiliary qubits are needed in the construction of the quantum
circuit.

$(vi)$ The functional quantum Fourier transform. The functional quantum
Fourier transform is really the quantum Fourier transform applying to a
non-regulation state subspace of the Hilbert space. Because the functional
quantum Fourier transform is related closely to the unitary operation of the
inversion function of a function it could not be generally constructed
efficiently for any function. Suppose that the function $f(x)$ is a periodic
function: $f(x)=f(x+r),$ here $r$ is the period of the function. Then the
functional quantum Fourier transform $Q_{rft}$ for the periodic function $%
f(x)$ may be defined as [36] 
\begin{equation}
Q_{rft}|f(l)\rangle =\frac{1}{\sqrt{r}}\stackrel{r-1}{\stackunder{k=0}{\sum }%
}\exp [i2\pi kl/r]|f(k)\rangle ,\text{ }k,l\in Z_{r}.  \label{6}
\end{equation}
It can be shown that the functional quantum Fourier transform $Q_{rft}$ can
be constructed efficiently if both the unitary operations for the periodic
function $f(x)$ and its inversion function $f(x)^{-1}$ in the variable value
range $Z_{r}$ can be built up efficiently. Suppose that the functional and
its inversion-functional unitary operations are defined by $V_{f}|x\rangle
|0\rangle =|x\rangle |f(x)\rangle $ and $V_{f^{-1}}|f(x)\rangle |0\rangle
=|f(x)\rangle |x\rangle $ for $x\in Z_{r},$ respectively. Then the unitary
sequence for the invertible periodic function $f(x)$ is $%
U_{f}=V_{f^{-1}}^{+}SV_{f}$ which satisfies $U_{f}|x\rangle =|f(x)\rangle $
for $x\in Z_{r},$ here any auxiliary qubits are dropped and $S$ is the SWAP
operation. Using the invertible-function unitary sequence $U_{f}$ the
functional quantum Fourier transform $Q_{rft}$ is related to the
conventional $r-$base quantum Fourier transform $Q_{rFT}$ by 
\[
Q_{rft}=U_{f}Q_{rFT}U_{f}^{+}. 
\]
Thus, the quantum circuit for the functional quantum Fourier transform $%
Q_{rft}$ can be efficiently constructed if there is an efficient quantum
circuit for the unitary operation $U_{f}$ of the invertible function $f(x).$

$(vii)$ The group operations of a cyclic group. A cyclic group $G$ can be
generated by a generator $g,$ $G=\langle g\rangle
=\{E,g,g^{2},...,g^{n_{r}-1}\}.$ If the generator $g$ is a unitary operator
which is denoted as $U_{g}$ here, then all the group elements of the cyclic
group $G$ are also unitary operators. When the unitary cyclic group
operation $U_{g}$ is applied to a cyclic group state the unitary
transformation is given by 
\[
U_{g}|g^{x}\func{mod}p\rangle =|g^{x+1}\func{mod}p\rangle . 
\]
The unitary operation of the cyclic group may be built up efficiently with
the help of the diagonal and anti-diagonal unitary operators [16]. Actually,
just like the modular multiplication unitary operation $U_{a,L}(\alpha )$
the cyclic group operation $U_{g}$ could also be constructed efficiently by
using the basic reversible logic operations [26, 27], but this construction
needs many extra auxiliary qubits. The cyclic group operation $U_{g}$ can
also be performed in a conditional form 
\[
U_{g}^{c}|a\rangle |g^{x}\func{mod}p\rangle =|a\rangle |g^{x+a}\func{mod}%
p\rangle . 
\]

With the help of these efficient unitary operations mentioned above an
efficient unitary sequence will be built up below, by which the index state $%
|s\rangle $ of the discrete logarithm can be generated from the modular
exponentiation state $|g^{s}\func{mod}p\rangle $.

The oracle unitary operation in the discrete logarithmic problem is the
usual conditional modular exponentiation operation $U_{f}=U_{b,g,p}^{c}(%
\alpha ,\beta ,\gamma ):$%
\[
U_{f}|x\rangle |y\rangle |b\rangle |g\rangle |0\rangle =|x\rangle |y\rangle
|b\rangle |g\rangle |f(x,y)\rangle ,\text{ }x,y\in Z_{N}. 
\]
The double-variable modular exponential function $f(x,y)$ is defined by 
\[
f(x,y)=b^{x}g^{y}\func{mod}p 
\]
where the integer $b=g^{s}\func{mod}p>0$ with the index $s\in Z_{p-1}$. The
Fermat little theorem (the Theorem 71 in Ref. [19]) shows that there holds $%
a^{p-1}\equiv 1\func{mod}p$ for a prime $p$ and any integer $a$ that is not
divided by the prime $p.$ In particular, for the integer $a=g,$ $b,$ or even 
$g^{z}\func{mod}p$ with $z=sx+y$ for any integers $x$ and $y$ there also
holds $a^{p-1}\equiv 1\func{mod}p$ since $g$ is a primitive root ($\func{mod}%
p$). Thus, the modular exponential function $f(x,y)$ is a periodic function
with the period $p-1$ by the Fermat little theorem. Since the periodic
function $f(x,y)$ satisfies $f(x,y)\equiv f_{1}(sx+y)=g^{sx+y}\func{mod}p,$ $%
f_{1}(z)=f_{1}(z+(p-1))$ and also $f(x,y)=f(x+l,y-ls)$ for any integer $l$
[25] the Fourier transform of the functional state $|f(x,y)\rangle $
therefore takes the form 
\begin{eqnarray}
|\widetilde{f}(l_{1},l_{2})\rangle &=&|\widetilde{f}(l_{2}s\func{mod}%
(p-1),l_{2})\rangle \delta ((l_{2}s-l_{1})\func{mod}(p-1))  \nonumber \\
&=&\delta ((l_{2}s-l_{1})\func{mod}(p-1))  \nonumber \\
&&\times \frac{1}{p-1}\stackrel{p-2}{\stackunder{x=0}{\sum }}\stackrel{p-2}{%
\stackunder{y=0}{\sum }}\exp [i2\pi l_{2}(sx+y)/(p-1)]|f(x,y)\rangle .
\label{7}
\end{eqnarray}
The indices $l_{1}$ and $l_{2}$ in the Fourier transform state $|\widetilde{f%
}(l_{1},l_{2})\rangle $ must satisfy the relation $(l_{2}s-l_{1})=0\func{mod}%
(p-1)$ for $l_{1},l_{2}=0,1,...,p-2$ due to the fact that $%
f(x,y)=f(x+l,y-ls) $. In terms of the Fourier transform states (7) the
functional state $|f(x,y)\rangle $ is expressed as 
\begin{equation}
|f(x,y)\rangle =\frac{1}{p-1}\stackrel{p-2}{\stackunder{l=0}{\sum }}\exp
[-i2\pi l(sx+y)/(p-1)]|\widetilde{f}(ls,l)\rangle .  \label{8}
\end{equation}
If one looks the function $f(x,y)$ as the single-variable periodic function $%
f_{1}(z)=g^{z}\func{mod}p$ with the variable $z=sx+y=0,1,...,p-2$, $%
f_{1}(z)=f_{1}(z+p-1),$ then one can express the functional state $%
|f(x,y)\rangle =|f_{1}(z)\rangle $ in terms of its Fourier transform states $%
\{|\widetilde{f}_{1}(l)\rangle \},$%
\[
|f_{1}(z)\rangle =\frac{1}{\sqrt{p-1}}\stackrel{p-2}{\stackunder{l=0}{\sum }}%
\exp [-i2\pi lz/(p-1)]|\widetilde{f}_{1}(l)\rangle . 
\]
By comparing it with equation (8) one can see that there holds the state
identity $|\widetilde{f}(ls,l)\rangle /\sqrt{p-1}=|\widetilde{f}%
_{1}(l)\rangle $ for $l=0,1,...,p-2$ and equation (7) is indeed the Fourier
transform of the functional state $|f_{1}(z)\rangle $ (its explanation can
be seen later).

The functional Fourier transform states (7) and the functional states (8)
will be used below in building up the unitary operation $V_{f^{-1}}$ of the
inversion function of the modular exponential function. There are many
auxiliary registers to be used in the construction of the unitary operation $%
V_{f^{-1}}$. The starting state in the construction may be taken as $|\Psi
_{0}\rangle =|\mathbf{R0}\rangle \bigotimes |g^{s}\func{mod}p\rangle .$ Here 
$|\mathbf{R0}\rangle =|0\rangle |0\rangle ...|0\rangle $ stands for the
library of auxiliary registers with the initial state $|0\rangle $ and
suppose that the register library stores sufficiently many registers to
supply to the coming quantum computation. The starting state is first
converted into the superposition by applying the conventional $(p-1)-$base
quantum Fourier transforms $Q_{(p-1)FT}$ to the first two registers,
respectively, which are supplied by the register library $|\mathbf{R0}%
\rangle $. Then the oracle unitary operation $U_{f}$ of the discrete
logarithm is applied to the first three registers, here the oracle unitary
operation $U_{f}$ uses the data $g$ and $b=g^{s}\func{mod}p.$ After the
oracle unitary operation $U_{f}$ the state of the quantum system is in the
state $|\Psi _{1}\rangle ,$ 
\begin{eqnarray*}
|\Psi _{0}\rangle &=&|\mathbf{R0}\rangle \bigotimes |g^{s}\func{mod}p\rangle
\equiv |\mathbf{R0}\rangle \bigotimes |0\rangle |0\rangle |g^{s}\func{mod}%
p\rangle \\
&&\stackrel{Q_{(p-1)FT}}{\rightarrow }|\mathbf{R0}\rangle \bigotimes \frac{1%
}{p-1}\stackrel{p-2}{\stackunder{x=0}{\sum }}\stackrel{p-2}{\stackunder{y=0}{%
\sum }}|x\rangle |y\rangle |g^{s}\func{mod}p\rangle \\
&&\stackrel{U_{f}}{\rightarrow }|\Psi _{1}\rangle =|\mathbf{R0}\rangle
\bigotimes \frac{1}{p-1}\stackrel{p-2}{\stackunder{x=0}{\sum }}\stackrel{p-2%
}{\stackunder{y=0}{\sum }}|x\rangle |y\rangle |f(x,y)\rangle |g^{s}\func{mod}%
p\rangle .
\end{eqnarray*}
The oracle unitary operation $U_{f}$ of the discrete logarithm is performed
in the conventional manner that the integers $g$ and $b=g^{s}\func{mod}p$
are first stored in auxiliary registers, the quantum computer reads the
integers $g$ and $b$ and values of the variables $x$ and $y$ in the first
two registers, then performs the functional operation $f(x,y)=b^{x}g^{y}%
\func{mod}p$ and puts the computing result in the third register which is
provided by the register library $|\mathbf{R0}\rangle $. Note that the data $%
b$ is already in the third register before the oracle unitary operation $%
U_{f}$ and in the fourth register after the oracle unitary operation, while
the known data $g$ can be stored in a temporary register beforehand and
after the operation $U_{f}$ it can be removed from the register. Using the
functional Fourier transform states (7) to express the functional state $%
|f(x,y)\rangle $ one obtains, by inserting equation (8) into the state $%
|\Psi _{1}\rangle ,$ 
\begin{eqnarray*}
|\Psi _{1}\rangle &=&|\mathbf{R0}\rangle \bigotimes \frac{1}{p-1}\stackrel{%
p-2}{\stackunder{l=0}{\sum }}\{[\frac{1}{\sqrt{p-1}}\stackrel{p-2}{%
\stackunder{x=0}{\sum }}\exp [-i2\pi lsx/(p-1)]|x\rangle ] \\
&&\bigotimes [\frac{1}{\sqrt{p-1}}\stackrel{p-2}{\stackunder{y=0}{\sum }}%
\exp [-i2\pi ly/(p-1)]|y\rangle ]|\widetilde{f}(ls,l)\rangle |g^{s}\func{mod}%
p\rangle \}.
\end{eqnarray*}
Now the conventional $(p-1)-$base quantum Fourier transforms $Q_{(p-1)FT}$
are applied again to the first two registers in the state $|\Psi _{1}\rangle
,$ respectively, then the quantum system is in the created state $|\Psi
_{2}\rangle $ after the SWAP unitary operation, 
\begin{eqnarray*}
&&|\Psi _{1}\rangle \stackrel{Q_{(p-1)FT}}{\rightarrow }|\mathbf{R0}\rangle
\bigotimes \frac{1}{p-1}\stackrel{p-2}{\stackunder{l=0}{\sum }}|ls\func{mod}%
(p-1)\rangle |l\rangle |\widetilde{f}(ls,l)\rangle |g^{s}\func{mod}p\rangle
\\
&&\stackrel{SWAP}{\rightarrow }|\Psi _{2}\rangle =|\mathbf{R0}\rangle
\bigotimes \frac{1}{p-1}\stackrel{p-2}{\stackunder{l=0}{\sum }}|l\rangle |ls%
\func{mod}(p-1)\rangle |\widetilde{f}(ls,l)\rangle |g^{s}\func{mod}p\rangle .
\end{eqnarray*}
The state $|\Psi _{2}\rangle $ contains the information of the index $s$ in
the last three registers. It is expected to extract the index $s$ from the
second register as the quantum states in other two registers are more
complicated. Therefore, the problem to be solved is how to extract the index 
$s$ from the state of the second register in the state $|\Psi _{2}\rangle $
and this is related to the construction of the unitary transformation $%
U_{s}: $ 
\[
|\Psi _{2}\rangle \stackrel{U_{s}}{\rightarrow }|\mathbf{R0}\rangle
\bigotimes \frac{C}{p-1}\stackrel{p-2}{\stackunder{l\geq 0}{\sum }}|l\rangle
|ls\func{mod}(p-1)\rangle |s\rangle |\widetilde{f}(ls,l)\rangle |g^{s}\func{%
mod}p\rangle , 
\]
where the index $l$ runs over only some specific values in the range $0\leq
l<p-1$ and $C$ is a normalization constant (see below). In the unitary
transformation $U_{s}$ the desired state transfer $|l\rangle |ls\func{mod}%
(p-1)\rangle |0\rangle \rightarrow |l\rangle |ls\func{mod}(p-1)\rangle
|s\rangle $ usually could not be achieved by the conventional inverse
multiplication operation $M_{p-1}^{+}(\alpha ,\beta ,\gamma )$. This is
because the function $f(s)=ls\func{mod}(p-1)$ does not have a one-to-one
correspondence to its variable $s$ for some integer values $l$ in the range $%
0\leq l<p-1$. Actually, it is possible that the inversion function $%
f(s)^{-1}\neq s$ if the integer $l$ is not coprime to $p-1.$ However, the
inversion function $f(s)^{-1}=s$ if the integer $l$ is coprime to $p-1,$ $%
i.e.,$ $(l,p-1)=1,$ and this is one of the two bases to achieve this unitary
state transfer and obtain the real index state $|s\rangle .$ It can be seen
that the state $|\Psi _{2}\rangle $ consists of $p-1$ orthogonal states with
index $l=0,1,...,p-2$. Among all the $(p-1)$ orthogonal states how many
orthogonal states have an index integer $l$ coprime to $(p-1)$? The question
can be answered by the Euler theorem in number theory (see the Theorem 72 in
reference [19]). As known in number theory [19], number for the positive
integers coprime to and not greater than $p-1$ is $\phi (p-1),$ where $\phi
(p-1)$ is the Euler totient function, and it is also known that the Euler
totient function $\phi (p-1)>\delta (p-1)/\log \log (p-1)$ for some constant 
$\delta .$ More exactly, if the integer $(p-1)$ has a prime factorization: $%
p-1=p_{1}^{a_{1}}p_{2}^{a_{2}}...p_{r}^{a_{r}},$ where $%
p_{1},p_{2},...,p_{r} $ are distinct primes, then $\phi
(p-1)=(p-1)\prod_{l=1}^{r}(1-p_{l}^{-1}).$ This shows that among the $p-1$
orthogonal states of the state $|\Psi _{2}\rangle $ there are $\phi (p-1)$
orthogonal states that have an index integer $l$ coprime to $p-1$. Thus, the
probability for all such orthogonal states in the state $|\Psi _{2}\rangle $
is $\phi (p-1)/(p-1)>\delta /\log \log (p-1).$ The probability is inversely
proportional to $\log \log (p-1)$ and hence is high even when the prime $p$
is very large. This is another basis to obtain the real index state $%
|s\rangle $. If the index integer $l$ is coprime to the integer $(p-1),$
there is a modular multiplication unitary operator $%
U_{l^{-1}}=U_{l,(p-1)}^{+}$ such that $U_{l^{-1}}|ls\func{mod}(p-1)\rangle
=|s\func{mod}(p-1)\rangle .$ Indeed, the unitary operation $U_{l^{-1}}$ can
generate the real index state $|s\rangle $ from the state $|ls\func{mod}%
(p-1)\rangle .$ But the unitary operation $U_{l^{-1}}$ does depend on the
integer $l,$ then it is clear that for the case $l\neq l^{\prime }$ the
unitary operation $U_{l^{-1}}$ does not generate the index state $|s\rangle $
from the state $|l^{\prime }s\func{mod}(p-1)\rangle ,$ that is, $%
U_{l^{-1}}|l^{\prime }s\func{mod}(p-1)\rangle \neq |s\func{mod}(p-1)\rangle $
if $l^{\prime }\neq l.$ Since all the index integers $l$ in the $(p-1)$
orthogonal states of the state $|\Psi _{2}\rangle $ are different it is
impossible to use a single unitary operation $U_{l^{-1}}$ to generate the
real index state $|s\rangle $ from these orthogonal states even if the index
integer $l$ for each of these states is coprime to $(p-1)$. In order to
generate the real index state $|s\rangle $ from the state $|\Psi _{2}\rangle 
$ the unitary transformation $U_{s}$ should be independent of any index
integer $l$. The conventional Euclidean algorithm [19] could be used to
construct the unitary transformation $U_{s}$. Suppose that the greatest
common divisor for the two integers $l$ and $(p-1)$ is $d_{l}$, i.e., $%
(l,p-1)=d_{l}$. The Euclidean algorithm can find efficiently two integers $%
a_{l}$ and $b_{l}$ such that the greatest common divisor $%
d_{l}=(l,p-1)=a_{l}l+b_{l}(p-1)$. Then $a_{l}l=d_{l}\func{mod}(p-1).$ If $%
d_{l}=1$ then $a_{l}l=1\func{mod}(p-1)$ and hence $a_{l}$ is the inverse
element of the integer $l$ ($\func{mod}(p-1)$). Using the Euclidean
algorithm the following unitary transformations can be obtained, 
\begin{eqnarray*}
&&|l\rangle |0\rangle |ls\func{mod}(p-1)\rangle |0\rangle \stackrel{GCD}{%
\rightarrow }|l\rangle |a_{l}\rangle |ls\func{mod}(p-1)\rangle |0\rangle \\
&&\stackrel{M_{p-1}(2,3,4)}{\rightarrow }|l\rangle |a_{l}\rangle |ls\func{mod%
}(p-1)\rangle |a_{l}ls\func{mod}(p-1)\rangle
\end{eqnarray*}
\begin{eqnarray*}
&=&|l\rangle |a_{l}\rangle |ls\func{mod}(p-1)\rangle |d_{l}s\func{mod}%
(p-1)\rangle \\
&&\stackrel{(GCD)^{+}}{\rightarrow }|l\rangle |0\rangle |ls\func{mod}%
(p-1)\rangle |d_{l}s\func{mod}(p-1)\rangle
\end{eqnarray*}
\[
=\left\{ 
\begin{array}{c}
|l\rangle |0\rangle |ls\func{mod}(p-1)\rangle |s\rangle ,\text{ if }d_{l}=1.
\\ 
|l\rangle |0\rangle |ls\func{mod}(p-1)\rangle |d_{l}s\func{mod}(p-1)\rangle ,%
\text{ if }d_{l}>1.
\end{array}
\right. 
\]
Here the Euclidean algorithm $GCD$ must be performed in a quantum parallel
form. This unitary transformation could be used to build up efficiently the
unitary transformation $U_{s}$ as the classical Euclidean algorithm can be
implemented in polynomial time $\thicksim O(\log ^{3}p).$ A quantum-version
extended Euclidean algorithm was given in Ref. [30b]. Another algorithm that
may be used to build up the unitary transformation $U_{s}$ is based on the
Euler theorem in number theory [19]. The Euler theorem (the Theorem 72 in
Reference [19]) states that if $(a,m)=1$, then $a^{\phi (m)}=1\func{mod}m.$
Thus, there holds $l^{\phi (p-1)}=1\func{mod}(p-1)$ for any integer $l$
coprime to $(p-1)$, i.e., $(l,p-1)=1$. But if $(l,p-1)\neq 1,$ the identity $%
l^{\phi (p-1)}=1\func{mod}(p-1)$ generally does not hold. Since the
computation for the modular exponentiation $l^{\phi (p-1)}\func{mod}(p-1)$
is simpler and efficient, it could be more convenient to use the modular
exponentiation operation to build up the unitary transformation $U_{s}$.
When the state $|\Psi _{2}\rangle $ is acted on by the conditional modular
exponentiation unitary operation $U_{\phi (p-1)-1,p-1}^{c}$ it will be
converted into the state $|\Psi _{3}\rangle ,$ 
\begin{eqnarray*}
&&|\Psi _{2}\rangle \stackrel{U_{\phi (p-1)-1,p-1}^{c}}{\rightarrow }|\Psi
_{3}\rangle =|\mathbf{R0}\rangle \bigotimes \frac{1}{p-1}\stackrel{p-2}{%
\stackunder{l=0}{\sum }}|l\rangle |ls\func{mod}(p-1)\rangle \\
&&\bigotimes |l^{\phi (p-1)}s\func{mod}(p-1)\rangle |\widetilde{f}%
(ls,l)\rangle |g^{s}\func{mod}p\rangle
\end{eqnarray*}
where the modular exponential function $l^{\phi (p-1)-1}\func{mod}(p-1)$ is
first computed by the conditional modular exponentiation operation $U_{\phi
(p-1)-1,p-1}^{c}$ in a quantum parallel form by using the integer $l$ in the
first register and then is put in a temporary register, then the function $%
l^{\phi (p-1)-1}\func{mod}(p-1)$ and the function $ls\func{mod}(p-1)$ in the
second register are multiplied with one another and the result is put in the
third register, and after these operations those states in the temporary
registers are removed unitarily. The state $|\Psi _{3}\rangle $ is written
as $|\Psi _{3}\rangle =|\Psi _{3s}\rangle +|\Psi _{3s^{\prime }}\rangle $
and the two orthogonal states $|\Psi _{3s}\rangle $ and $|\Psi _{3s^{\prime
}}\rangle $ are given respectively by 
\begin{eqnarray*}
|\Psi _{3s}\rangle &=&|\mathbf{R0}\rangle \bigotimes \frac{1}{p-1}\stackrel{%
p-2}{\stackunder{(l,p-1)=1}{\sum }}|l\rangle |ls\func{mod}(p-1)\rangle \\
&&\bigotimes |s\rangle |\widetilde{f}(ls,l)\rangle |g^{s}\func{mod}p\rangle ,
\\
|\Psi _{3s^{\prime }} &\rangle =&|\mathbf{R0}\rangle \bigotimes \frac{1}{p-1}%
\stackrel{p-2}{\stackunder{(l,p-1)>1}{\sum }}|l\rangle |ls\func{mod}%
(p-1)\rangle \\
&&\bigotimes |s^{\prime }\rangle |\widetilde{f}(ls,l)\rangle |g^{s}\func{mod}%
p\rangle ,
\end{eqnarray*}
where the sum with symbol $(l,p-1)=1$ means that the index $l$ takes those
integers less than and coprime to the integer $(p-1)$ and the sum with $%
(l,p-1)>1$ for the index $l$ runs over those integers less than and not
coprime to the integer $(p-1)$, the index $s^{\prime }=l^{\phi (p-1)}s\func{%
mod}(p-1)$ for $(l,p-1)>1$ (this also includes $l=0)$ and the index $%
s=l^{\phi (p-1)}s\func{mod}(p-1)$ by the Euler theorem that $l^{\phi (p-1)}=1%
\func{mod}(p-1)$ if $l$ is coprime to $p-1.$ Generally, the index $s^{\prime
}\neq s$. It is known that the computational complexity for the modular
exponentiation operation is $\thicksim O(\log ^{3}p)$ and hence the
conditional modular exponentiation unitary operation $U_{\phi
(p-1)-1,p-1}^{c}$ may be implemented in polynomial time $\thicksim O(\log
^{3}p)$. Now there are the desired state $|\Psi _{3s}\rangle $ which
contains the real index state $|s\rangle $ and the undesired state $|\Psi
_{3s^{\prime }}\rangle $ which does not have the index state $|s\rangle $ in
the state $|\Psi _{3}\rangle .$ Obviously, the probability for the desired
state $|\Psi _{3s}\rangle $ in the state $|\Psi _{3}\rangle $ is $\phi
(p-1)/(p-1)$ and hence the probability for the real index state $|s\rangle $
in the state $|\Psi _{3}\rangle $ is $\phi (p-1)/(p-1)>\delta /\log \log
(p-1)$. It is necessary to remove unitarily the undesired state $|\Psi
_{3s^{\prime }}\rangle $ from the state $|\Psi _{3}\rangle $ or to convert
it into the desired state $|\Psi _{3s}\rangle $ by a unitary transformation
so that the real index state $|s\rangle $ can be obtained from the desired
state $|\Psi _{3s}\rangle $ in a high probability $(\thicksim 1)$.

Here gives a simple method to convert unitarily the whole state $|\Psi
_{3}\rangle $ into the desired state $|\Psi _{3s}\rangle .$ This method is
similar to the amplitude amplification method [6, 30a]. It uses simply two
unitary operations, one is the inversion operation for the desired state $%
|\Psi _{3s}\rangle ,$%
\[
U(|\Psi _{3s}\rangle )=\exp \{-i\pi (|\Psi _{3s}\rangle \langle \Psi
_{3s}|)\} 
\]
and another is simply taken as 
\begin{eqnarray*}
U(|\Psi _{3}\rangle ) &=&\exp \{-i\pi (|\Psi _{3}\rangle \langle \Psi
_{3}|)\} \\
&=&\exp \{-i\pi (|\Psi _{3s}\rangle +|\Psi _{3s^{\prime }}\rangle )(\langle
\Psi _{3s}|+\langle \Psi _{3s^{\prime }}|)\}.
\end{eqnarray*}
Firstly, the inversion for the state $|\Psi _{3s}\rangle $ can be achieved
efficiently. Because $g$ is a primitive root ($\func{mod}p$), it has the
inverse element $g^{-1}=g^{p-2}\func{mod}p$ such that $g^{-1}g=1\func{mod}p.$
Then by making the conditional cyclic group operation $U_{g^{-1}}^{c}$ one
obtains the following state transformation: 
\[
U_{g^{-1}}^{c}|s^{\prime }\rangle |g^{s}\func{mod}p\rangle |0\rangle
=|s^{\prime }\rangle |g^{s}\func{mod}p\rangle |g^{-s^{\prime }+s}\func{mod}%
p\rangle ,s,s^{\prime }\in Z_{p-1}. 
\]
Here the operation result is put in the last register. Therefore, there
holds the unitary transformation: 
\[
U_{g^{-1}}^{c}|s^{\prime }\rangle |g^{s}\func{mod}p\rangle |0\rangle
=\left\{ 
\begin{array}{c}
|s\rangle |g^{s}\func{mod}p\rangle |1\rangle ,\text{ if }s^{\prime }=s \\ 
|s^{\prime }\rangle |g^{s}\func{mod}p\rangle |g^{-s^{\prime }+s}\func{mod}%
p\rangle ,\text{ if }s^{\prime }\neq s
\end{array}
\right. 
\]
Because the state $|1\rangle $ is orthogonal to these states $|g^{-s^{\prime
}+s}\func{mod}p\rangle $ for any indices $s^{\prime }\neq s,$ one can make
the selective inversion operation $C_{1}(\pi )=\exp (-i\pi D_{1})$ to invert
the state $|1\rangle ,$ while leaving these states $|g^{-s^{\prime }+s}\func{%
mod}p\rangle $ with $s^{\prime }\neq s$ unchanged. If now the conditional
cyclic group operation $U_{g^{-1}}^{c}$ acts on the state $|\Psi _{3}\rangle
,$ then only the desired state $|\Psi _{3s}\rangle $ generates the state $%
|1\rangle $ because it contains the index state $|s\rangle $, while the
state $|\Psi _{3s^{\prime }}\rangle $ produces the states $|g^{-s^{\prime
}+s}\func{mod}p\rangle $ with $s^{\prime }\neq s.$ After the unitary
operation $U_{g^{-1}}^{c}$ the selective inversion operation $C_{1}(\pi )$
is applied to the register whose state is either $|1\rangle $ or $%
|g^{-s^{\prime }+s}\func{mod}p\rangle ,$ then only the state $%
U_{g^{-1}}^{c}|\Psi _{3s}\rangle $ is inverted, while the state $%
U_{g^{-1}}^{c}|\Psi _{3s^{\prime }}\rangle $ keeps unchanged. After the
selective inversion operation $C_{1}(\pi )$ the states $U_{g^{-1}}^{c}|\Psi
_{3s}\rangle $ and $U_{g^{-1}}^{c}|\Psi _{3s^{\prime }}\rangle $ are
returned to the states $|\Psi _{3s}\rangle $ and $|\Psi _{3s^{\prime
}}\rangle ,$ respectively, by applying the inverse unitary operation $%
(U_{g^{-1}}^{c})^{+}.$ The inversion for the state $|\Psi _{3s}\rangle $
therefore is achieved, while the state $|\Psi _{3s^{\prime }}\rangle $ keeps
unchanged. Another unitary operation $U(|\Psi _{3}\rangle )$ is generated
from the oracle unitary operation: $U_{os}(\theta )=\exp \{-i\theta (|%
\mathbf{R0}\rangle |g^{s}\func{mod}p\rangle \langle g^{s}\func{mod}p|\langle 
\mathbf{R0}|)\}$ with $\theta =\pi .$ It is shown above that the state $|%
\mathbf{R0}\rangle |g^{s}\func{mod}p\rangle $ can be efficiently converted
into the state $|\Psi _{3}\rangle =|\Psi _{3s}\rangle +|\Psi _{3s^{\prime
}}\rangle $ by a sequence of unitary operations which may be simply denoted
as $U_{T}(|\Psi _{3}\rangle ).$ Then $|\mathbf{R0}\rangle |g^{s}\func{mod}%
p\rangle \stackrel{U_{T}(|\Psi _{3}\rangle )}{\rightarrow }|\Psi _{3}\rangle 
$ and the unitary operation $U(|\Psi _{3}\rangle )$ can be expressed as $%
U(|\Psi _{3}\rangle )=U_{T}(|\Psi _{3}\rangle )U_{os}(\pi )U_{T}^{+}(|\Psi
_{3}\rangle ).$ The unitary operation sequence that converts the state $%
|\Psi _{3}\rangle $ into the desired state $|\Psi _{3s}\rangle $ then is
given simply by 
\[
R(m)=[U(|\Psi _{3}\rangle )C(|\Psi _{3s}\rangle )]^{m}, 
\]
where the iterative number $m$ takes $\thicksim $ $O(\sqrt{\log \log (p-1)})$
so that the state $|\Psi _{3}\rangle $ is converted in a high probability ($%
\thicksim 1$) into the desired state $|\Psi _{3s}\rangle ,$ this is because
the probability for the desired state $|\Psi _{3s}\rangle $ in the state $%
|\Psi _{3}\rangle $ is $\phi (p-1)/(p-1)>\delta /\log \log (p-1)$. Thus,
under the unitary operation sequence $R(m)$ the state $|\Psi _{3}\rangle $
is converted completely into the desired state $|\Psi _{3s}\rangle ,$ 
\begin{eqnarray*}
|\Psi _{3}\rangle \stackrel{R(m)}{\rightarrow }|\Psi _{3s}\rangle &=&|%
\mathbf{R0}\rangle \bigotimes \frac{C}{p-1}\stackrel{p-2}{\stackunder{%
(l,p-1)=1}{\sum }}|l\rangle |ls\func{mod}(p-1)\rangle \\
&&\bigotimes |s\rangle |\widetilde{f}(ls,l)\rangle |g^{s}\func{mod}p\rangle ,
\end{eqnarray*}
where $C$ is a normalization constant, $C=\sqrt{(p-1)/\phi (p-1)}$. Now all
the orthogonal states in the state $|\Psi _{3s}\rangle $ have the index
state $|s\rangle .$ The state $|ls\func{mod}(p-1)\rangle $ in the second
register in the state $|\Psi _{3s}\rangle $ can be removed unitarily by
making an inverse multiplication operation $M_{p-1}^{+}(1,3,2)$ on the state 
$|\Psi _{3s}\rangle .$ After the index state $|s\rangle $ in the third
register in the state $|\Psi _{3s}\rangle $ is moved to the last register,
in which the index state $|s\rangle $ will be kept to the end, the state $%
|\Psi _{3s}\rangle $ is changed to the state $|\Psi _{4s}\rangle :$%
\[
|\Psi _{3s}\rangle \stackrel{M_{p-1}^{+}(1,3,2)}{\rightarrow }\stackrel{SWAP%
}{\rightarrow }|\Psi _{4s}\rangle 
\]
\[
=|\mathbf{R0}\rangle \bigotimes \frac{C}{p-1}\stackrel{p-2}{\stackunder{%
(l,p-1)=1}{\sum }}|l\rangle |\widetilde{f}(ls,l)\rangle |g^{s}\func{mod}%
p\rangle |s\rangle . 
\]
By inserting the inverse Fourier transform state $|\widetilde{f}%
(ls,l)\rangle $ (7) the state $|\Psi _{4s}\rangle $ can be rewritten as 
\begin{eqnarray*}
|\Psi _{4s}\rangle &=&|\mathbf{R0}\rangle \bigotimes \frac{C}{p-1}\frac{1}{%
p-1}\stackrel{p-2}{\stackunder{x_{1}=0}{\sum }}\stackrel{p-2}{\stackunder{%
x_{2}=0}{\sum }}\stackrel{p-2}{\stackunder{(l,p-1)=1}{\sum }}\exp [i2\pi
l(sx_{1}+x_{2})/(p-1)] \\
&&\times |l\rangle |f(x_{1},x_{2})\rangle |g^{s}\func{mod}p\rangle |s\rangle
.
\end{eqnarray*}
Since the functional state $|f(x_{1},x_{2})\rangle =|g^{sx_{1}+x_{2}}\func{%
mod}p\rangle =|f_{1}(sx_{1}+x_{2})\rangle ,$ there are only $p-1$ functional
states $|f(x_{1},x_{2})\rangle $ to be independent. However, there are $%
(p-1)\times (p-1)$ functional states $|f(x_{1},x_{2})\rangle $ in the state $%
|\Psi _{4s}\rangle ,$ then not all these $(p-1)\times (p-1)$ functional
states are independent. Actually, the state $|\Psi _{4s}\rangle $ can be
reduced to the simple form $|\Psi _{5s}\rangle :$%
\begin{eqnarray*}
|\Psi _{5s}\rangle &=&|\mathbf{R0}\rangle \bigotimes \frac{C}{p-1}\stackrel{%
p-2}{\stackunder{z=0}{\sum }}\stackrel{p-2}{\stackunder{(l,p-1)=1}{\sum }}%
\exp [i2\pi lz/(p-1)] \\
&&\times |l\rangle |f_{1}(z)\rangle |g^{s}\func{mod}p\rangle |s\rangle .
\end{eqnarray*}
Why can the state $|\Psi _{4s}\rangle $ be written as the simple form $|\Psi
_{5s}\rangle ?$ There are totally $(p-1)\times (p-1)$ different index pairs $%
(x_{1},x_{2})$ in the state $|\Psi _{4s}\rangle $ since the indices $x_{1},$ 
$x_{2}=0,1,...,p-2.$ Now for each given $z=(sx_{1}+x_{2})\func{mod}(p-1)$
for $z=0,1,...,p-2$ there are $(p-1)$ different index pairs $(x_{1},x_{2})$
to satisfy the same equation $z=(sx_{1}+x_{2})\func{mod}(p-1),$ while for
all these $(p-1)$ pairs of indices $(x_{1},x_{2})$ the functional states $%
|f(x_{1},x_{2})\rangle $ take the same one: $|f_{1}(z)\rangle $ and the
phase factor $\exp [i2\pi l(sx_{1}+x_{2})/(p-1)]$ also are the same as $\exp
[i2\pi lz/(p-1)]$. These $(p-1)$ different index pairs $(x_{1},x_{2})$ that
fulfill the same equation: $z=(sx_{1}+x_{2})\func{mod}(p-1)$ may be taken as 
$(x_{1},$ $(z-sx_{1})\func{mod}(p-1))$ for $x_{1}=0,1,...,p-2.$ Thus, taking 
$x_{1}=0,1,...,p-2$ and $z=0,1,...,p-2$ generates just all possible $%
(p-1)\times (p-1)$ different index pairs $(x_{1},x_{2}).$ Then in the state $%
|\Psi _{4s}\rangle $ the sums over the indices $x_{1}$ and $x_{2}$ may be
carried out in such a way that the sum for the index $x_{1}$ is first to run
over the $(p-1)$ different index pairs $(x_{1},$ $(z-sx_{1})\func{mod}(p-1))$
for $x_{1}=0,1,...,p-2$ and for any given $z=(sx_{1}+x_{2})\func{mod}(p-1),$
this sum will generate a factor of $(p-1)$\ as the same functional states $%
|f(x_{1},x_{2})\rangle $ and the same phase factors $\exp [i2\pi
l(sx_{1}+x_{2})/(p-1)]$ in the state $|\Psi _{4s}\rangle $ are taken for
these $(p-1)$ index pairs, then the sum for the index $z$ is carried out for 
$z=0,1,...,p-2$, and hence the state $|\Psi _{4s}\rangle $ can be written as
the simple state $|\Psi _{5s}\rangle $. Now one can also understand why the
Fourier transform state $|\widetilde{f}_{1}(l)\rangle =|\widetilde{f}%
(ls,l)\rangle /\sqrt{p-1}$ for $l=0,1,...,p-2$ (see before).

Now observe the state $|\Psi _{5s}^{^{\prime }}\rangle $ and a series of
unitary transformations: 
\begin{eqnarray*}
|\Psi _{5s}^{^{\prime }}\rangle &=&|\mathbf{R0}\rangle \bigotimes \frac{1}{%
p-1}\stackrel{p-2}{\stackunder{z=0}{\sum }}\stackrel{p-2}{\stackunder{l=0}{%
\sum }}\exp [i2\pi lz/(p-1)] \\
&&\times |l\rangle |f_{1}(z)\rangle |g^{s}\func{mod}p\rangle |s\rangle \\
&&\stackrel{Q_{(p-1)FT}^{+}}{\rightarrow }|\mathbf{R0}\rangle \bigotimes 
\frac{1}{\sqrt{p-1}}\stackrel{p-2}{\stackunder{z=0}{\sum }}|z\rangle |g^{z}%
\func{mod}p\rangle |g^{s}\func{mod}p\rangle |s\rangle \\
&&\stackrel{(U_{g,p}^{c})^{+}}{\rightarrow }|\mathbf{R0}\rangle \bigotimes 
\frac{1}{\sqrt{p-1}}\stackrel{p-2}{\stackunder{z=0}{\sum }}|z\rangle |g^{s}%
\func{mod}p\rangle |s\rangle \\
&&\stackrel{Q_{(p-1)FT}^{+}}{\rightarrow }|\mathbf{R0}\rangle \bigotimes
|g^{s}\func{mod}p\rangle |s\rangle .
\end{eqnarray*}
It can be seen that by making the inverse Fourier transform, the inverse
modular exponentiation operation $(U_{g,p}^{c})^{+},$ and again the inverse
Fourier transform the state $|\Psi _{5s}^{^{\prime }}\rangle $ is changed to
the state $|\mathbf{R0}\rangle \bigotimes |g^{s}\func{mod}p\rangle |s\rangle
.$

However, the state $|\Psi _{5s}\rangle $ is different from the state $|\Psi
_{5s}^{^{\prime }}\rangle $ in that the sum for the index $l$\ in the state $%
|\Psi _{5s}\rangle $ runs over only those integers that are less than and
coprime to the integer $(p-1)$. By making the inverse Fourier transform on
the state $|l\rangle $ in the first register the state $|\Psi _{5s}\rangle $
is changed to the state $|\Psi _{6s}\rangle ,$%
\[
|\Psi _{6s}\rangle =|\mathbf{R0}\rangle \bigotimes \frac{1}{p-1}\stackrel{p-2%
}{\stackunder{z=0}{\sum }}\stackrel{p-2}{\stackunder{z^{\prime }=0}{\sum }}%
h(z,z^{\prime })|z^{\prime }\rangle |f_{1}(z)\rangle |g^{s}\func{mod}%
p\rangle |s\rangle . 
\]
The trigonometrical sum $h(z,z^{\prime })$ is given by 
\[
h(z,z^{\prime })=\frac{1}{\sqrt{\phi (p-1)}}\stackrel{p-2}{\stackunder{%
(l,p-1)=1}{\sum }}\exp [i2\pi l(z-z^{\prime })/(p-1)] 
\]
where the sum for the index $l$ runs over only those integers less than and
coprime to the integer $(p-1)$. Obviously, the trigonometrical sum $h(z,z)=%
\sqrt{\phi (p-1)}$ if the index $z^{\prime }=z,$ for the number of the
integers less than and coprime to the integer $(p-1)$ is $\phi (p-1).$ Then
the state $|\Psi _{6s}\rangle $ can be rewritten as the sum of the two
terms: 
\begin{eqnarray*}
|\Psi _{6s}\rangle &=&|\mathbf{R0}\rangle \bigotimes \frac{\sqrt{\phi (p-1)}%
}{p-1}\stackrel{p-2}{\stackunder{z=0}{\sum }}|z\rangle |g^{z}\func{mod}%
p\rangle |g^{s}\func{mod}p\rangle |s\rangle \\
&&+|\mathbf{R0}\rangle \bigotimes \frac{1}{p-1}\stackrel{p-2}{\stackunder{%
z\neq z^{\prime },z,z^{\prime }=0}{\sum }}h(z,z^{\prime })|z^{\prime
}\rangle |g^{z}\func{mod}p\rangle |g^{s}\func{mod}p\rangle |s\rangle .
\end{eqnarray*}
By making the inverse modular exponentiation operation $(U_{g,p}^{c})^{+}$
on the first two registers the state $|\Psi _{6s}\rangle $ is transferred to
the state $|\Psi _{7s}\rangle :$ 
\begin{eqnarray*}
|\Psi _{7s}\rangle &=&|\mathbf{R0}\rangle \bigotimes \frac{\sqrt{\phi (p-1)}%
}{p-1}\stackrel{p-2}{\stackunder{z=0}{\sum }}|z\rangle |1\rangle |g^{s}\func{%
mod}p\rangle |s\rangle \\
&&+|\mathbf{R0}\rangle \bigotimes \frac{1}{p-1}\stackrel{p-2}{\stackunder{%
z\neq z^{\prime },z,z^{\prime }=0}{\sum }}h(z,z^{\prime })|z^{\prime
}\rangle |g^{z-z^{\prime }}\func{mod}p\rangle |g^{s}\func{mod}p\rangle
|s\rangle .
\end{eqnarray*}
Since the index $z^{\prime }\neq z,$ the state $|g^{z-z^{\prime }}\func{mod}%
p\rangle \neq |1\rangle $ and hence the two terms in the state $|\Psi
_{7s}\rangle $ are orthogonal to one another. Evidently, the first term in
the state $|\Psi _{7s}\rangle $ has a total probability $\phi (p-1)/(p-1)$
which is greater than $\delta /\log \log (p-1)$ for some constant $\delta .$
Again using the amplitude amplification method the second term in the state $%
|\Psi _{7s}\rangle $ can be converted into the first term in a high
probability $(\thicksim 1)$ and the iterative number in the amplitude
amplification process to achieve this complete state conversion needs $%
\thicksim O(\sqrt{\log \log (p-1)}).$ This time the selective inversion
operation is applied to the state $|1\rangle $ in the second register in the
state $|\Psi _{7s}\rangle $ and another unitary operation for the amplitude
amplification process is just the unitary operator $\exp \{-i\pi (|\Psi
_{7s}\rangle \langle \Psi _{7s}|)\}$ which can be also built up efficiently
because the state $|\Psi _{7s}\rangle $ itself can be generated efficiently
from the initial state $|\Psi _{0}\rangle $, as shown in the state-transfer
process above. After the state $|\Psi _{7s}\rangle $ is changed to its first
term completely, an inverse Fourier transform on the state $|z\rangle $ in
the first register and the state transfer $F_{1}^{+}:$ $|1\rangle
\rightarrow |0\rangle $ in the second register change the first term to the
desired state ultimately, 
\begin{eqnarray*}
|\Psi _{7s}\rangle &\rightarrow &|\mathbf{R0}\rangle \bigotimes \frac{1}{%
\sqrt{p-1}}\stackrel{p-2}{\stackunder{z=0}{\sum }}|z\rangle |1\rangle |g^{s}%
\func{mod}p\rangle |s\rangle \\
&&\stackrel{Q_{(p-1)FT}^{+}}{\rightarrow }\stackrel{F_{1}^{+}}{\rightarrow }|%
\mathbf{R0}\rangle \bigotimes |g^{s}\func{mod}p\rangle |s\rangle .
\end{eqnarray*}
Obviously, the whole unitary transformation process above really performs a
unitary transformation that firstly converts the starting state $|\Psi
_{0}\rangle =|\mathbf{R0}\rangle \bigotimes |g^{s}\func{mod}p\rangle
|0\rangle $ to the state $|\Psi _{3}\rangle ,$ then to the state $|\Psi
_{7s}\rangle ,$ and finally to the desired state $|\mathbf{R0}\rangle
\bigotimes |g^{s}\func{mod}p\rangle |s\rangle .$ Evidently, this is an
efficient unitary transformation process. This unitary operation sequence is
just the unitary operation $V_{f^{-1}}$ of the inversion function of the
modular exponential function $f(s)=g^{s}\func{mod}p$ if the register library 
$|\mathbf{R0}\rangle $ is dropped. Once the inversion-functional unitary
operation $V_{f^{-1}}$ is obtained the unitary operation $U_{\log }(g)$ of
the discrete logarithmic function $s=\log _{g}f(s)$ can be set up by $%
U_{\log }(g)=V_{f}^{+}SV_{f^{-1}}$.

If the starting state is a superposition, $|\Psi _{0}\rangle =\sum_{s}\alpha
_{s}|\mathbf{R0}\rangle \bigotimes |g^{s}\func{mod}p\rangle ,$ then it is
required that the unitary operator $U(|\Psi _{0}\rangle )=\exp \{-i\theta
(|\Psi _{0}\rangle \langle \Psi _{0}|)\}$ be efficiently constructed so that
the unitary operation $U(|\Psi _{3}\rangle ),$ $etc.,$ can be efficiently
built up with the unitary operator $U(|\Psi _{0}\rangle ).$ In this case the
superposition $|\Psi _{0}\rangle $ can be efficiently converted into the
superposition $|\Psi _{f}\rangle =\sum_{s}\alpha _{s}|\mathbf{R0}\rangle
\bigotimes |g^{s}\func{mod}p\rangle |s\rangle .$ For the quantum discrete
logarithmic problem the integer $b=g^{s}\func{mod}p$ is given beforehand and
hence the oracle unitary operation $\overline{U}_{os}(\theta )=\exp
[-i\theta \overline{D}_{s}(g)]$ can be constructed efficiently in advance.
Note that here the data $b$ is used to prepare the unitary operation instead
of a quantum state. Then using the above unitary operation sequence $%
V_{f^{-1}}$ the initial known state $|\Psi _{0}\rangle =|\mathbf{R0}\rangle
\bigotimes |g^{s}\func{mod}p\rangle $ can be efficiently converted into the
state $|\mathbf{R0}\rangle \bigotimes |g^{s}\func{mod}p\rangle |s\rangle .$
Furthermore, by using directly the discrete logarithmic unitary operation $%
U_{\log }(g)$ the initial known state $|\Psi _{0}\rangle $ can be
efficiently transferred to the index state $|\mathbf{R0}\rangle \bigotimes
|s\rangle $ and a quantum measurement on the index state will output
directly the complete information of the index $s$ of the integer $b=g^{s}%
\func{mod}p.$\newline
\newline
{\large 4. The efficient state transformations among the cyclic group state
subspaces}

Once it\ is obtained the unitary operation $U_{\log }(g)$ of the discrete
logarithmic function $x=\log _{g}f(x)$ with $f(x)=g^{x}\func{mod}p$, one may
further use it to prepare some useful auxiliary oracle unitary operations $%
U_{os^{\prime }}(\theta )$ where the index $s^{\prime }\neq s$ generally and
the index $s$ is of the oracle unitary operation $U_{os}(\theta )=\exp
[-i\theta D_{s}(g)]$. The process to generate the auxiliary oracle unitary
operation $U_{os^{\prime }}(\theta )$ with index $s^{\prime }=js$ from the
basic oracle unitary operation $U_{os}(\theta )$ is related to the state
unitary transformation $V_{js}$: 
\[
|\mathbf{R0}\rangle \bigotimes |g^{s}\func{mod}p\rangle \stackrel{V_{js}}{%
\rightarrow }|\mathbf{R0}\rangle \bigotimes |g^{js}\func{mod}p\rangle . 
\]
In the classical irreversible computation the modular exponential function $%
g^{js}\func{mod}p$ can be efficiently computed for any given integers $j$
and $b=g^{s}\func{mod}p$ [21], but it may not be so easy in the quantum
search problem to generate unitarily the state $|g^{js}\func{mod}p\rangle $
from the state $|g^{s}\func{mod}p\rangle $ for any given integer $j.$ If the
integer $j$ is coprime to the integer $(p-1)$, then there is an efficient
unitary transformation such that $U_{j,p-1}(\alpha )|s\rangle =|js\func{mod}%
(p-1)\rangle $ and the unitary transformation $V_{js}$ therefore can be
achieved efficiently with the help of the unitary operation $U_{\log }(g)$
of the discrete logarithm. Hence the auxiliary oracle unitary operation $%
U_{ojs}(\theta )$ can be efficiently generated from the oracle unitary
operation $U_{os}(\theta ).$ However, in order to simplify the quantum
search problem in the cyclic group state space one had better convert the
marked state into a small subspace of the cyclic group state space. Then the
auxiliary oracle unitary operation $U_{ojs}(\theta )$ usually is specific
one and the integer $j$ takes only some specific positive integer values
that are usually not coprime to the integer $(p-1)$. How can such an
auxiliary oracle unitary operation $U_{ojs}(\theta )$ be constructed from
the oracle unitary operation $U_{os}(\theta )$?

Evidently, the following unitary transformations can be achieved efficiently
for any integer $j$: 
\begin{eqnarray*}
&&|\mathbf{R0}\rangle \bigotimes |g^{s}\func{mod}p\rangle \stackrel{U_{\log
}(g)}{\rightarrow }|\mathbf{R0}\rangle \bigotimes |s\rangle \stackrel{F_{j}}{%
\rightarrow }|\mathbf{R0}\rangle \bigotimes |j\rangle |s\rangle \\
&&\stackrel{M_{p-1}(\alpha ,\beta ,\gamma )}{\rightarrow }|\mathbf{R0}%
\rangle \bigotimes |j\rangle |s\rangle |js\func{mod}(p-1)\rangle
\end{eqnarray*}
where the unitary transformation $F_{j}|0\rangle =|j\rangle $ for any known
integer $j$ can be built up efficiently. If the integer $j$ is coprime to $%
p-1,$ then a further unitary transformation sequence can be made: 
\begin{eqnarray*}
&&|\mathbf{R0}\rangle \bigotimes |j\rangle |s\rangle |js\func{mod}%
(p-1)\rangle |0\rangle \\
&&\stackrel{U_{\phi (p-1)-1,p-1}^{c}(1,3,4)}{\rightarrow }|\mathbf{R0}%
\rangle \bigotimes |j\rangle |s\rangle |js\func{mod}(p-1)\rangle |s\rangle \\
&&\stackrel{COPY(4,2)}{\rightarrow }|\mathbf{R0}\rangle \bigotimes |j\rangle
|js\func{mod}(p-1)\rangle |s\rangle \\
&&\stackrel{U_{\phi (p-1)-1,p-1}^{c}(1,2,3)^{+}}{\rightarrow }|\mathbf{R0}%
\rangle \bigotimes |j\rangle |js\func{mod}(p-1)\rangle \\
&&\stackrel{F_{j}^{+}}{\rightarrow }|\mathbf{R0}\rangle \bigotimes |js\func{%
mod}(p-1)\rangle \stackrel{U_{\log }^{+}(g)}{\rightarrow }|\mathbf{R0}%
\rangle \bigotimes |g^{js}\func{mod}p\rangle .
\end{eqnarray*}
Therefore, the state unitary transformation $V_{js}$ can be achieved too by
a more complicated way. However, from these detailed state unitary
transformations one may see more clearly why the state unitary
transformation $V_{js}$ is not easy to be constructed if the integer $j$ is
not coprime to the integer $(p-1)$.

If the integer $j$ is not coprime to the integer $(p-1),$ that is, $%
(j,p-1)>1 $, then situation becomes much more complicated. Firstly, the
state transformation $|j\rangle |js\func{mod}(p-1)\rangle |0\rangle
\rightarrow |j\rangle |js\func{mod}(p-1)\rangle |s\rangle $ for any index $%
s\in Z_{p-1}$ usually could not be unitary. This is related to the problem
whether there exists a unique inversion function of the function $f(x)=jx%
\func{mod}(p-1)$ or not for any index variable $x\in Z_{p-1}$. Since the
function $f(x)$ may not be a one-to-one function corresponding to its
variable $x\in Z_{p-1}$ if the integer $j$ is not coprime to $(p-1),$ the
inversion-functional operation $f(x)^{-1}$ therefore may not be unitary in
the variable value range: $x\in Z_{p-1}$. In the same argument the state
transformation $|g^{js}\func{mod}p\rangle |0\rangle \rightarrow |g^{js}\func{%
mod}p\rangle |s\rangle $ for any $s\in Z_{p-1}$ usually could not be unitary
if $(j,p-1)>1$, although the state transformation $|g^{js}\func{mod}p\rangle
|0\rangle \rightarrow |g^{js}\func{mod}p\rangle |js\func{mod}(p-1)\rangle $
is unitary. More generally, there could not be a single unitary
transformation for any integer $j\in Z_{p-1}$ such that $|j\rangle |js\func{%
mod}(p-1)\rangle |0\rangle \rightarrow |j\rangle |js\func{mod}(p-1)\rangle
|s\rangle $ for any given index $s,$ as shown in section 3$.$ These may be
best understood with the knowledge of number theory [19]. Suppose that one
is given a set of the integers $j=a_{k}$ and $js\func{mod}(p-1)=b_{k}$ for $%
k=1,2,...,r$ to reproduce uniquely the index $s.$ This problem is really
equivalent to solving the congruence system: 
\begin{equation}
a_{k}x=b_{k}\func{mod}(p-1),k=1,2,...,r,  \label{9}
\end{equation}
where the integers $\{a_{k}\}$ may not be coprime to $p-1$ and evidently $%
x=s $ is a solution to the congruence system. First consider a single
congruence, for example, the $k-$th congruence: $a_{k}x\func{mod}(p-1)=b_{k}$%
. Denote that the greatest common divisor between $a_{k}$ and $p-1$ is $%
d_{k}=(a_{k},p-1).$ Then the single $k-$th congruence has exactly $d_{k}$
solutions [19, 20] as $d_{k}$ is a divisor of the integer $b_{k}=a_{k}s\func{%
mod}(p-1)$ (i.e. $d_{k}|b_{k})$ for $k=1,2,...,r.$ If now the integer $a_{k}$
is not coprime to the integer $(p-1)$, that is, $d_{k}>1$, then there are $%
d_{k}$ different index values $s$ to satisfy the same $k-$th congruence,
indicating that there is not a single unitary state transformation: $%
|a_{k}\rangle |a_{k}s\func{mod}(p-1)\rangle |0\rangle \rightarrow
|a_{k}\rangle |a_{k}s\func{mod}(p-1)\rangle |s\rangle $ for any index $s\in
Z_{p-1}.$

Now consider the whole congruence system (9). Obviously, the congruence
system is solvable. Note that $d_{k}$ divides the integers $(p-1)$, $a_{k}$,
and $b_{k}$. Denote the integer $m_{k}=(p-1)/d_{k}$, $\widetilde{a}%
_{k}=a_{k}/d_{k},$ and $\widetilde{b}_{k}=b_{k}/d_{k}\equiv \widetilde{a}%
_{k}s\func{mod}m_{k}.$ Then the Theorem 57 in reference [19] shows that the
congruence system is equivalent to the simpler one: $\widetilde{a}_{k}x=%
\widetilde{b}_{k}\func{mod}m_{k},$ $k=1,2,...,r.$ Since $(\widetilde{a}%
_{k},m_{k})=1$ there exists an inverse element $h_{k}$ of $\widetilde{a}_{k}$
such that $h_{k}\widetilde{a}_{k}=1\func{mod}m_{k},$ the congruence system $%
\widetilde{a}_{k}x=\widetilde{b}_{k}\func{mod}m_{k},$ $k=1,2,...,r,$ then
can be further reduced to the standard one: 
\begin{equation}
x=c_{k}\func{mod}m_{k},\text{ }k=1,2,....,r,  \label{10}
\end{equation}
where the coefficients $c_{k}=h_{k}\widetilde{b}_{k}.$ Now the Chinese
remainder theorem [19, 20] shows that if $m_{1},m_{2},...,m_{r}$ are coprime
in pair, i.e., $(m_{i},m_{j})=1$ for $1\leq i<j\leq r,$ then the standard
congruence system (10) has a unique solution $(\func{mod}m),$ 
\begin{equation}
x=(n_{1}M_{1}c_{1}+n_{2}M_{2}c_{2}+...+n_{r}M_{r}c_{r})\func{mod}m,
\label{11}
\end{equation}
where $m=m_{1}m_{2}...m_{r}=m_{1}M_{1}=m_{2}M_{2}=...=m_{r}M_{r}$ and the
inverse element $n_{k}$ of $M_{k}$ ($\func{mod}m_{k}$) satisfies $%
n_{k}M_{k}=1\func{mod}m_{k}$ for $k=1,2,...,r$ because $(m_{k},M_{k})=1$.
Hence using the efficient Euclidean algorithm [19] the integer $n_{k}$ is
determined from the known integers $M_{k}$ and $m_{k}$ for $k=1,2,...,r$.
The solution $x$ of equation (11) is really the index $s$ if the index $s$
is bounded on by $0\leq s<m$ because the solution $x$ is unique ($\func{mod}%
m).$ However, the index $s$ really belongs to the integer set $%
Z_{p-1}=\{0,1,...,p-2\}.$ Then the solution $x$ could not be the real index $%
s$ if $m<p-1,$ for example, it could occur that $s=x+m$ for $0\leq s<p-1$.
In order that the solution $x$ of equation (11) is exactly the real index $s$
the integer $m$ should be equal to or greater than $(p-1)$. In fact, it is
better to take the integer $m$ exactly as the integer $p-1,$ that is, $%
m=p-1, $ as the situation is related closely to the prime factorization of
the integer $p-1$ and the structure of the cyclic group $S(C_{p-1}),$ as
shown in section 2. Now consider this specific case that the integer $%
m=(p-1) $ and $(p-1)$ has the prime factorization$:$ $%
(p-1)=p_{1}^{a_{1}}p_{2}^{a_{2}}...p_{r}^{a_{r}}$ ($p_{k}$ are distinct
primes). Take $a_{k}=(p-1)/p_{k}^{a_{k}}=M_{k}$ and $b_{k}=a_{k}s\func{mod}%
(p-1)=M_{k}s\func{mod}(p-1)$. Thus, $d_{k}=(a_{k},p-1)=M_{k}$ and $%
m_{k}=(p-1)/d_{k}=p_{k}^{a_{k}}.$ Then $\widetilde{a}_{k}=1$ and $\widetilde{%
b}_{k}=s\func{mod}m_{k}.$ Moreover, $(p-1)=m_{1}m_{2}...m_{r}=m_{i}M_{i},$
and $(m_{i},m_{j})=1$, $(m_{i},M_{i})=1$, for $1\leq i<j\leq r.$ Clearly, $%
h_{k}=1$ and $c_{k}=s\func{mod}m_{k}.$ Thus, the solution (11) is further
reduced to the form 
\begin{equation}
x=(n_{1}M_{1}c_{1}+n_{2}M_{2}c_{2}+...+n_{r}M_{r}c_{r})\func{mod}(p-1).
\end{equation}
Now the solution $x$ of equation (12) is just the real index $s$ and the
vector $\{c_{k}\}$ is just the index vector $\{s_{k}\}$ in the equation (3)
in section 2. Actually, in comparison with the equation (3) in section 2 one
now sees that the equation (12) is just the equation (3), showing once again
that this solution $x$ is just the real index $s$. Therefore, the Chinese
remainder theorem [19, 20] ensures that any index state $|s\rangle $ with $%
0\leq s<p-1$ can be exactly expressed as 
\begin{eqnarray}
|s\rangle &\equiv &|(n_{1}M_{1}s_{1}+n_{2}M_{2}s_{2}+...+n_{r}M_{r}s_{r})%
\func{mod}(p-1)\rangle  \nonumber \\
&\equiv &|(n_{1}M_{1}+n_{2}M_{2}+...+n_{r}M_{r})s\func{mod}(p-1)\rangle ,
\label{13}
\end{eqnarray}
where the identity $M_{k}s_{k}\equiv M_{k}s\func{mod}(p-1)$ has been used
for $k=1,2,...,r$.

The index state identity (13) could be helpful to prepare some specific
auxiliary oracle unitary operations in the additive cyclic group state space 
$S(Z_{p-1})$. Now it can be shown below that the index state $|s\rangle $
can be converted unitarily into a tension product of the $r$ states $\{|s%
\func{mod}m_{k}\rangle \}$ or $\{|M_{k}s\func{mod}(p-1)\rangle \}$ for $%
k=1,2,...,r$ in the $r$ different registers. Firstly, by simply applying the
reversible modular arithmetic operation $MOD(m_{k})$ on the index state $%
|s\rangle $ one obtains 
\[
|\mathbf{R0\rangle }\bigotimes |s\rangle \stackrel{MOD(m_{k})}{\rightarrow }%
|\Phi _{0}\rangle =|\mathbf{R0\rangle }\bigotimes |s\rangle |s\func{mod}%
m_{k}\rangle . 
\]
The reversible modular arithmetic operation can be thought of as a specific
reversible modular addition operation. Evidently, the state $|s\func{mod}%
m_{k}\rangle $ $\in S(Z_{m_{k}})$ and here $0\leq s\func{mod}m_{k}<m_{k}$
for $k=1,2,...,r$. Repeating the reversible modular arithmetic operation $r$
times for $k=1,2,...,r$ one arrives at the state $|\Phi _{1}\rangle :$%
\begin{eqnarray*}
|\mathbf{R0\rangle }\bigotimes |s\rangle &\rightarrow &|\Phi _{1}\rangle =|%
\mathbf{R0\rangle }\bigotimes |s\rangle \bigotimes |s\func{mod}m_{1}\rangle
\\
&&\bigotimes |s\func{mod}m_{2}\rangle \bigotimes ...\bigotimes |s\func{mod}%
m_{r}\rangle .
\end{eqnarray*}
Here each state $|s\func{mod}m_{k}\rangle =|s_{k}\rangle $ occupies one
register for $k=1,2,...,r$. Now substituting the state identity (13) for the
index state $|s\rangle $ the state $|\Phi _{1}\rangle $ is expressed as 
\begin{eqnarray*}
|\Phi _{1}\rangle &=&|\mathbf{R0\rangle }\bigotimes
|(n_{1}M_{1}s_{1}+n_{2}M_{2}s_{2}+...+n_{r}M_{r}s_{r})\func{mod}(p-1)\rangle
\\
&&\bigotimes |s_{1}\rangle \bigotimes |s_{2}\rangle \bigotimes ...\bigotimes
|s_{r}\rangle .
\end{eqnarray*}
In order to remove unitarily the composite state $|\sum_{k}n_{k}M_{k}s_{k}%
\func{mod}(p-1)\rangle $ in the state $|\Phi _{1}\rangle $ one needs to
perform a series of the modular multiplication operations $M_{p-1}(\alpha
,\beta ,\gamma )$ and inverse modular addition operations $%
ADD_{p-1}^{+}(\alpha ,\beta )$ on the state $|\Phi _{1}\rangle ,$ for
example, 
\begin{eqnarray*}
&&|\Phi _{1}\rangle \stackrel{F_{n_{1}M_{1}}}{\rightarrow }\ \ \stackrel{%
M_{p-1}(\alpha _{1},\beta _{1},\gamma _{1})}{\rightarrow } \\
&&|\mathbf{R0\rangle }\bigotimes |n_{1}M_{1}\func{mod}(p-1)\rangle
|n_{1}M_{1}s_{1}\func{mod}(p-1)\rangle \\
&&\bigotimes |(n_{1}M_{1}s_{1}+n_{2}M_{2}s_{2}+...+n_{r}M_{r}s_{r})\func{mod}%
(p-1)\rangle \\
&&\bigotimes |s_{1}\rangle \bigotimes |s_{2}\rangle \bigotimes ...\bigotimes
|s_{r}\rangle \\
&&\stackrel{ADD_{p-1}(\alpha _{1},\beta _{1})^{+}}{\rightarrow }|\mathbf{%
R0\rangle }\bigotimes |n_{1}M_{1}\func{mod}(p-1)\rangle |n_{1}M_{1}s_{1}%
\func{mod}(p-1)\rangle \\
&&\bigotimes |(n_{2}M_{2}s_{2}+n_{3}M_{3}s_{3}+...+n_{r}M_{r}s_{r})\func{mod}%
(p-1)\rangle \\
&&\bigotimes |s_{1}\rangle \bigotimes |s_{2}\rangle \bigotimes ...\bigotimes
|s_{r}\rangle \\
&&\stackrel{M_{p-1}^{+}(\alpha _{1},\beta _{1},\gamma _{1})}{\rightarrow }\
\ \stackrel{F_{n_{1}M_{1}}^{+}}{\rightarrow } \\
&&|\mathbf{R0\rangle }\bigotimes
|(n_{2}M_{2}s_{2}+n_{3}M_{3}s_{3}+...+n_{r}M_{r}s_{r})\func{mod}(p-1)\rangle
\\
&&\bigotimes |s_{1}\rangle \bigotimes |s_{2}\rangle \bigotimes ...\bigotimes
|s_{r}\rangle .
\end{eqnarray*}
The unitary transformation process in the example is stated below. The state 
$|n_{1}M_{1}\func{mod}(p-1)\rangle $ is first created by the unitary
transformation: $F_{n_{1}M_{1}}|0\rangle =|n_{1}M_{1}\func{mod}(p-1)\rangle
, $ then the modular multiplication operation $M_{p-1}(\alpha _{1},\beta
_{1},\gamma _{1})$ acts on both the states $|n_{1}M_{1}\func{mod}%
(p-1)\rangle $ and $|s_{1}\rangle $ to generate the state $|n_{1}M_{1}s_{1}%
\func{mod}(p-1)\rangle ,$ and then the modular subtraction operation or the
inverse modular addition operation $ADD_{p-1}^{+}(\alpha _{1},\beta _{1})$
acts on both the state $|n_{1}M_{1}s_{1}\func{mod}(p-1)\rangle $ and the
composite state $|\sum_{k}n_{k}M_{k}s_{k}\func{mod}(p-1)\rangle $ so that
the composite state is changed to the state $%
|(n_{2}M_{2}s_{2}+n_{3}M_{3}s_{3}+...+n_{r}M_{r}s_{r})\func{mod}(p-1)\rangle
.$ After these unitary transformations the unitary operations $%
M_{p-1}^{+}(\alpha ,\beta ,\gamma )$ and $F_{n_{1}M_{1}}^{+}$ are used to
convert the states $|n_{1}M_{1}\func{mod}(p-1)\rangle $ and $|n_{1}M_{1}s_{1}%
\func{mod}(p-1)\rangle $ back to the states $|0\rangle .$ Clearly, the whole
unitary transformation process really cancels the term $n_{1}M_{1}s_{1}\func{%
mod}(p-1)$ in the composite state $|\sum_{k}n_{k}M_{k}s_{k}\func{mod}%
(p-1)\rangle $ of the state $|\Phi _{1}\rangle .$ If this unitary
transformation process is repeated $r$ times with different unitary
operations $F_{n_{k}M_{k}},$ $M_{p-1}(\alpha _{k},\beta _{k},\gamma _{k}),$
and $ADD_{p-1}^{+}(\alpha _{k},\beta _{k})$ for $k=1,2,...,r,$ then the
composite state $|\sum_{k}n_{k}M_{k}s_{k}\func{mod}(p-1)\rangle $ is
ultimately converted into the state $|0\rangle $ in the state $|\Phi
_{1}\rangle .$ Therefore, it is shown that with the help of the state
identity (13) the index state $|\mathbf{R0\rangle }\bigotimes |s\rangle $
can be efficiently converted into the state $|\Phi _{2}\rangle :$ 
\begin{eqnarray*}
|\mathbf{R0\rangle }\bigotimes |s\rangle &\rightarrow &|\Phi _{2}\rangle =|%
\mathbf{R0\rangle }\bigotimes |s\func{mod}m_{1}\rangle \\
&&\bigotimes |s\func{mod}m_{2}\rangle \bigotimes ...\bigotimes |s\func{mod}%
m_{r}\rangle .
\end{eqnarray*}
The state $|\Phi _{2}\rangle $ is a tension product of the $r$ states $\{|s%
\func{mod}m_{k}\rangle \}$ in the $r$ different registers. In an analogue
way, the index state $|\mathbf{R0\rangle }\bigotimes |s\rangle $ also can be
efficiently converted into a tension product of the $r$ states $\{|M_{k}s%
\func{mod}(p-1)\rangle \}$ in the $r$ different registers, 
\begin{eqnarray*}
|\mathbf{R0\rangle }\bigotimes |s\rangle &\rightarrow &|\Phi _{3}\rangle =|%
\mathbf{R0\rangle }\bigotimes |M_{1}s\func{mod}(p-1)\rangle \\
&&\bigotimes |M_{2}s\func{mod}(p-1)\rangle \bigotimes ...\bigotimes |M_{r}s%
\func{mod}(p-1)\rangle .
\end{eqnarray*}
Since there is the identity $M_{k}s_{k}\equiv M_{k}s\func{mod}(p-1)$ for $%
k=1,2,...,r$, the state $|M_{k}s\func{mod}(p-1)\rangle =|M_{k}s_{k}\func{mod}%
(p-1)\rangle .$ By using the inverse discrete logarithmic unitary operation $%
U_{\log }^{+}(g)$ the state $|M_{k}s_{k}\func{mod}(p-1)\rangle $ can be
converted into the state $|(g^{M_{k}})^{s_{k}}\func{mod}p\rangle $ which
belongs to the state subspace $S(C_{p_{k}^{a_{k}}})$ of the cyclic subgroup $%
C_{p_{k}^{a_{k}}}.$ On the other hand, using the inverse discrete
logarithmic unitary operation $U_{\log }^{+}(g^{M_{k}})$ with the
logarithmic base $g^{M_{k}}$ the state $|s\func{mod}m_{k}\rangle ,$ i.e., $%
|s_{k}\rangle ,$ can also be converted to the same state $%
|(g^{M_{k}})^{s_{k}}\func{mod}p\rangle .$ These results show that the
auxiliary oracle unitary operation $\exp \{-i\theta |(g^{M_{k}})^{s_{k}}%
\func{mod}p\rangle \langle (g^{M_{k}})^{s_{k}}\func{mod}p|\}$ of the
multiplicative cyclic group state subspace $S(C_{p_{k}^{a_{k}}})$ can be
efficiently built out of the auxiliary oracle unitary operation $\exp
\{-i\theta (|s\func{mod}m_{k}\rangle \langle s\func{mod}m_{k}|)\}$ of the
additive cyclic group state subspace $S(Z_{m_{k}})$ or the auxiliary oracle
unitary operation $\exp \{-i\theta (|M_{k}s\func{mod}(p-1)\rangle \langle
M_{k}s\func{mod}(p-1)|)\}$ of the additive cyclic group state space $%
S(Z_{p-1})$.

Now consider the multiplicative cyclic group state space $S(C_{p-1})$.
Suppose that the prime factors of the integer $%
(p-1)=p_{1}^{a_{1}}p_{2}^{a_{2}}...p_{r}^{a_{r}}$ are ordered in magnitude: $%
p_{1}^{a_{1}}<p_{2}^{a_{2}}<...<p_{r}^{a_{r}}$ and $p_{r}^{a_{r}}\thicksim
O(\log p).$ Then $m_{1}<m_{2}<...<m_{r}$ and $M_{1}>M_{2}>...>M_{r}.$ As
shown in section 2, the cyclic group $C_{p-1}$ with order $p-1$ is the
direct product of $r$ factor cyclic subgroups: $C_{p-1}=C_{p_{1}^{a_{1}}}%
\times C_{p_{2}^{a_{2}}}\times ...\times C_{p_{r}^{a_{r}}}.$ Each such
cyclic subgroup $C_{p_{k}^{a_{k}}}$ corresponds to a state subspace $%
S(C_{p_{k}^{a_{k}}})$ with dimension $p_{k}^{a_{k}}$ of the cyclic group
state space $S(C_{p-1})$. For convenience, denote $S(m_{k})\equiv
S(C_{p_{k}^{a_{k}}})$ with $m_{k}=p_{k}^{a_{k}}.$ It can be proven that the
state $|g^{sM_{k}}\func{mod}p\rangle $ is in the state subspace $S(m_{k})$
for any index integer $s.$ This is because the generator and the order of
the cyclic subgroup $C_{p_{k}^{a_{k}}}$ is $g^{M_{k}}$ and $m_{k},$
respectively, then there holds the state identity $|g^{sM_{k}}\func{mod}%
p\rangle =|(g^{M_{k}})^{s\func{mod}m_{k}}\func{mod}p\rangle $ for any index $%
s,$ while the latter state $|(g^{M_{k}})^{s_{k}}\func{mod}p\rangle $ with
the index $s_{k}=s\func{mod}m_{k}$ is just in the state subspace $S(m_{k}).$
Since the dimensional size of the cyclic group state subspace $S(m_{k})$ is
just the order $m_{k}$ of the subgroup $C_{p_{k}^{a_{k}}}$ and $%
m_{1}<m_{2}<...<m_{r},$ then the state $|g^{sM_{1}}\func{mod}p\rangle $ is
in the smallest state subspace $S(m_{1})$, the state $|g^{sM_{2}}\func{mod}%
p\rangle $ in the second smallest subspace $S(m_{2})$, ..., and the state $%
|g^{sM_{r}}\func{mod}p\rangle $ in the largest subspace $S(m_{r})$ of the $r$
state subspaces $\{S(m_{k})\}$. It follows from the equation (2) in section
2 that every state of the cyclic group state space $S(C_{p-1})$ can be
expressed as 
\begin{equation}
|g^{s}\func{mod}p\rangle \equiv |(g^{M_{1}})^{n_{1}s_{1}}\times
(g^{M_{2}})^{n_{2}s_{2}}\times ...\times (g^{M_{r}})^{n_{r}s_{r}}\func{mod}%
p\rangle  \label{14}
\end{equation}
The state identity (14) plays a similar role to the state identity (13) in
decomposing any state of the cyclic group state space $S(C_{p-1})$ as a
tension product of the states of the state subspaces $\{S(m_{k})\}$ of the
factor cyclic subgroups $\{C_{p_{k}^{a_{k}}}\}.$ By the modular
exponentiation operation the state $|(g^{M_{k}})^{s}\func{mod}p\rangle $ of
the state subspace $S(m_{k})$ can be generated from the cyclic group state $%
|g^{s}\func{mod}p\rangle ,$ 
\begin{eqnarray*}
|\mathbf{R0\rangle }\bigotimes |g^{s}\func{mod}p\rangle &\rightarrow &|\Phi
_{4}\rangle =|\mathbf{R0\rangle }\bigotimes |g^{s}\func{mod}p\rangle
|(g^{M_{k}})^{s}\func{mod}p\rangle \\
&=&|\mathbf{R0\rangle }\bigotimes |g^{s}\func{mod}p\rangle
|(g^{M_{k}})^{s_{k}}\func{mod}p\rangle .
\end{eqnarray*}
Repeating this modular exponentiation operation $r$ times for $k=1,2,...,r$
the state $|\mathbf{R0\rangle }\bigotimes |g^{s}\func{mod}p\rangle $ is
converted into the state $|\Phi _{5}\rangle ,$ 
\begin{eqnarray*}
|\mathbf{R0\rangle }\bigotimes |g^{s}\func{mod}p\rangle &\rightarrow &|\Phi
_{5}\rangle =|\mathbf{R0\rangle }\bigotimes |g^{s}\func{mod}p\rangle
\bigotimes |(g^{M_{1}})^{s_{1}}\func{mod}p\rangle \\
&&\bigotimes |(g^{M_{2}})^{s_{2}}\func{mod}p\rangle \bigotimes ...\bigotimes
|(g^{M_{r}})^{s_{r}}\func{mod}p\rangle .
\end{eqnarray*}
By using the state identity (14) and the modular exponentiation, the modular
multiplication, and the COPY operation the state $|g^{s}\func{mod}p\rangle $
in the state $|\Phi _{5}\rangle $ can be removed unitarily and hence the
state $|\mathbf{R0\rangle }\bigotimes |g^{s}\func{mod}p\rangle $ can be
efficiently converted into a tension product of the $r$ states $%
\{|(g^{M_{k}})^{s_{k}}\func{mod}p\rangle \}$ of the $r$ different subspaces $%
\{S(m_{k})\}$ in the $r$ different registers$,$ 
\begin{eqnarray*}
|\mathbf{R0\rangle }\bigotimes |g^{s}\func{mod}p\rangle &\rightarrow &|\Phi
_{6}\rangle =|\mathbf{R0\rangle }\bigotimes |(g^{M_{1}})^{s_{1}}\func{mod}%
p\rangle \\
&&\bigotimes |(g^{M_{2}})^{s_{2}}\func{mod}p\rangle \bigotimes ...\bigotimes
|(g^{M_{r}})^{s_{r}}\func{mod}p\rangle .
\end{eqnarray*}
This unitary transformation is stated below. The states $%
\{|(g^{M_{k}})^{n_{k}s_{k}}\func{mod}p\rangle \}$ are first generated
efficiently from the states $\{|(g^{M_{k}})^{s_{k}}\func{mod}p\rangle \}$ by
the modular exponentiation operations in temporary registers in the state $%
|\Phi _{5}\rangle $ because the integers $\{n_{k}\}$ are known, as shown
before. Then by the modular multiplication operations the state $%
|\prod_{k}(g^{M_{k}})^{n_{k}s_{k}}\func{mod}p\rangle $ is created
efficiently from these states $\{|(g^{M_{k}})^{n_{k}s_{k}}\func{mod}p\rangle
\}.$ The state identity (14) shows that the state $%
|\prod_{k}(g^{M_{k}})^{n_{k}s_{k}}\func{mod}p\rangle $ is just the state $%
|g^{s}\func{mod}p\rangle .$ Then using the COPY operation the state $|g^{s}%
\func{mod}p\rangle $ can be removed from the state $|\Phi _{5}\rangle .$
After these unitary operations those states $%
|\prod_{k}(g^{M_{k}})^{n_{k}s_{k}}\func{mod}p\rangle $ and $%
\{|(g^{M_{k}})^{n_{k}s_{k}}\func{mod}p\rangle \}$ in temporary registers are
returned back to the state $|0\rangle $ and therefore the state $|\Phi
_{6}\rangle $ is obtained. Note that these states $|(g^{M_{j}})^{s}\func{mod}%
p\rangle $ for different index $j$ in the state $|\Phi _{6}\rangle $ belong
to different subspaces $\{S(m_{j})\}$ and also different registers. It has
been shown that any unknown state can be efficiently transferred to a larger
state subspace from a small subspace in the Hilbert space [16]. Then the
state $|(g^{M_{j}})^{s}\func{mod}p\rangle $ which is in the subspace $%
S(m_{j})$ with the dimensional size $m_{j}$ may be efficiently transferred
to a larger subspace $S(m_{k})$ with dimensional size $m_{k}>m_{j}.$ Since
the dimensional size $m_{k}$ for any subspace $S(m_{k})$ is $\thicksim
O(\log p)$ and there hold $0\leq s_{k}<m_{k}$ and $m_{1}<m_{2}<...<m_{r}$,
the unitary operation for the state transfer $|(g^{M_{j}})^{s}\func{mod}%
p\rangle \rightarrow |(g^{M_{k}})^{s}\func{mod}p\rangle $ for $1\leq j<k\leq
r$ always can be constructed efficiently [16]. Now the state transfer is
carried out from a small subspace $S(m_{k})$ ($k\neq r)$ to the largest
subspace $S(m_{r}),$ that is, $|(g^{M_{k}})^{s_{k}}\func{mod}p\rangle
\rightarrow |(g^{M_{r}})^{s_{k}}\func{mod}p\rangle $ for $k=1,2,...,r-1,$
then the state $|\Phi _{6}\rangle $ will be directly changed to a tension
product of the $r$ states $\{|(g^{M_{r}})^{s_{k}}\func{mod}p\rangle ,$ $%
k=1,2,...,r\}$ of the largest subspace $S(m_{r})$ in the $r$ different
registers respectively: 
\begin{eqnarray*}
|\Phi _{6}\rangle &\rightarrow &|\Phi _{7}\rangle =|\mathbf{R0\rangle }%
\bigotimes |(g^{M_{r}})^{s_{1}}\func{mod}p\rangle \\
&&\bigotimes |(g^{M_{r}})^{s_{2}}\func{mod}p\rangle \bigotimes ...\bigotimes
|(g^{M_{r}})^{s_{r}}\func{mod}p\rangle ,
\end{eqnarray*}
where the state transfers can be performed in a parallel manner in the first 
$r-1$ registers of the state $|\Phi _{6}\rangle .$ The state $|\Phi
_{7}\rangle $ shows that any state $|g^{s}\func{mod}p\rangle $ of the cyclic
group state space $S(C_{p-1})$ can be efficiently converted into a tension
product of the $r$ cyclic group states of the largest subspace $S(m_{r})$.
If the index state $|s\rangle $ is unknown, then in the state $|\Phi
_{7}\rangle $ all these states $\{|(g^{M_{r}})^{s_{k}}\func{mod}p\rangle
,k=1,2,...,r\}$ are also unknown and they carry the complete information of
the index state $|s\rangle $. Evidently, if the initial index state $%
|s\rangle $ or the initial cyclic group state $|g^{s}\func{mod}p\rangle $ is
replaced with a superposition, then the above state transformations work as
well.

For the discrete logarithmic problem it is much simple to generate unitarily
the auxiliary oracle unitary operation $\overline{U}_{ojs}(\theta )=\exp
[-i\theta \overline{D}_{js}(g)]$ with the diagonal operator $\overline{D}%
_{js}(g)=|\mathbf{R0\rangle }\langle \mathbf{R0|}\bigotimes |g^{js}\func{mod}%
p\rangle \langle g^{js}\func{mod}p|$ and $j=M_{k}$ or even $j=(p-1)/p_{k}$
from the basic oracle unitary operation $\overline{U}_{os}(\theta )=\exp
[-i\theta \overline{D}_{s}(g)]$ in polynomial time. Actually, this can be
achieved directly by the state transformation: $|\mathbf{R0\rangle }%
\bigotimes |g^{s}\func{mod}p\rangle \rightarrow |\Phi _{4}\rangle $ without
using any state identity (13) or (14). This is because $(i)$ the integer $%
b=g^{s}\func{mod}p$ is given beforehand and hence the oracle unitary
operation $\overline{U}_{os}(\theta )$ can be efficiently constructed in
advance, and $(ii)$ the known state $|g^{s}\func{mod}p\rangle $ in the state 
$|\Phi _{4}\rangle $ can be efficiently converted to the state $|0\rangle .$
Therefore, using the auxiliary oracle unitary operation $\overline{U}%
_{ojs}(\theta )$ and the standard quantum search algorithm one can solve
efficiently the discrete logarithmic problem in polynomial time if the
dimensional size $m_{k}$ for every cyclic group state subspace $S(m_{k})$ is 
$\thicksim O(\log p).$ This quantum discrete logarithmic algorithm is
similar to the classical counterpart [21]. By combining with the quantum
discrete logarithmic algorithm in section 3 this algorithm will obtain much
more speedup.

However, the quantum search problem is much harder than the discrete
logarithmic problem. The auxiliary oracle unitary operations corresponding
to the states $|\Phi _{2}\rangle $ and $|\Phi _{7}\rangle $ still may be
unsuitable for the quantum search task, for these factor states $\{|s\func{%
mod}m_{k}\rangle \}$ in the state $|\Phi _{2}\rangle $ or $%
\{|(g^{M_{r}})^{s_{k}}\func{mod}p\rangle \}$ in the state $|\Phi _{7}\rangle 
$ that carry the complete information of the index state $|s\rangle $ are in
the $r$ different registers and this makes the search space too large for
the quantum search problem. There are two possible schemes to solve this
problem. One scheme is to compress unitarily all these $r$ states in the $r$
different registers into one register only in the state $|\Phi _{2}\rangle $
or $|\Phi _{7}\rangle ,$ and this scheme will lead to that the quantum
search space is limited to the largest cyclic group state subspace $%
S(Z_{m_{r}})$ or $S(m_{r}).$ Since the dimension of the state subspace $%
S(Z_{m_{r}})$ or $S(m_{r})$ is $m_{r}\thicksim O(\log p)$ the quantum search
process may be implemented efficiently in these state subspaces. Another is
to keep only one desired state but remove unitarily the other $r-1$ states
in the state $|\Phi _{2}\rangle $ or $|\Phi _{7}\rangle .$ For example, one
may let all those states $|(g^{M_{r}})^{s_{j}}\func{mod}p\rangle $ for $%
j\neq k$ return unitarily to the known state $|0\rangle $ but only the
desired state $|(g^{M_{r}})^{s_{k}}\func{mod}p\rangle $ be retained in the
state $|\Phi _{7}\rangle $. It could be better that the two schemes are used
together. In next section a possible algorithm is proposed on a universal
quantum computer to further reduce the quantum search space for the state $%
|\Phi _{7}\rangle $ in the multiplicative cyclic group state $S(C_{p-1})$,
while the reduction for the quantum search space on the basis of the state $%
|\Phi _{2}\rangle $ in the additive cyclic group state space $S(Z_{p-1})$ is
left in the future work. \newline
\newline
{\large 5. An efficient reduction for the quantum search space on an ideal
universal quantum computer}

A universal quantum computer [29, 38, 40] should be capable of computing any
recursive function in mathematics and any computational process on it obeys
the unitary quantum dynamics in physics. Now a quantum computational program
based on the reversible computation [26, 27] is designed to transform some
states $\{|(g^{M_{r}})^{s_{k}}\func{mod}p\rangle \}$ back to the known state 
$|0\rangle $ but keep the desired state in the state $|\Phi _{7}\rangle $.
This quantum program $Q_{p}$ may run on a universal quantum computer [29,
38, 40]. It is given by 
\[
|n_{h}\rangle =|0\rangle 
\]
\[
|b_{h}\rangle =|0\rangle 
\]
\[
\text{For }i=1\text{ to }m_{r} 
\]
\[
\text{If }|g_{r}(y)\rangle =|1\rangle \text{ then }|b_{h}\rangle \rightarrow
|b_{h}+1\rangle \text{ end if} 
\]
\[
\text{When }|g_{r}(y)\rangle =|1\rangle ,\text{ Do }|g_{r}(y)\rangle
|n_{h}\rangle =|1\rangle |0\rangle \rightarrow |0\rangle |0\rangle ,\text{ }%
|n_{h}\rangle =|0\rangle \rightarrow |1\rangle ,\text{ halt} 
\]
\[
\text{If }|b_{h}\rangle =|0\rangle \text{ then} 
\]
\[
U_{g^{M_{r}}}|f_{r}(x)\rangle |g_{r}(y)\rangle 
\]
\[
U_{r}|f_{r}(x)\rangle |g_{r}(y)\rangle 
\]
\[
\text{else }U_{g^{M_{r}}}|f_{r}(x)\rangle |g_{r}(y)\rangle \text{ end if} 
\]
\[
\text{end for.} 
\]
The quantum program $Q_{p}$ can be really written as $Q_{p}=\{Q_{u}%
\}^{m_{r}} $ in which the basic operational unit $Q_{u}$ is repeated to
execute $m_{r}$ times. The basic operational unit $Q_{u}$ may be formally
expressed as $Q_{u}=\{U_{r}^{c}U_{g^{M_{r}}}P^{c}\}$, here the operation $%
P^{c}$ executes the two statements:\ $^{\prime \prime }$If $|g_{r}(y)\rangle
=|1\rangle $ then $|b_{h}\rangle \rightarrow |b_{h}+1\rangle $ end if$%
^{\prime \prime }$ and $^{\prime \prime }$When $|g_{r}(y)\rangle =|1\rangle
, $ Do $|g_{r}(y)\rangle |n_{h}\rangle =|1\rangle |0\rangle \rightarrow
|0\rangle |0\rangle ,$ $|n_{h}\rangle =|0\rangle \rightarrow |1\rangle ,$
halt$^{\prime \prime },$ the operation $U_{r}^{c}$ performs conditionally
the unitary operation $U_{r}$ if the branch-control state $|b_{h}\rangle
=|0\rangle $, and the operation $U_{g^{M_{r}}}$ performs the unitary cyclic
group operation of the cyclic subgroup $C_{p_{r}^{a_{r}}}.$ The state $%
|n_{h}\rangle $ is the halting state of the quantum program and belongs to
an independent two-dimensional state space $\{|0\rangle ,|1\rangle \}$. The
branch-control state $|b_{h}\rangle $ belongs to a larger and independent
state space $\{|0\rangle ,$ $|1\rangle ,$ $|2\rangle ,$ $...\}$ instead of a
simple two-dimensional state space. The index $i$ ($1\leq i\leq m_{r}$)
stands for number of the basic operational unit $Q_{u}$ to have been already
executed. In the quantum program the functions $f_{r}(x)$ and $g_{r}(x)$ are 
$f_{r}(x)=g_{r}(x)=(g^{M_{r}})^{x}\func{mod}p$ for $0\leq x<m_{r}.$ Both the
functions are periodic functions, $f_{r}(x)=f_{r}(x+m_{r})$ and $%
g_{r}(y)=g_{r}(y+m_{r}),$ and they also satisfy $f_{r}(x)=g_{r}(x)=1$ for $%
x=0\func{mod}m_{r}.$ In the quantum program the cyclic group operation $%
U_{g^{M_{r}}}$ acts on only the state $|f_{r}(x)\rangle ,$%
\[
U_{g^{M_{r}}}|f_{r}(x)\rangle |g_{r}(y)\rangle =|f_{r}(x+1)\rangle
|g_{r}(y)\rangle , 
\]
while the state transformation of the unitary operation $U_{r}$ is defined
by 
\begin{equation}
U_{r}|f_{r}(x)\rangle |g_{r}(y)\rangle =\left\{ 
\begin{array}{c}
|f_{r}(x)\rangle |g_{r}(y)\rangle ,\text{ if }x+y\neq 0\func{mod}m_{r}. \\ 
|f_{r}(x)\rangle |1\rangle ,\text{ if }x+y=0\func{mod}m_{r}.
\end{array}
\right.  \label{15}
\end{equation}
Note that for any given indices $x$ and $y$ ($0\leq x,y<m_{r}$) there is a
unique index $i$ $(1\leq i\leq m_{r})$ such that $x+y+i=0\func{mod}m_{r}.$
Therefore, there is a unique index $i$ $(1\leq i\leq m_{r})$ such that the
state $|f_{r}(x+i)\rangle |g_{r}(y)\rangle $ can be changed to the state $%
|f_{r}(x+i)\rangle |1\rangle $ for given indices $x$ and $y$ by the unitary
operation $U_{r}$ in the quantum program.

In order to explain clearly how the quantum program $Q_{p}$ works the
statement $^{\prime \prime }$When $|g_{r}(y)\rangle =|1\rangle ,$ Do $%
|g_{r}(y)\rangle |n_{h}\rangle =|1\rangle |0\rangle \rightarrow |0\rangle
|0\rangle ,$ $|n_{h}\rangle =|0\rangle \rightarrow |1\rangle ,$ halt$%
^{\prime \prime }$ which involves in the halting protocol of quantum Turing
machine [29] in the quantum program is not considered temporarily. The
quantum program starts at the initial state $|b_{h}=0\rangle
|f_{r}(x)\rangle |g_{r}(y)\rangle $ of the quantum system of a universal
quantum computer. The program first checks whether the state $%
|g_{r}(y)\rangle $ is $|1\rangle $ or not. If yes, then the branch-control
state $|b_{h}\rangle =|0\rangle $ is changed to the state $|1\rangle ,$
otherwise it keeps unchanged. If the branch-control state $|b_{h}\rangle $
is not $|0\rangle ,$ then the program performs only the cyclic group
operation $U_{g^{M_{r}}}$, otherwise ($|b_{h}\rangle =|0\rangle $) it
executes another unitary operation sequence, that is, it executes first the
cyclic group operation $U_{g^{M_{r}}}$ and then the unitary operation $%
U_{r}. $ At the end of the step ($i=1$) the quantum system is either $(a)$
in the state $|b_{h}=1\rangle |f_{r}(x+1)\rangle |1\rangle $ if the initial
state $|g_{r}(y)\rangle =|1\rangle $ or $(b)$ in the state $|b_{h}=0\rangle
|f_{r}(x+1)\rangle |1\rangle $ if the initial state $|g_{r}(y)\rangle \neq
|1\rangle $ but $x+y+1=0\func{mod}m_{r}$ or $(c)$ in the state $%
|b_{h}=0\rangle |f_{r}(x+1)\rangle |g_{r}(y)\rangle $ if the initial state $%
|g_{r}(y)\rangle \neq |1\rangle $ and $x+y+1\neq 0\func{mod}m_{r}.$
Therefore, at next step ($i=2$) the three situations need to be considered,
respectively. For the case $(a),$ since the state $|g_{r}(y)\rangle
=|1\rangle $ and the branch-control state $|b_{h}\rangle =|1\rangle $ the
program performs only the cyclic group operation $U_{g^{M_{r}}}$ which
converts the state $|n_{h}=1\rangle |f_{r}(x+1)\rangle |1\rangle $ into the
state $|n_{h}=1\rangle |f_{r}(x+2)\rangle |1\rangle .$ Evidently, once the
state $|g_{r}(y)\rangle $ is transformed to the state $|1\rangle $ and then
the state $|b_{h}\rangle $ to the state $|1\rangle ,$ the two states $%
|g_{r}(y)\rangle $ and $|b_{h}\rangle $ are kept at the state $|1\rangle $
in following steps and even to the end of the program, and hence the program
performs only the cyclic group operation $U_{g^{M_{r}}}$ to the end ($%
i=m_{r} $). Then at the end the quantum system is in the state $%
|b_{h}=1\rangle |f_{r}(x+m_{r})\rangle |1\rangle =|b_{h}=1\rangle
|f_{r}(x)\rangle |1\rangle .$ For the case $(b),$ since the state $%
|g_{r}(y)\rangle =|1\rangle $, then the branch-control state $|b_{h}\rangle
=|0\rangle $ is changed to $|1\rangle ,$ that is, $|b_{h}=0\rangle
|f_{r}(x+1)\rangle |1\rangle $ is transformed to $|b_{h}=1\rangle
|f_{r}(x+1)\rangle |1\rangle $ which will be further changed to the state $%
|b_{h}=1\rangle |f_{r}(x)\rangle |1\rangle $ at the end of the program, as
explained in the case $(a)$. For the case $(c)$, just like at the end of the
step ($i=1$), at the end of the step ($i=2$) there are also three situations
to be considered again and these situations can be analyzed in a similar way
given in the step ($i=1$). The analysis shows that when the program is at
the $k-$th step $(i=k)$ such that $x+y+k=0\func{mod}m_{r},$ the quantum
system is changed from the state $|b_{h}=0\rangle |f_{r}(x+k-1)\rangle
|g_{r}(y)\rangle $ with $|g_{r}(y)\rangle \neq |1\rangle $ at the beginning
to the state $|b_{h}=0\rangle |f_{r}(x+k)\rangle |1\rangle $ at the end of
the $k-$th step by the unitary operation $U_{r}$. At the following step ($%
i=k+1$) the branch-control state $|b_{h}\rangle =|0\rangle $ is transformed
to the state $|1\rangle .$ Then starting from the step $(i=k+1)$ the quantum
system is acted on only by the cyclic group operation $U_{g^{M_{r}}}$ and
this action continues to the end of the program. The final state ($i=m_{k}$)
of the quantum system therefore is $|b_{h}=1\rangle |f_{r}(x+m_{r})\rangle
|1\rangle =|b_{h}=1\rangle |f_{r}(x)\rangle |1\rangle .$ Thus, after
execution of the whole quantum program one time the input state $%
|b_{h}=0\rangle |f_{r}(x)\rangle |g_{r}(y)\rangle $ is changed to the output
state $|b_{h}=1\rangle |f_{r}(x)\rangle |1\rangle .$

However, there is a precondition for the quantum program to work as stated
above that once the state $|g_{r}(y)\rangle $ is changed to the state $%
|1\rangle $ by the unitary operation $U_{r},$ the branch-control state $%
|b_{h}\rangle =|0\rangle $ is changed to the state $|1\rangle $ and \textit{%
since then the branch-control state }$|b_{h}\rangle =|1\rangle $\textit{\ is
kept unchanged to the end of the program}. This precondition may be achieved
by the statement: $^{\prime \prime }$When $|g_{r}(y)\rangle =|1\rangle ,$ Do 
$|g_{r}(y)\rangle |n_{h}\rangle =|1\rangle |0\rangle \rightarrow |0\rangle
|0\rangle ,$ $|n_{h}\rangle =|0\rangle \rightarrow |1\rangle ,$ halt$%
^{\prime \prime }$ in the program. This statement is executed after the
branch-control state $|b_{h}\rangle =|0\rangle $ is changed to the state $%
|1\rangle .$ The statement shows that once the state $|g_{r}(y)\rangle $
goes to the state $|1\rangle ,$ the state $|g_{r}(y)\rangle |n_{h}\rangle
=|1\rangle |0\rangle $ is changed to the state $|0\rangle |0\rangle $ which
means that the state $|g_{r}(y)\rangle =|1\rangle $ is changed to the state $%
|0\rangle $ conditionally when the halting state $|n_{h}\rangle =|0\rangle ,$
then the halting state $|n_{h}\rangle =|0\rangle $ is changed to the state $%
|1\rangle ,$ and \textit{since then the halting state }$|n_{h}\rangle
=|1\rangle $\textit{\ is kept unchanged to the end of the program} which is
executed by the instruction $^{\prime \prime }$halt$^{\prime \prime }$ of
the statement. There are three operations in the statement, the first is the
unitary operation $U_{h}:|g_{r}(y)\rangle |n_{h}\rangle =|1\rangle |0\rangle
\leftrightarrow |0\rangle |0\rangle ,$ the second is the trigger pulse $%
P_{c} $ on the halting qubit$:|n_{h}\rangle =|0\rangle \leftrightarrow
|1\rangle ,$ and the last operation $T(n):^{\prime \prime }$halt$^{\prime
\prime },$ which could involve in the unitary nondemolition measurement
operation on the halting qubit [29, 40], will kept the halting qubit at the
state $|n_{h}\rangle =|1\rangle $ unchanged until the end of the program. It
can be shown that if the halting state $|n_{h}\rangle =|1\rangle $ can be
kept unchanged, then the branch-control state $|b_{h}\rangle =|1\rangle $
can also be kept unchanged. Suppose that at the $i-$th step of the program
the state $|g_{r}(y)\rangle $ goes to the state $|1\rangle ,$ then at the $%
(i+1)- $th step the state $|b_{h}\rangle $ goes to the state $|1\rangle $
which will stop the unitary operation $U_{r}$ later, and then the state $%
|g_{r}(y)\rangle =|1\rangle $ is changed to the state $|0\rangle $ and the
halting state $|n_{h}\rangle $ enters the state $|1\rangle .$ Note that the
cyclic group operation $U_{g^{M_{r}}}$ does not affect the state $%
|g_{r}(y)\rangle $ and the unitary operation $U_{r}$ now is halted. Now at
the $(i+2)-$th step the conditional unitary operation $U_{b}:|b_{h}\rangle
\rightarrow |b_{h}+1\rangle $ does not change the state $|b_{h}\rangle
=|1\rangle $ because the state $|g_{r}(y)\rangle =|0\rangle ,$ and the
unitary operation $U_{h}:|g_{r}(y)\rangle |n_{h}\rangle =|1\rangle |0\rangle
\leftrightarrow |0\rangle |0\rangle $ also has not net effect on the quantum
system because the state $|g_{r}(y)\rangle |n_{h}\rangle =|1\rangle
|1\rangle $ now. Though the unitary operation $P_{c}:|n_{h}\rangle
=|0\rangle \leftrightarrow |1\rangle $ may change the halting state $%
|n_{h}\rangle =|1\rangle $ back to the state $|0\rangle ,$ but the halting
state $|n_{h}\rangle =|1\rangle $ is prevented by the halting operation $%
T(i+2)$ from the action of the unitary operation $P_{c}$ so that it still
keeps at the same state $|1\rangle $ at the step, and this is the key point
for the whole quantum program. Thus, from the $(i+2)-$th step to the end of
the program the halting state is kept at the state $|1\rangle $ and hence
the branch-control state is also kept at the state $|1\rangle $. Obviously,
when the whole quantum program includes the statement: $^{\prime \prime }$%
When $|g_{r}(y)\rangle =|1\rangle ,$ Do $|g_{r}(y)\rangle |n_{h}\rangle
=|1\rangle |0\rangle \rightarrow |0\rangle |0\rangle ,$ $|n_{h}\rangle
=|0\rangle \rightarrow |1\rangle ,$ halt$^{\prime \prime },$ the output
state $|n_{h}\rangle |b_{h}\rangle |f_{r}(x)\rangle |g_{r}(y)\rangle $ is $%
|1\rangle |1\rangle |f_{r}(x)\rangle |0\rangle $ if the input state is $%
|0\rangle |0\rangle |f_{r}(x)\rangle |g_{r}(y)\rangle .$

One might ask one question: is the unitarity of the quantum program
destroyed?, because there are different input states $|0\rangle |0\rangle
|f_{r}(x)\rangle |g_{r}(y)\rangle $ for different states $|g_{r}(y)\rangle ,$
but the quantum program obtains the same output state $|1\rangle |1\rangle
|f_{r}(x)\rangle |0\rangle .$ Actually, there is a different index $i$ ($%
1\leq i\leq m_{r}$) such that $(x+y+i)=0\func{mod}m_{r}$ for a different
input state $|0\rangle |0\rangle |f_{r}(x)\rangle |g_{r}(y)\rangle $ where
the state $|f_{r}(x)\rangle $ may be fixed. Then there is a different time
(e.g., the $i-$th step) for the state $|g_{r}(y)\rangle $ to go to the state 
$|1\rangle $ and for the halting operation $T(i)$ to act on the quantum
system. In effect the halting operation $T(i)$ acting on the quantum system
at different time $i$ is equivalent to that the quantum program in a
different unitary operation acts on the input state. According as the
universal quantum computer model [29], the halting state $|n_{h}\rangle $
should be periodically observed from the outside in a unitary and
nondemolition form so that once the halting state is found at the state $%
|1\rangle $ the halting operation $T(i)$ starts to act on the quantum system
of the quantum computer. Before the halting operation $T(i)$ takes an action
the quantum system has already been made a unitary transformation $U(i)$
which is clearly dependent on the time $i$. Obviously, this unitary
transformation generally is different if the halting operation $T(i)$ takes
an action at a different time, while for the current quantum program this is
clearly correct as well. Therefore, the same output state $|1\rangle
|1\rangle |f_{r}(x)\rangle |0\rangle $ is obtained from different input
state $|0\rangle |0\rangle |f_{r}(x)\rangle |g_{r}(y)\rangle $ by a
different unitary transformation in the current quantum program. Although
different input states can not be converted to the same output state by a
same unitary transformation, they are admitted to change to the same output
state by different unitary transformations! Therefore, the quantum program
keeps its unitarity.

The key point to make the quantum program $Q_{p}$ work as stated above is
that the halting protocol of quantum Turing machine is available and must be
unitary. The unitarity for the halting protocol of quantum Turing machine is
crucial for the quantum program when it is used to solve the quantum search
problem based on the quantum unitary dynamics. Unlike the conventional
measurement operation in quantum computation where the measurement operation
usually could not be unitary and some information could loss during the
measurement operation but these usually do not much affect the final
computing results, the current halting operation must be unitary which
contains the unitary nondemolition measurement operation since it could
carry some information of the input state, as shown before, while the
information could be necessary because in theory the inverse halting
operation which contains the inverse unitary process of the nondemolition
measurement operation could be necessary for solving quantum search problem
based on the unitary quantum dynamics.

The quantum program $Q_{p}$ is really assumed to run on an ideal universal
quantum computer which has the unitary halting protocol of quantum Turing
machine. Obviously, this program is trivial and could be irreversible if it
runs on a conventional classical computer, but it could be simulated
efficiently by the reversible computation [26, 27, 28]. The quantum program
could also be efficiently performed on a quantum Turing machine (QTM) [29,
37, 40], as analyzed above. In fact, in a quantum Turing machine one may set
directly the halting state $|n_{h}\rangle $ in the program to be the QTM
halting-control state to control the quantum program. Once the state $%
|g_{r}(y)\rangle $ is $|1\rangle $ in the program the halting state $%
|n_{h}\rangle =|0\rangle $ is changed to the state $|1\rangle ,$ then the
program stops performing the operational branch consisting of the two
unitary operations $U_{g^{M_{r}}}$ and $U_{r}$ but turns to perform another
operational branch of a single cyclic group operation $U_{g^{M_{r}}}$ to the
end ($i=m_{r}$), which ensures that the whole process of the program is
unitary, as pointed out in [41c]. However, there hides a basic assumption
that \textit{any input state of the quantum program is a single basis state}%
. This basic assumption could ensure that the halting protocol of quantum
Turing machine could be made available and unitary for the quantum program
on an ideal universal quantum computer [29, 40, 41a-41d].

However, if the input state of the quantum program is a superposition $%
\sum_{s}\alpha _{s}|n_{h}=0\rangle |b_{h}=0\rangle |f_{r}(x(s))\rangle
|g_{r}(y(s))\rangle ,$ there seems to be a question whether the halting
protocol can be available and unitary or not on a quantum Turing machine
[41a-41d] when the quantum program is run on the QTM machine. This is
because in this situation there are many operational branches to be executed
simultaneously, and one does not known in advance when the state $%
|g_{r}(y(s))\rangle $ is changed to the state $|1\rangle $ and actually for
different index value $s$ there may be a different time (the index $i$, $%
1\leq i\leq m_{r}$) for the state $|g_{r}(y(s))\rangle $ to go to the state $%
|1\rangle $ in the program, although for any index value $s$ the quantum
program always stops at the same time when the index $i=m_{r}$. At present
there is not a satisfactory halting protocol on a quantum Turing machine
when the input state is a superposition. The detail discussion relevant to
the halting problem of quantum Turing machine for this situation can be seen
in Refs. [41a-41d]. However, it has been shown [29, 40, 41a-41d] that there
is an acceptable halting protocol of quantum Turing machine which may be
made unitary if the input state is limited to be any single basis state on a
quantum Turing machine. Then there should not be any problem to run the
quantum program in a unitary form on a quantum Turing machine if its input
state is limited to be a single basis state. One therefore concludes that if
there existed a universal quantum Turing machine (UQTM) that on it any
computational process obeys the unitary quantum dynamics in physics and it
is capable of computing any computable functions in mathematics such as any
recursive functions which of course include the current one computed by the
quantum program, then such a universal quantum Turing machine could run the
current quantum program in a unitary form when the input state is limited to
be any single basis state for the program.

Though the quantum program could work on a universal quantum Turing machine
and it has been shown that a quantum circuit model is equivalent to a
universal quantum Turing machine in computation [39], it is still a
challenge to construct an efficient quantum circuit for the quantum program.
From the point of view of a quantum circuit model [38] the situation may be
different. A quantum circuit model usually does not use any halting protocol
and its input state can be either a single basis state or a superposition.
However, in order to achieve the same result as the quantum program run on a
universal quantum Turing machine, the quantum circuit model should be really
able to simulate faithfully and efficiently the quantum program and
especially the unitary halting protocol of quantum Turing machine used in
the program. According to the definition (15) of the unitary operation $%
U_{r} $ the unitary operation $U_{k}$ for $k=1,2,...,r$ can be generally
defined by 
\[
|(g^{M_{k}})^{x}\func{mod}p\rangle |(g^{M_{k}})^{-x}\func{mod}p\rangle
\leftrightarrow |(g^{M_{k}})^{x}\func{mod}p\rangle |1\rangle ,\text{ }0\leq
x\leq m_{k}-1, 
\]
\begin{eqnarray*}
|(g^{M_{k}})^{x}\func{mod}p\rangle |(g^{M_{k}})^{y}\func{mod}p\rangle
&\leftrightarrow &|(g^{M_{k}})^{x}\func{mod}p\rangle |(g^{M_{k}})^{y}\func{%
mod}p\rangle , \\
\text{ }x+y &\neq &0\func{mod}m_{k};\text{ }0\leq x,y\leq m_{k}-1,
\end{eqnarray*}
while the conditional unitary operation $U_{k}^{c}$ is defined as 
\begin{eqnarray*}
&&U_{k}^{c}|b_{h}\rangle |(g^{M_{k}})^{x}\func{mod}p\rangle |(g^{M_{k}})^{y}%
\func{mod}p\rangle \\
&=&\left\{ 
\begin{array}{c}
|b_{h}\rangle U_{k}(|(g^{M_{k}})^{x}\func{mod}p\rangle |(g^{M_{k}})^{y}\func{%
mod}p\rangle ),\text{ if }b_{h}=0. \\ 
|b_{h}\rangle |(g^{M_{k}})^{x}\func{mod}p\rangle |(g^{M_{k}})^{y}\func{mod}%
p\rangle ,\text{ if }b_{h}\neq 0.
\end{array}
\right.
\end{eqnarray*}
The unitary operations $U_{k}$ and $U_{k}^{c}$ always can be built up
efficiently since the dimension of the cyclic group state subspace $%
S(m_{k})=\{|(g^{M_{k}})^{x}\func{mod}p\rangle \}$ is $m_{k}$ and $%
m_{k}\thicksim O(\log p)$. The conditional unitary operation $U_{k}^{c}$ is
dependent on the branch-control state $|b_{h}\rangle $. When the
branch-control state $|b_{h}\rangle \neq |0\rangle $ the conditional unitary
operation $U_{k}^{c}$ does not act on the state $|(g^{M_{k}})^{x}\func{mod}%
p\rangle |(g^{M_{k}})^{y}\func{mod}p\rangle $ for any indices $x$ and $y$.
If the unitary operator $U_{k}$ can be written as $U_{k}=\exp [-iH_{k}]$
with the Hamiltonian $H_{k}$, then it is clear that $U_{k}^{c}=\exp
[-i(|b_{h}=0\rangle \langle b_{h}=0|)\bigotimes H_{k}].$ As pointed out
before, the key point for the quantum circuit is to simulate faithfully the
unitary halting protocol of quantum Turing machine. Since the quantum
circuit does not use the halting qubit, one may use an isolated two-level
state control subspace to replace it. Denote the isolated two-level state
subspace as $\{|c\rangle ,|0\rangle \}$. The two states in the subspace $%
\{|c\rangle ,|0\rangle \}$ are not in the cyclic group state subspace $%
S(m_{r})$ but still belong to the Hilbert space $\{|Z_{p}\rangle \}$. The
control unit of the quantum circuit that simulates the halting protocol
consists of a conditional trigger pulse and a conditional state-locking
pulse. The conditional trigger pulse $P_{t}$ is designed to change the state 
$|g_{r}(y)\rangle $ to the state $|c\rangle $ of the control subspace when
the state $|g_{r}(y)\rangle $ is the state $|1\rangle .$ It may be defined
by 
\[
P_{t}|b_{h}\rangle |f_{r}(x)\rangle |g_{r}(y)\rangle =\left\{ 
\begin{array}{c}
|b_{h}\rangle |f_{r}(x)\rangle |c\rangle ,\text{ if }g_{r}(y)=1 \\ 
|b_{h}\rangle |f_{r}(x)\rangle |g_{r}(y)\rangle ,\text{ if }g_{r}(y)\neq 1
\end{array}
\right. 
\]
Note that the conditional trigger pulse $P_{t}$ is different from that
trigger pulse $P_{c}$ in the quantum program $Q_{p}$. The conditional
trigger pulse connects the state $|c\rangle $ of the control subspace to the
state $|1\rangle $ of the cyclic group state subspace $S(m_{r})$. The
time-dependent state-locking pulse $P_{SL}^{c}(\{\varphi _{i}(t)\}),$ where $%
\{\varphi _{i}(t)\}$ are time-dependent control parameters, can be only
applied to the control subspace and does not affect any other states in the
quantum system. Then the state-locking pulse does not make any net effect on
the quantum system if the quantum system is not in the control subspace.
Therefore, the conditional state-locking pulse can take an action on the
quantum system only when the quantum system goes to the states of the
control subspace. The ideal conditional state-locking pulse $%
P_{SL}^{c}(\{\varphi _{i}(t)\})$ could be defined by 
\[
P_{SL}^{c}(\{\varphi _{i}(t)\})|b_{h}\rangle |f_{r}(x)\rangle
|g_{r}(y)\rangle =|b_{h}\rangle |f_{r}(x)\rangle |g_{r}(y)\rangle ,\text{ }%
t<t_{0}^{-}, 
\]
\[
P_{SL}^{c}(\{\varphi _{i}(t)\})|b_{h}\rangle |f_{r}(x)\rangle |c\rangle
=|b_{h}\rangle |f_{r}(x)\rangle |0\rangle ,\text{ }t_{0}\leq t\leq
t_{0}+\Delta t_{0}, 
\]
\[
P_{SL}^{c}(\{\varphi _{i}(t)\})|b_{h}\rangle |f_{r}(x)\rangle |0\rangle
=|b_{h}\rangle |f_{r}(x)\rangle |0\rangle ,\text{ }t>t_{0}+\Delta t_{0}, 
\]
where $t_{0}$ is the time at which the state $|c\rangle $ is generated
completely by the trigger pulse $P_{t}$ and evidently there are $m_{r}$
different times $t_{0}$ at most for the quantum circuit $Q_{c}$ (see below), 
$\Delta t_{0}$ is the interval that the state $|c\rangle $ is converted
completely into the state $|0\rangle $ and it is shorter than the interval
to execute the statement: $^{\prime \prime }$While $|g_{r}(y)\rangle
=|1\rangle ,$ Do $P_{t}:|g_{r}(y)\rangle =|1\rangle \rightarrow |c\rangle ,$ 
$P_{SL}^{c}:|c\rangle \rightarrow |0\rangle ^{\prime \prime }$ (see the
quantum program $Q_{c}$ below). Here also assume that during the period ($%
t_{0}^{-}=t_{0}-\delta t_{0}\leq t<t_{0})$ of the trigger pulse $P_{t}$ the
state-locking pulse has a negligible effect on the quantum system. The
unitary transformation shows that after the state $|c\rangle $ is changed to
the state $|0\rangle $, the state $|0\rangle $ is kept unchanged by the
state-locking pulse and hence it will not change as the time. The
conditional trigger pulse $P_{t}$ instructs what time the conditional
state-locking pulse $P_{SL}^{c}(\{\varphi _{i}(t)\})$ starts to take an
action on the quantum system because before the trigger pulse $P_{t}$
changes the state $|1\rangle $ of the cyclic group state subspace $S(m_{r})$
to the state $|c\rangle $ of the control subspace the quantum system is not
in the control subspace and hence the state-locking pulse has not a net
effect on the quantum system. On the other hand, the conditional trigger
pulse $P_{t}$ can change the state $|1\rangle $ of the cyclic group state
subspace $S(m_{r})$ to the state $|c\rangle $ of the control subspace only
when the state $|g_{r}(y)\rangle $ goes to the state $|1\rangle .$
Therefore, the conditional state-locking pulse $P_{SL}^{c}(\{\varphi
_{i}(t)\})$ can take an action on the quantum system only after the state $%
|g_{r}(y)\rangle $ goes to the state $|1\rangle .$ In effect the conditional
state-locking pulse will replace the halting operation $T(i)$ of the quantum
program $Q_{p}$ to control the quantum circuit $Q_{c},$ as can be seen
below. This is because when the state $|c\rangle $ is changed to the state $%
|0\rangle $ of the control subspace and then the state $|0\rangle $ is
locked by the state-locking pulse, the quantum system really leaves the
state $|c\rangle $ and hence the trigger pulse $P_{t}$ is no longer to take
an action on the quantum system. Using the conditional unitary operation $%
U_{b}:|g_{r}(y)=1\rangle |b_{h}\rangle \rightarrow |g_{r}(y)=1\rangle
|b_{h}+1\rangle $, the conditional unitary operation $U_{r}^{c},$ and the
cyclic group operation $U_{g^{M_{r}}}$ as well as the conditional trigger
pulse $P_{t}$ and the conditional state-locking pulse $P_{SL}^{c}$ a
possible unitary quantum circuit that simulates faithfully and efficiently
the quantum program $Q_{p}$ could be constructed by 
\[
Q_{c}=\{P_{SL}^{c}:OFF\}\{U_{r}^{c}U_{g^{M_{r}}}P_{t}U_{b}\}^{m_{r}}%
\{P_{SL}^{c}:ON\}. 
\]
In fact, given any input basis state this quantum circuit in theory is
exactly equivalent to the following quantum program $Q_{c}$: 
\[
\text{State-Locking Pulse}:ON 
\]
\[
|b_{h}\rangle =|0\rangle 
\]
\[
\text{For }i=1\text{ to }m_{r} 
\]
\[
\text{If }|g_{r}(y)\rangle =|1\rangle \text{ then }|b_{h}\rangle
=|b_{h}+1\rangle \text{ end if} 
\]
\[
\text{While }|g_{r}(y)\rangle =|1\rangle ,\text{ Do }P_{t}:|g_{r}(y)\rangle
=|1\rangle \rightarrow |c\rangle ,\text{ }P_{SL}^{c}:|c\rangle \rightarrow
|0\rangle 
\]
\[
\text{If }|b_{h}\rangle =|0\rangle \text{ then} 
\]
\[
U_{g^{M_{r}}}|f_{r}(x)\rangle |g_{r}(y)\rangle 
\]
\[
U_{r}|f_{r}(x)\rangle |g_{r}(y)\rangle 
\]
\[
\text{else }U_{g^{M_{r}}}|f_{r}(x)\rangle |g_{r}(y)\rangle \text{ end if} 
\]
\[
\text{end for} 
\]
\[
\text{State-Locking Pulse}:OFF. 
\]
The sole difference from the previous one $Q_{p}$ is that the halting qubit $%
\{|n_{h}\rangle \}$ of the quantum program $Q_{p}$ is replaced with the
two-level state subspace $\{|c\rangle ,|0\rangle \}$ in the quantum program $%
Q_{c}.$ Here the input state of the quantum circuit $Q_{c}$ is still limited
to be a single basis state, although a quantum circuit does not limit any
input state. In theory the output state of the quantum circuit $Q_{c}$ is $%
|b_{h}=1\rangle |f_{r}(x)\rangle |0\rangle $ if the input state is the
single basis state $|b_{h}=0\rangle |f_{r}(x)\rangle |g_{r}(y)\rangle .$ The
quantum program $Q_{c}$ shows that the state-locking pulse $P_{SL}^{c}$ is
first applied to the quantum system at the beginning of the quantum circuit.
Because the quantum system may not be in the control subspace at the
beginning, the state-locking pulse does not make an action on the quantum
system, but it keeps applying and does not start to act on the quantum
system until the quantum system goes to the state $|c\rangle ,$ and only at
the end of the quantum circuit the state-locking pulse is switched off.

The performance of the quantum circuit $Q_{c}$ usually may be mainly
dependent on the state-locking pulse $P_{SL}^{c}(\{\varphi _{i}(t)\})$. The
real unitary transformation during the period of the state-locking pulse
applying to the quantum system should be generally written as\newline
\[
P_{SL}^{c}(\{\varphi _{i}(t)\})|b_{h}\rangle |f_{r}(x)\rangle
|g_{r}(y)\rangle =|b_{h}\rangle |f_{r}(x)\rangle |g_{r}(y)\rangle ,\text{ }%
t<t_{0}^{-}, 
\]
\begin{eqnarray*}
&&P_{SL}^{c}(\{\varphi _{i}(t)\})|b_{h}\rangle |f_{r}(x)\rangle |c\rangle \\
&=&|b_{h}\rangle |f_{r}(x)\rangle (\varepsilon (t,t_{0})|c\rangle
+e^{-i\gamma (t,t_{0})}\sqrt{1-|\varepsilon (t,t_{0})|^{2}}|0\rangle ),\text{
}t\geq t_{0},
\end{eqnarray*}
where $\gamma (t,t_{0})$ is a phase factor and the absolute amplitude value $%
|\varepsilon (t,t_{0})|$ is zero in theory when the time $t>t_{0}+\Delta
t_{0}$ for every time $t_{0}$. Hereafter the absolute amplitude value $%
|\varepsilon (t,t_{0})|$ is referred to the one with the time $%
t>t_{0}+\Delta t_{0}.$ The real amplitude value $|\varepsilon (t,t_{0})|$
may be dependent on the real physical process of the quantum circuit. The
amplitude value $|\varepsilon (t,t_{0})|$ measures how close the quantum
circuit $Q_{c}$ is to the quantum program $Q_{p}$, the closer the amplitude
value $|\varepsilon (t,t_{0})|$ to zero, the closer the quantum circuit $%
Q_{c}$ to the quantum program $Q_{p}$. The quantum circuit $Q_{c}$ is really
equivalent to the quantum program $Q_{p}$ when the amplitude value $%
|\varepsilon (t,t_{0})|=0$ exactly for every time $t_{0}$, but this could be
possibly achieved only in an ideal case. However, the amplitude value $%
|\varepsilon (t,t_{0})|$ could not be always equal to zero for every time $%
t_{0}$ if the input state of the quantum circuit is a superposition. This is
one reason why the input state of the quantum circuit $Q_{c}$ is still
limited to be a single basis state, although the input state is allowed to
be any state such as a superposition in the quantum circuit. Therefore, the
quantum circuit $Q_{c}$ is really an approximation to the ideal quantum
program $Q_{p}$ in a real physical process. In practice the conditional
state-locking pulses need to be designed so that the amplitude value $%
|\varepsilon (t,t_{0})|$ is as close zero as possible for every time $t_{0}.$
Hence this involves in quantum control in technique. The conditional
state-locking pulse generally could be an amplitude- and phase-modulation
time-dependent pulse. A better choice for the state-locking pulse could be
an adiabatic pulse.

If there existed a universal quantum computer that in computation obeys the
unitary quantum dynamics in physics and is capable of computing any
computable functions in mathematics such as any recursive functions, then
such an ideal universal quantum computer would be enough powerful to solve
efficiently the quantum search problem in the cyclic group state space.
Actually, by taking the basis state $|n_{h}=0\rangle |b_{h}=0\rangle |\Phi
_{7}\rangle $ as the input state of the quantum program $Q_{p}$ and setting
the function $f_{r}(x)=(g^{M_{r}})^{x}\func{mod}p$ with $x=s_{1}$ and $%
g_{r}(y)=(g^{M_{r}})^{y}\func{mod}p$ with $y=s_{2},$ after executing one
time the quantum program the output state is given by 
\begin{eqnarray*}
|n_{h} &=&0\rangle |b_{h}=0\rangle \bigotimes |(g^{M_{r}})^{s_{1}}\func{mod}%
p\rangle |(g^{M_{r}})^{s_{2}}\func{mod}p\rangle \bigotimes |\Phi
_{7}^{^{\prime }}\rangle \\
\stackrel{Q_{p}}{\rightarrow }|\Phi _{8}\rangle &=&|1\rangle |1\rangle
\bigotimes |(g^{M_{r}})^{s_{1}}\func{mod}p\rangle |0\rangle \bigotimes |\Phi
_{7}^{^{\prime }}\rangle ,
\end{eqnarray*}
where the state $|\Phi _{7}\rangle =|(g^{M_{r}})^{s_{1}}\func{mod}p\rangle
|(g^{M_{r}})^{s_{2}}\func{mod}p\rangle \bigotimes |\Phi _{7}^{^{\prime
}}\rangle $ and the state $|\Phi _{7}^{^{\prime }}\rangle $ is given by 
\[
|\Phi _{7}^{^{\prime }}\rangle =|\mathbf{R0\rangle }\bigotimes
|(g^{M_{r}})^{s_{3}}\func{mod}p\rangle \bigotimes ...\bigotimes
|(g^{M_{r}})^{s_{r}}\func{mod}p\rangle . 
\]
This is because only the state $|(g^{M_{r}})^{s_{1}}\func{mod}p\rangle
\bigotimes |(g^{M_{r}})^{s_{2}}\func{mod}p\rangle $ in the first two
registers of the state $|\Phi _{7}\rangle $ is made the unitary
transformation by the quantum program $Q_{p},$ while the state $|\Phi
_{7}^{^{\prime }}\rangle $ in other registers of the state $|\Phi
_{7}\rangle $ keeps unchanged, and the output state of the quantum program
is $|1\rangle |1\rangle |f_{r}(x)\rangle |0\rangle $ if the input state is $%
|0\rangle |0\rangle |f_{r}(x)\rangle |g_{r}(y)\rangle ,$ as shown before.
This unitary transformation removes the state $|(g^{M_{r}})^{s_{2}}\func{mod}%
p\rangle $ in the second register of the state $|\Phi _{7}\rangle .$ Next
step is to remove unitarily the state $|(g^{M_{r}})^{s_{3}}\func{mod}%
p\rangle $ in the third register of the state $|\Phi _{7}\rangle .$ First,
both the branch-control state $|b_{h}\rangle =|1\rangle $ and the halting
state $|n_{h}\rangle =|1\rangle $ are changed back to the state $|0\rangle $
in the state $|\Phi _{8}\rangle $ and the state $|0\rangle $ in the register
four of the state $|\Phi _{8}\rangle $ is absorbed by the register library $|%
\mathbf{R0\rangle .}$ After these operations the state $|\Phi _{8}\rangle $
is changed to the state $|\Phi _{9}\rangle :$ 
\[
|\Phi _{9}\rangle =|n_{h}=0\rangle |b_{h}=0\rangle \bigotimes
|(g^{M_{r}})^{s_{1}}\func{mod}p\rangle |(g^{M_{r}})^{s_{3}}\func{mod}%
p\rangle \bigotimes |\Phi _{8}^{^{\prime }}\rangle 
\]
where the state $|\Phi _{8}^{^{\prime }}\rangle =|\mathbf{R0\rangle }%
\bigotimes |(g^{M_{r}})^{s_{4}}\func{mod}p\rangle \bigotimes ...\bigotimes
|(g^{M_{r}})^{s_{r}}\func{mod}p\rangle .$ Now taking the state $|\Phi
_{9}\rangle $ as the input state of the quantum program $Q_{p}$ and setting
the function $f_{r}(x)=(g^{M_{r}})^{x}\func{mod}p$ with $x=s_{1}$ and $%
g_{r}(y)=(g^{M_{r}})^{y}\func{mod}p$ with $y=s_{3},$ the unitary
transformation of the quantum program removes the state $|(g^{M_{r}})^{s_{3}}%
\func{mod}p\rangle $ of the state $|\Phi _{9}\rangle .$ In an analogue way,
by setting the fixed function $f_{r}(x)=(g^{M_{r}})^{x}\func{mod}p$ with $%
x=s_{1}$ and the function $g_{r}(y)=(g^{M_{r}})^{y}\func{mod}p$ with $%
y=s_{k} $ for $k=2,3,...,r,$ respectively, and then repeating $r-1$ times
the application of the quantum program $Q_{p},$ the states $%
|(g^{M_{r}})^{s_{k}}\func{mod}p\rangle $ with $k=2,3,..,r$ are one by one
removed unitarily from the state $|\Phi _{7}\rangle $ and ultimately the
state $|\Phi _{7}\rangle $ is transformed to the desired state $|\mathbf{%
R0\rangle }\bigotimes |(g^{M_{r}})^{s_{1}}\func{mod}p\rangle ,$ where the
branch-control state $|0\rangle $ and the halting state $|0\rangle $ are
also absorbed by the register library. This transformation may also be
carried out in a parallel manner. In an analogue way, one may obtain the
desired state $|\mathbf{R0\rangle }\bigotimes |(g^{M_{r}})^{s_{k}}\func{mod}%
p\rangle $ from the state $|\Phi _{7}\rangle $ for $k=1,2,...,r,$
respectively. Once the unitary state transformation $|\mathbf{R0\rangle }%
\bigotimes |g^{s}\func{mod}p\rangle \rightarrow |\mathbf{R0\rangle }%
\bigotimes |(g^{M_{r}})^{s_{k}}\func{mod}p\rangle $ is efficiently achieved
for $k=1,2,...,r$, the auxiliary oracle unitary operation $\overline{U}%
_{os_{k}}(\theta )=\exp [-i\theta \overline{D}_{s_{k}}(g^{M_{r}})]$ with the
quantum-state diagonal operator $\overline{D}_{s_{k}}(g^{M_{r}})=$ $|\mathbf{%
R0\rangle }\langle \mathbf{R0|}\bigotimes |(g^{M_{r}})^{s_{k}}\func{mod}%
p\rangle \langle (g^{M_{r}})^{s_{k}}\func{mod}p|$ can be efficiently built
out of the oracle unitary operation $\overline{U}_{os}(\theta ).$ This
auxiliary oracle unitary operation is applied only to the cyclic group state
subspace $S(m_{r})$. The state $|\mathbf{R0\rangle }\bigotimes
|(g^{M_{r}})^{s_{k}}\func{mod}p\rangle $ may be transferred to the register
of the search space by a $SWAP$ operation so as to obtain the auxiliary
oracle unitary operation $\overline{U}_{os_{k}}(\theta )$ which is applied
only to the search space with dimension $m_{r}\thicksim O(\log p)$. Note
that the register of the search space in which the index vector $\{s_{k}\}$
is determined may be different from all those registers in the state $|\Phi
_{7}\rangle .$ \newline
\newline
{\Large 6. An efficient quantum search process in the cyclic group state
subspaces}

When the auxiliary oracle unitary operation $\overline{U}_{os_{k}}(\theta )$
with $k=1,2,..,r$ is obtained the quantum search process to find the index $%
s_{k}$ can be efficiently constructed. As shown in the previous section 5,
the initial state for the quantum search process should be limited to be a
single basis state because both the input states of the quantum program $%
Q_{p}$ and the quantum circuit $Q_{c}$ are limited to be a single basis
state. Therefore, the standard quantum search algorithm which usually starts
at a superposition will not be used here to determine the index vector $%
\{s_{k}\}$. Because the quantum search space now is limited to the cyclic
group state subspace $S(m_{k})$ with dimensional size $m_{k}\thicksim O(\log
p),$ one may use every basis state of the cyclic group state subspace $%
S(m_{k})$ as the initial state of the quantum search process without
changing essentially the computational complexity of the quantum search
process. For convenience, now the oracle unitary operation $\overline{U}%
_{os_{k}}(\theta )$ acting on a basis state of the search space $S(m_{r})$
can be rewritten as 
\begin{equation}
\overline{U}_{os_{k}}(\theta )|(g^{M_{r}})^{x}\func{mod}p\rangle =\left\{ 
\begin{array}{c}
\exp (-i\theta )|(g^{M_{r}})^{x}\func{mod}p\rangle ,\text{ if }x=s_{k}, \\ 
|(g^{M_{r}})^{x}\func{mod}p\rangle ,\text{ if }x\neq s_{k},
\end{array}
\right.  \label{16}
\end{equation}
where the register library $|\mathbf{R0\rangle }$ is dropped without
confusion. On the other hand, the basis state $|(g^{M_{r}})^{x}\func{mod}%
p\rangle $ with $0\leq x<m_{r}$ can also be expressed in terms of the binary
dynamical parameter $\{b_{k}^{x}\}$ (see sections 2.1 and 2.2), 
\[
|(g^{M_{r}})^{x}\func{mod}p\rangle =\stackrel{n}{\stackunder{k=1}{\bigotimes 
}}(\frac{1}{2}T_{k}+b_{k}^{x}S_{k}). 
\]
The dynamical parameters $\{b_{k}^{x}\}$ can be determined conveniently
below for a given integer $(g^{M_{r}})^{x}\func{mod}p$ and will be used
later in the construction of the quantum search process. The integer $%
(g^{M_{r}})^{x}\func{mod}p$ is first expressed in terms of the usual binary
representation: 
\begin{equation}
(g^{M_{r}})^{x}\func{mod}%
p=a_{n}2^{n-1}+a_{n-1}2^{n-2}+...+a_{2}2^{1}+a_{1}2^{0},  \label{17}
\end{equation}
where the qubit number $n=[\log _{2}p]+1$ and $a_{k}=0$ or $1.$ Then the
dynamical parameter $b_{k}^{x}$ is given by $b_{k}^{x}=(1-2a_{k})$ for $%
k=1,2,..,n.$ Since the oracle unitary operation $\overline{U}%
_{os_{k}}(\theta )$ can generate a phase factor $\exp (-i\theta )$ only for
the marked state $|(g^{M_{r}})^{s_{k}}\func{mod}p\rangle $ but nothing for
any other states of the search space $S(m_{r}),$ as shown in (16), one can
only use this phase factor to distinguish the marked state $%
|(g^{M_{r}})^{s_{k}}\func{mod}p\rangle $ from any other states of the search
space. This search process to find the marked state can be made efficient
due to the fact that the dimension of the search space $S(m_{r})$ is $%
m_{k}\thicksim O(\log p).$ Here, an efficient quantum search process is
suggested to find the marked state in the search space. It is based on the
use of the multiple-quantum unitary operators [42] in the $n-$qubit quantum
spin system ($n=[\log _{2}p]+1$) whose Hilbert space contains the search
space $S(m_{r})$.

A particularly important multiple-quantum transition to be used in the
quantum search process is the highest-order quantum transition in the $n-$
qubit quantum spin system. The highest-order quantum transition is defined
as the transition between the ground state $|00...0\rangle $ and the highest
excited state $|11...1\rangle $ of the $n-$qubit spin system. In an $n-$%
qubit spin system the highest order of quantum transition is $\pm n$ [43]
and the Hermitian highest-order quantum operators $Q_{nx}$ and $Q_{ny}$ may
be defined by 
\begin{equation}
Q_{nx}=\frac{1}{2}%
(I_{1}^{+}I_{2}^{+}......I_{n}^{+}+I_{1}^{-}I_{2}^{-}......I_{n}^{-}),
\label{18}
\end{equation}
and 
\begin{equation}
Q_{ny}=\frac{1}{2i}%
(I_{1}^{+}I_{2}^{+}......I_{n}^{+}-I_{1}^{-}I_{2}^{-}......I_{n}^{-}),
\label{19}
\end{equation}
where the operators $I_{k}^{\pm }=I_{kx}\pm iI_{ky}$ for $k=1,2,...,n$. The
highest-order quantum unitary operators are defined by $U_{n\mu }(\theta
)=\exp (-i2\theta Q_{n\mu })$ with $\mu =x,y.$ They can induce an $n-$order
quantum transition only between the ground state $|00...0\rangle $ and the
highest excited state $|11...1\rangle $ of the Hilbert space of the $n-$%
qubit spin system, but they do not induce any other order quantum transition
between any pair of quantum states of the spin system different from the
pair of the ground state and the highest excited state. This is because the
transition matrix elements $\langle k|Q_{n\mu }|r\rangle =\langle r|Q_{n\mu
}|k\rangle ^{*}=0$ $(\mu =x,y)$ for any computational base $|k\rangle $ and $%
|r\rangle $ of the spin system other than the ground state $|00...0\rangle $
or the highest excited state $|11...1\rangle $. Since $I_{k}^{+}|0_{l}%
\rangle =0,$ $I_{k}^{+}|1_{l}\rangle =\delta _{kl}|0_{k}\rangle ,$ $%
I_{k}^{-}|0_{l}\rangle =\delta _{kl}|1_{k}\rangle ,$ and $%
I_{k}^{-}|1_{l}\rangle =0$ [43] for $k,l=1,2,...,n,$ the $n-$order quantum
operator $Q_{ny}$ acting on the ground state (the highest excited state)
creates the highest excited state (the ground state), 
\[
2Q_{ny}|00...0\rangle =i|11...1\rangle 
\]
and 
\[
2Q_{ny}|11...1\rangle =-i|00...0\rangle . 
\]
Then it is easy to turn out that there are the unitary transformations when
the $n-$order quantum unitary operator $U_{ny}(\theta )=\exp (-i2\theta
Q_{ny})$ acts on the ground state and the highest excited state,
respectively, 
\begin{equation}
\exp (-i2\theta Q_{ny})|00...0\rangle =\cos \theta |00...0\rangle +\sin
\theta |11...1\rangle ,\text{ }(-\pi \leq \theta \leq \pi )  \label{20}
\end{equation}
and 
\begin{equation}
\exp (-i2\theta Q_{ny})|11...1\rangle =\cos \theta |11...1\rangle -\sin
\theta |00...0\rangle ,\text{ }(-\pi \leq \theta \leq \pi ).  \label{21}
\end{equation}
In particular, when $\theta =\pi /4$ the equally weighted superposition of
the ground state and the highest excited state is obtained from (20), 
\begin{equation}
|\Psi _{0n}\rangle =\exp (-i\frac{\pi }{2}Q_{ny})|00...0\rangle =\frac{1}{%
\sqrt{2}}(|00...0\rangle +|11...1\rangle ).  \label{22}
\end{equation}

The efficient quantum circuit for the highest-order quantum unitary operator 
$U_{ny}(\theta )$ is constructed below. By using the quantum-state diagonal
operator $D_{0}$ [15] the $n-$order quantum operator $Q_{ny}$ may be
expressed as 
\begin{equation}
2iQ_{ny}=[D_{0},2^{n}I_{1x}I_{2x}...I_{nx}].  \label{23}
\end{equation}
On the other hand, the operator $Q_{ny}$ can also be written as 
\begin{equation}
2iQ_{ny}=(-i)\exp (i\varphi I_{z})[D_{0},2^{n}I_{1x}I_{2x}...I_{nx}]_{+}\exp
(-i\varphi I_{z})  \label{24}
\end{equation}
with $n\varphi =\pi /2.$ The relation (24) can be proved below. Since there
holds the unitary transformation: $\exp (-i\varphi I_{kz})I_{k}^{\pm }\exp
(i\varphi I_{kz})=\exp (\mp i\varphi )I_{k}^{\pm }$ [43] it follows from
(18) and (19) that there exists the unitary transformation when the unitary
operator $\exp (-i\varphi I_{z})=\exp [-i\varphi \sum_{k=1}^{n}I_{kz}$ $]$
with $n\varphi =\pi /2$ acts on the $n-$order quantum operator $Q_{ny}$, 
\begin{eqnarray*}
&&\exp (-i\varphi I_{z})Q_{ny}\exp (i\varphi I_{z}) \\
&=&\frac{1}{2i}[\exp (-in\varphi )I_{1}^{+}I_{2}^{+}......I_{n}^{+}-\exp
(in\varphi )I_{1}^{-}I_{2}^{-}......I_{n}^{-}] \\
&=&-\frac{1}{2}[D_{0},2^{n}I_{1x}I_{2x}...I_{nx}]_{+}=-Q_{nx}
\end{eqnarray*}
Obviously, the relation (24) can be obtained directly from this unitary
transformation. There is a general unitary transformation identity for the
selective rotation operation $C_{t}(\theta )$ [15]$,$%
\begin{eqnarray}
C_{t}(\theta )\rho C_{t}(\theta )^{-1} &=&\rho -(1-\cos \theta )[\rho
,D_{t}]_{+}+i\sin \theta [\rho ,D_{t}]  \nonumber \\
&&+2(1-\cos \theta )D_{t}\rho D_{t}.  \label{25}
\end{eqnarray}
Taking $\rho =2^{n}I_{1x}I_{2x}...I_{nx},$ $D_{t}=D_{0},$ and $\theta =\pi ,$
and noting that there holds the operator identity $%
D_{t}2^{n}I_{1x}I_{2x}...I_{nx}D_{t}=0$ for any index $t,$ one obtains the
following relation from the identity (25), 
\begin{eqnarray}
\lbrack D_{0},2^{n}I_{1x}I_{2x}...I_{nx}]_{+} &=&\frac{1}{2}%
\{2^{n}I_{1x}I_{2x}...I_{nx}  \nonumber \\
&&-C_{0}(\pi )2^{n}I_{1x}I_{2x}...I_{nx}C_{0}(\pi )^{-1}\}.
\end{eqnarray}
With the help of the relations (24) and (26) and the Trotter-Suzuki formula
[44] the quantum circuit for the highest-order quantum unitary operator $%
U_{ny}(\theta )$ can be constructed efficiently by 
\[
U_{ny}(\theta )=\exp (-i2\theta Q_{ny})
\]
\begin{equation}
=\exp (i\varphi I_{z})\{C_{0}(\pi )GC_{0}{}(\pi )^{-1}G^{-1}\}^{m}\exp
(-i\varphi I_{z})+O(m^{-1})  \label{27}
\end{equation}
where the unitary operation $G=\exp (-i\theta 2^{n-1}I_{1x}I_{2x}...I_{nx}/m)
$ can be decomposed efficiently into a sequence of one- and two-qubit
quantum gates [15]. Note that the norms $||D_{0}||=1$ and $%
||2^{n}I_{1x}I_{2x}...I_{nx}||=1.$ For a modest integer $m$ the
decomposition (27) converges quickly.

With the help of the unitary transformations of (20) and (21) of the
highest-order quantum unitary operator $U_{ny}(\theta )$ one can set up two
quantum circuits to judge whether a known quantum state is just the solution
of the quantum search problem or not in polynomial time. One quantum circuit 
$U_{0n}(\pi )$ is constructed with the selective inversion operation $%
C_{t}(\pi )$ and the highest-order quantum unitary operator $U_{ny}(\theta
), $ 
\begin{eqnarray*}
U_{0n}(\pi )|00...0\rangle &=&\exp (i\frac{1}{2}\pi Q_{ny})C_{t}(\pi )\exp
(-i\frac{1}{2}\pi Q_{ny})|00...0\rangle \\
&=&\left\{ 
\begin{array}{l}
\ \ |11...1\rangle ,\text{ if }t=0 \\ 
-|11...1\rangle ,\text{ if }t=N-1 \\ 
\ \ |00...0\rangle ,\text{ if }t\neq 0,N-1
\end{array}
\right.
\end{eqnarray*}
where $N=2^{n}$. The quantum circuit $U_{0n}(\pi )$ acting on the ground
state $|00...0\rangle $ induces the highest-order quantum transition only
when the selective inversion operation $C_{t}(\pi )$ with $t=0$ or $N-1$ is
applied to either the ground state $|00...0\rangle $ or the highest excited
state $|11...1\rangle ,$ while for any other selective inversion operation $%
C_{t}(\pi )$ with $t\neq 0$ and $N-1$ which is applied to neither the ground
state nor the highest excited state the quantum circuit $U_{0n}(\pi )$
induces no transition from the ground state to the highest excited state.

Generally, the quantum circuit $U_{0n}(\theta )$ with a general selective
rotation operation $C_{t}(\theta )$ $(-\pi \leq \theta \leq \pi )$ acting on
the ground state induces the $n-$order quantum transition with a transition
probability dependent on the rotation angle $\theta ,$%
\begin{eqnarray*}
&&\exp (i\frac{1}{2}\pi Q_{ny})C_{t}(\theta )\exp (-i\frac{1}{2}\pi
Q_{ny})|00...0\rangle \\
&=&\left\{ 
\begin{array}{l}
P_{+}|00...0\rangle +P_{-}|11...1\rangle \newline
,\text{ if }t=0 \\ 
P_{+}|00...0\rangle -P_{-}|11...1\rangle ,\text{ if }t=N-1 \\ 
|00...0\rangle ,\text{ if }t\neq 0,N-1
\end{array}
\right.
\end{eqnarray*}
with $P_{\pm }=\frac{1}{2}(1\pm \exp (-i\theta )),$ but the quantum circuit
does not induce any quantum transition when the selective rotation operation 
$C_{t}(\theta )\neq $ $C_{0}(\theta )$ and $C_{N-1}(\theta )$. When $%
C_{t}(\theta )=C_{0}(\theta )$ or $C_{N-1}(\theta )$ the unitary operation $%
U_{0n}(\theta )$ does induce the highest order quantum transition with the
transition probability: 
\[
P_{0n}(\theta )=|P_{-}|^{2}=\frac{1}{2}(1-\cos \theta ). 
\]
The transition probability $P_{0n}(\theta )\geq 0.5$ when $\pi /2\leq
|\theta |\leq \pi .$

Using the total quantum circuit $U_{0n}(\theta )|00...0\rangle $ $(\pi
/2\leq |\theta |\leq \pi )$ which includes the initial state, i.e., the
ground state, one can know whether the quantum state $|t\rangle $ is one of
the two states: the ground state and the highest excited state or any other
quantum state of the Hilbert space. If the quantum state $|t\rangle $ is
either the ground state or the highest excited state, then one need use
further another quantum circuit $U_{0n}^{\prime }$ to determine certainly
the quantum state $|t\rangle $ to be the ground state or the highest excited
state, 
\begin{eqnarray*}
U_{0n}^{\prime }|00...0\rangle &\equiv &\exp (i\frac{1}{2}\pi
Q_{ny})C_{0}(\pi /2)C_{t}(-\pi /2)\exp (-i\frac{1}{2}\pi
Q_{ny})|00...0\rangle \\
&=&\left\{ 
\begin{array}{l}
\ |00...0\rangle ,\text{ if }t=0 \\ 
i|11...1\rangle ,\text{ if }t=N-1
\end{array}
\right. .
\end{eqnarray*}
If the quantum state $|t\rangle $ is the highest excited state, which means
that $C_{t}(-\pi /2)=C_{N-1}(-\pi /2),$ then there is an $n-$order quantum
transition from the ground state to the highest excited state under the
action of the unitary operation $U_{0n}^{\prime }$ on the ground state,
otherwise there is not such an $n-$order quantum transition and the ground
state keeps unchanged. Now it is easy to judge if an unknown state $%
|t\rangle $ is the state $|00...0\rangle \}$ or the state $|11...1\rangle $
or any other state of the Hilbert space by using first the quantum circuit $%
U_{0n}(\pi )|00...0\rangle $ and then $U_{0n}^{\prime }|00...0\rangle .$

It is well known in computational complexity that an NP-hard problem is hard
to be solved on a classical computer, but whether a given solution is just
the real solution to the NP problem or not can be efficiently checked
computationally. This fact is also true on a quantum computer. How to
confirm whether a given state is the solution to the quantum search problem
on a quantum computer? Suppose that the marked state $|s\rangle $ is the
real solution to the quantum search problem and the oracle unitary operation
of the marked state is $C_{s}(\theta )$. For a given quantum state $%
|r\rangle $ one knows its dynamical parameter vector $\{a_{k}^{r}\},$ an
example can be seen in equation (17). One first sets up an auxiliary oracle
unitary operation $C_{t}(\theta )=U_{or}C_{s}(\theta )U_{or}^{+}:$ 
\[
U_{or}C_{s}(\theta )U_{or}^{+}=\left\{ 
\begin{array}{l}
C_{0}(\theta ),\text{ if }|r\rangle =|s\rangle \\ 
C_{t}(\theta )\text{ }(t\neq 0),\text{ if }|r\rangle \neq |s\rangle
\end{array}
\right. 
\]
where the known unitary operator $U_{or}$ that depends upon the dynamical
parameter vector $\{a_{k}^{r}\}$ is given by [15a], 
\[
U_{or}=\stackrel{n}{\stackunder{k=1}{\prod }}\{\exp (i\pi I_{kx}/2)\exp
(-i\pi a_{k}^{r}I_{kx}/2)\}. 
\]
Then using the quantum circuit $U_{0n}(\pi )|00...0\rangle $ one knows
whether the auxiliary oracle unitary operation $C_{t}(\theta )$ is just $%
C_{0}(\theta )$ or $C_{N-1}(\theta )$ or any other one. If $C_{t}(\theta
)\neq C_{0}(\theta )$ and $C_{N-1}(\theta ),$ then the quantum state $%
|r\rangle $ is not the real solution $|s\rangle $ to the quantum search
problem. If $C_{t}(\theta )=C_{0}(\theta )$ or $C_{N-1}(\theta ),$ then the
quantum circuit $U_{0n}^{\prime }|00...0\rangle $ is further used to judge
whether $C_{t}(\theta )=C_{0}(\theta )$ or $C_{t}(\theta )=C_{N-1}(\theta ).$
If $C_{t}(\theta )=C_{N-1}(\theta ),$ then the state $|r\rangle $ is not the
solution $|s\rangle $. But if $C_{t}(\theta )=C_{0}(\theta )$ one knows
certainly the quantum state $|r\rangle $ is just the solution $|s\rangle .$
Therefore, in polynomial time one can confirm whether a given quantum state
is just the solution to the quantum search problem.

Both the ground state $|00...0\rangle $ and the highest excited state $%
|11...1\rangle $ of the Hilbert space of the $n-$qubit spin system with $%
n=[\log _{2}p]+1$ do not belong the search space $S(m_{k}).$ This is clear
that the ground state $|00...0\rangle $ is not contained in the
multiplicative cyclic group state space $S(C_{p-1})$, as shown in section
2.1. On the other hand, the prime $p$ is less than $2^{n}$ with $n=[\log
_{2}p]+1,$ that is, $p\leq 2^{n}-1,$ then $p-1\leq 2^{n}-2,$ which means
that every cyclic group state $|g^{x}\func{mod}p\rangle $ of the state space 
$S(C_{p-1})$ corresponds one-to-one to its own integer $g^{x}\func{mod}p\in
Z_{p}^{+}$ which is never greater than $2^{n}-2,$ while the highest excited
state $|11...1\rangle $ stands for the number $2^{n}-1.$ Therefore, the
cyclic group state space $S(C_{p-1})$ does not contain the state $%
|11...1\rangle .$ By checking the quantum program $Q_{p}$ and the quantum
circuit $Q_{c}$ in section 5 one can see that the state $|00...0\rangle $
has been used by the program $Q_{p}$ in the unitary transformation: $%
|g_{r}(y)\rangle |n_{h}\rangle =|1\rangle |0\rangle \leftrightarrow
|0\rangle |0\rangle $ and by the quantum circuit $Q_{c}$ as the control
state $|0\rangle $ of the control subspace $\{|c\rangle ,|0\rangle \}$ with $%
|c\rangle \neq |11...1\rangle $, but that state $|00...0\rangle $ is not in
the current search space $S(m_{r})$, while the state $|11...1\rangle $ of
the Hilbert space that contains the search space $S(m_{r})$ is never used by
both the program and the quantum circuit. Therefore, there hold the unitary
transformations: $Q_{p}|00...0\rangle =|00...0\rangle $ and $%
Q_{p}|11...1\rangle =|11...1\rangle $ in the search space $S(m_{r})$. This
is also in agreement with the fact that the oracle unitary operation $%
\overline{U}_{os_{k}}(\theta )$ does not make an effect on both the states.
The quantum circuit $U_{0n}(\pi )|00...0\rangle $ now can be modified so
that it can be used to determine the index $s_{k}$ of the oracle unitary
operation $\overline{U}_{os_{k}}(\theta ).$ Obviously, the superposition $%
|\Psi _{0n}\rangle $ of (22) is not in the search space $S(m_{r})$ and not
affected by the quantum program $Q_{p}.$ Now the ground state $%
|00...0\rangle $ in the superposition $|\Psi _{0n}\rangle $ is changed to
the state $|1\rangle $ by the unitary operation $F_{1}$ and further changed
to the cyclic group state $|(g^{M_{r}})^{x}\func{mod}p\rangle $ by the
cyclic group operation $(U_{g^{M_{r}}})^{x},$%
\begin{eqnarray*}
|\Psi _{0n}\rangle &=&\frac{1}{\sqrt{2}}(|00...0\rangle +|11...1\rangle ) \\
\stackrel{F_{1}}{\rightarrow }\stackrel{(U_{g^{M_{r}}})^{x}}{\rightarrow }%
|\Psi _{1n}\rangle &=&\frac{1}{\sqrt{2}}(|(g^{M_{r}})^{x}\func{mod}p\rangle
+|11...1\rangle ).
\end{eqnarray*}
The unitary operation $F_{1}$ and the cyclic group operation $%
(U_{g^{M_{r}}})^{x}$ do not affect the highest-level state $|11...1\rangle $%
. If now the superposition $|\Psi _{1n}\rangle $ is taken as the input state
of the oracle unitary operation $\overline{U}_{os_{k}}(\pi ),$ then in
effect the input state is essentially a single basis state for the oracle
unitary operation $\overline{U}_{os_{k}}(\pi )$ and also for the quantum
program $Q_{p}.$ Since the highest-level state $|11...1\rangle $ is not in
the search space $S(m_{r})$ and also not affected by the quantum program,
there is only the single basis state $|(g^{M_{r}})^{x}\func{mod}p\rangle $
in the state $|\Psi _{1n}\rangle $ that the quantum program can take an
action, although the state $|\Psi _{1n}\rangle $ is a superposition of two
states. Now it is applied the oracle unitary operation $\overline{U}%
_{os_{k}}(\pi )$ to the state $|\Psi _{1n}\rangle .$ Note that only the
basis state $|(g^{M_{r}})^{x}\func{mod}p\rangle $ in the state $|\Psi
_{1n}\rangle $ is affected by the oracle unitary operation. If the index $%
s_{k}=x$ then the state $|(g^{M_{r}})^{x}\func{mod}p\rangle $ is inverted by
the oracle unitary operation $\overline{U}_{os_{k}}(\pi )$, otherwise the
state $|\Psi _{1n}\rangle $ keeps unchanged. After these unitary operations
the state $|(g^{M_{r}})^{x}\func{mod}p\rangle $ is changed back to the
ground state $|0\rangle $ by the inverse operations $%
[(U_{g^{M_{r}}})^{x}]^{+}$ and $F_{1}^{+}.$ At the final step the inverse $%
n- $order quantum unitary operation $\exp (i\frac{1}{2}\pi Q_{ny})$ is
applied so that it can be judged whether the index $x=s_{k}$ or not by the
quantum measurement. The final result is given by 
\begin{eqnarray*}
Q(x,s_{k})|00...0\rangle &=&\exp (i\frac{1}{2}\pi
Q_{ny})F_{1}^{+}[(U_{g^{M_{r}}})^{x}]^{+}\overline{U}_{os_{k}}(\pi ) \\
&&\times (U_{g^{M_{r}}})^{x}F_{1}\exp (-i\frac{1}{2}\pi Q_{ny})|00...0\rangle
\\
&=&\left\{ 
\begin{array}{l}
\ \ |11...1\rangle ,\text{ if }x=s_{k} \\ 
\ \ |00...0\rangle ,\text{ if }x\neq s_{k}
\end{array}
\right.
\end{eqnarray*}
The quantum measurement is carried out on the highest-level state $%
|11...1\rangle .$ Given the oracle unitary operation $\overline{U}%
_{os_{k}}(\pi )$ one can try $m_{r}$ different index values $%
x=0,1,...,m_{r}-1$ at most with the quantum circuit $Q(x,s_{k})|00...0%
\rangle $ to find the index $s_{k}$ due to the fact that $0\leq s_{k}<m_{r}.$
If the highest-level state $|11...1\rangle $ is measured in a high
probability ($\thicksim 1)$, then the corresponding index value $x$ is just
the index $s_{k}.$ Again it is pointed out that the input state of the
quantum program $Q_{p}$ is essentially limited to be a single basis state
during the quantum search process. When the index values $\{s_{k}\}$ are
obtained one may use the index identity (3) or (12) to compose the index $s$
and hence the marked state $|s\rangle $ is found ultimately for the quantum
search problem in the cyclic group state space. \newline
\newline
{\large 7. Discussion}

In the paper an oracle-based quantum dynamical method has been set up to
solve the quantum search problem in the cyclic group state space of the
Hilbert space of an $n-$qubit pure-state quantum system. The main attempt is
to make use of the symmetric properties and structures of groups to help
solving a general unstructured quantum search problem in the Hilbert space.

It is known that the hardness to solve an unstructured quantum search
problem by a standard quantum search algorithm mainly originates from the
low efficiency to amplify the amplitude of the marked state in the Hilbert
space by the oracle unitary operation associated with other known quantum
operations. This low amplitude-amplification efficiency results in that a
standard quantum search algorithm generally can have only a square speedup
over the best known classical counterparts. In order to break through the
square speedup limitation it is necessary to develop other type of quantum
search algorithms. The quantum dynamical method [15] may be a better choice,
for it allows a parameterization description for an unknown quantum state
such as the marked state and its oracle unitary operation in the Hilbert
space of the $n-$qubit quantum system. Since the oracle unitary operation
corresponds one-to-one to the unknown marked state, with the help of the
parameterization description the quantum dynamical method makes it possible
to manipulate at will the evolution process of the marked state in the
quantum system and hence it also makes it possible to manipulate at will the
oracle unitary operation. The quantum dynamical method is different from the
standard quantum search algorithm in that any quantum state of the Hilbert
space can be described completely by a set of dynamical parameters and hence
the quantum searching for the marked state can be indirectly achieved by
determining the set of dynamical parameters which describe completely the
marked state instead by directly measuring the marked state. Therefore,
amplification of amplitude of the marked state and the direct measurement on
the marked state to obtain the complete information of the marked state,
both are the key components of a standard quantum search algorithm, may not
be necessary in the quantum dynamical method. In the quantum dynamical
method the quantum measurement to output the computing results may be
carried out on those states that carry the information of the marked state,
while the complete information of the marked state can be further extracted
from these computing results. In the paper the binary dynamical
representation for a quantum state in the Hilbert space of an $n-$qubit
quantum system is generalized to a general multi-base dynamical
representation for a quantum state in a cyclic group state space and the
quantum dynamical method therefore is extended to solve the quantum search
problem in the cyclic group state space of the Hilbert space.

A cyclic group state space of the Hilbert space of an $n-$qubit quantum
system carries the symmetric property and structure of the cyclic group. A
quantum search process may be affected greatly by the symmetric property and
structure of the cyclic group if the quantum search is performed in the
cyclic group state space. It is known that the amplitude-amplification
efficiency for the marked state by the oracle unitary operation associated
with other known unitary operations generally is inversely proportional to
the square root of the dimensional size of the search space of the quantum
search problem and this low efficiency results in the square speedup
limitation for a standard quantum search algorithm. There is naturally a
possible scheme to bypass this speedup limitation that the search space of
the problem is limited to a small subspace of the Hilbert space so that this
speedup limitation becomes less important or even unimportant in the quantum
search problem. Therefore, it is a challenge how to reduce efficiently the
search space from the whole Hilbert space to its small subspaces in the
unstructured quantum search problem in the Hilbert space. It has been shown
that the symmetric property and structure in spin space of an $n-$qubit spin
system may be helpful for this reduction of search space. In the paper it is
made a further emphasis and generalization for the idea that the symmetric
property and structure of a quantum system or even a group may be employed
to speed up the quantum search process through the scheme of the
search-space reduction. A cyclic group is one of the simplest groups and its
symmetric property and structure has been studied in detail and extensively.
Therefore, it could be simplest and most convenient to exploit the symmetric
property and structure of a cyclic group to help solving the quantum search
problem in the cyclic group state space of the Hilbert space.

The reversible mathematical-logic operations have been used extensively in
quantum computation. They may be generally thought of as selective unitary
operations in a quantum system and have be employed in the construction of
quantum search processes in the cyclic group state space. A large advantage
for the type of unitary operations is that the time evolution process of a
quantum state in a complex multi-qubit quantum system may be traced more
easily under the action of the mathematical-logic operations. However, in
order to be reversible and unitary a logic operation in mathematics usually
needs to consume much more extra auxiliary qubits with respect to those
unitary operators quantum physically. Since the dimensional size of the
Hilbert space of a quantum system increases exponentially as the qubit
number, it must be careful to use the mathematical-logic operations in
solving a quantum search problem, otherwise these extra auxiliary qubits
could lead to a large search space for the quantum search problem and make
the quantum search process degraded. On the other hand, the conventional
unitary operators, propagators, operations, or quantum gates in a quantum
system in physics usually need not any extra auxiliary qubits except those
artificial conditional unitary operations which usually need only few extra
qubits to help to achieve some specific conditional operations instead of
their unitarity. The time evolution process of a quantum state in a
multi-qubit quantum system generally is complex and is not easy to trace
under the action of the type of unitary operations. However, there is a
general rule that any unknown quantum state can be efficiently transferred
to a larger subspace from a small subspace in the Hilbert space of the
multi-qubit quantum system. Through this general rule one could set up the
connection between the Hilbert space of the $n-$qubit quantum system and its
cyclic group state space for an unstructured quantum search problem.

It has been shown that if there existed a universal quantum computer that in
computation obeys the unitary quantum dynamics in physics and is capable of
computing any computable functions in mathematics such as any recursive
functions, then such a universal quantum computer would be enough powerful
to solve efficiently the quantum search problem in the cyclic group state
space. There seems to be a question whether such an ideal universal quantum
computer existed or not. This question is due to the argument that a
universal quantum computer could not have a satisfactory halting protocol
when its input state is a superposition. However, as far as the present
quantum search process in the multiplicative cyclic group state space is
concerned, there seems not to be such a question because the input state in
the quantum search process can be strictly limited to be a single basis
state. An ideal quantum program, which is a key component of the present
quantum search process, is designed for the efficient reduction of quantum
search space for the quantum search problem. It has been shown in theory
that this quantum program could be run unitarily on an ideal universal
quantum computer when its input state is strictly limited to be a single
basis state and hence it could be used to solve efficiently the quantum
search problem in the cyclic group state space. Moreover, a quantum circuit
is also designed to simulate efficiently the ideal quantum program. The key
point for the quantum circuit is to use the state-locking pulse and the
two-level control subspace to simulate efficiently the unitary halting
protocol of the quantum program. Although at present a state-locking pulse
that is continuously applied to a quantum system during the whole period of
the quantum circuit is not popularly used in quantum computation, a large
number of similar techniques have been used extensively in the conventional
NMR experiments [43]. Obviously, it is necessary to further investigate in
detail the quantum circuit in some important problems such as how to design
a state-locking pulse with a better performance and how the state-locking
pulse affects the practical computational complexity of the quantum circuit
and the whole quantum search process. Evidently, it is possible to design
simpler quantum program and quantum circuit than the present ones to solve
the quantum search problem in the cyclic group state space.

With the help of the symmetric property and structure of a cyclic group and
the Chinese remainder theorem in number theory any quantum state in the
cyclic group state space can be efficiently converted into a tension product
of the states of the cyclic group state subspaces of the cyclic group state
space. There are the relations among these states of the cyclic group state
subspaces through the Chinese remainder theorem. These relations are
important and may be further employed to develop efficient quantum search
methods in the cyclic group state space in the future work. \newline
$\newline
${\large References}\newline
* E\_mail address for the author: miaoxijia@yahoo.com.\newline
1. (a) S.A.Cook, The P versus NP problem, http://www.cs.toronto.edu/ \symbol{%
126}sacook, 2000; (b) M.R.Garey and D.S.Johnson, Computers and
Intractability: A guide to the theory of NP-completeness, Freeman and
Company, New York, 1979; (c) C.Papadimtriou, Computational Complexity,
Addison-Wesley, 1994. \newline
2. L.K.Grover, Quantum mechanics helps in searching for a needle in a
haystack, Phys.Rev.Lett. 79, 325 (1997);\newline
3. C.H.Bennett, E.Bernstein, G.Brassard, and U.Vazirani, Strengths and
weaknesses of quantum computing, http://arxiv.org/abs/quant-ph/9701001 (1997)%
\newline
4. E.Farhi and S.Gutmann, Analog analogue of digital quantum computation,
Phys.Rev. A 57, 2403 (1998); \newline
5. E.Farhi, J.Goldstone, S.Gutmann, and M.Sipser, Quantum computation by
adiabatic evolution, http://arxiv.org/abs/quant-ph/0001106 (2000)\newline
6. G.Brassard, P.Hoyer, M.Mosca, and A.Tapp, Quantum amplitude amplification
and estimation, http://arxiv.org/abs/quant-ph/0005055 (2000) \newline
7. N.J.Cerf, L.K.Grover, and C.P.Williams, Nested quantum search and
NP-complete problems, Phys. Rev. A 61, 2303 (2000) (see also:
quant-ph/9806078)\newline
8. T. Hogg, A framework for structured quantum search, http://arxiv.org
/abs/quant-ph/9701013 (1997).\newline
9. C.Zalka, Grover quantum searching algorithm is optimal, Phys.Rev. A 60,
27462751 (1999) (see also: quant-ph/9711070) \newline
10. E.Biham, O.Biham, D.Biron, M.Grass, D.A.Lidar, and D.Shapira, Analysis
of Generalized Grover's Quantum search algorithms using recursion equations,
http://arxiv.org/abs/quant-ph/0010077 (2000) \newline
11. N.Shenvi, J.Kempe, K.Whaley, Quantum random-walk search algorithm,
Phys.Rev. A, 67, 052307 (2003)\newline
12. A.Childs and J.Goldstone, Spatial search by quantum walk,

http://arxiv.org/abs/quant-ph/0306054 (2003) \newline
13. W. van Dam, M.Mosca, and U.Vazirani, How powerful is adiabatic quantum
computation?, http://arxiv.org/abs/quant-ph/0206003 (2002)\newline
14. B.Robert, H.Buhrman, R.Cleve, M.Mosca, and R.De Wolf, Quantum lower
bounds by polynomials, Proceedings of 39th Annual Symposium on Foundations
of Computer Science, pp. 352 (1998)\newline
15. (a) X.Miao, Universal construction for the unsorted quantum search
algorithms, http://arxiv.org/abs/quant-ph/0101126 (2001)

(b) X.Miao, Solving the quantum search problem in polynomial time on an NMR
quantum computer, http://arxiv.org/abs/quant-ph/0206102 (2002)\newline
16. X.Miao, Efficient multiple-quantum transition processes in an $n-$qubit
spin system, http://arxiv.org/abs/quant-ph/0411046 (2004)\newline
17. R. Jozsa, Quantum algorithms and the Fourier transform,

http://arxiv.org/abs/quant-ph/9707033\newline
18. H.Kurzweil and B.Stellmacher, \textit{An introduction to the theory of
finite groups}, Springer-Verlag, New York, 2004\newline
19. G.H.Hardy and E.M.Wright, \textit{An introduction to the theory of
numbers}, 5th. ed., Oxford Science Press, 1979.\newline
20. W.J.Leveque, \textit{Fundamentals of number theory}, Dover Publications,
Inc., New York, 1996\newline
21. S.C.Pohlig and M.E.Hellman, An improved algorithm for computing
logarithms over $GF(p)$ and its cryptographic significance, IEEE
transactions on information theory, IT-24, 106 (1978)\newline
22. P.W.Shor, Polynomial-time algorithms for prime factorization and
discrete logarithms on a quantum computer, SIAM J.Comput. 26, 1484 (1997),
also see: Proc. 35th Annual Symposium on Foundations of Computer Science,
IEEE Computer Society, Los Alamitos, CA, pp.124 (1994)\newline
23. D.Beckman, A.N.Chari, S.Devabhaktuni, and J.Preskill, Efficient networks
for quantum factoring, Phys. Rev. A 54, 1034 (1996)\newline
24. V.Vedral, A.Barenco, and A.Ekert, Quantum networks for elementary
arithmetic operations, Phys. Rev. A 54, 147 (1996)\newline
25. M.A.Nielsen and I.L.Chuang, \textit{Quantum computation and quantum
information}, Chapter 5, Cambridge University Press, 2000\newline
26. C.H.Bennett, Logical reversibility of computation, IBM J. Res. Develop.
17, 525 (1973)\newline
27. C.H.Bennett, Time/space trade-offs for reversible computation, SIAM J.
Comput. 18, 766 (1989)\newline
28. Y.Levine and A.T.Sherman, A note on Bennett$^{\prime }$s time-space
tradeoff for reversible computation, SIAM J.Comput. 19, 673 (1990)\newline
29. D.Deutsch, Quantum theory, the Church-Turing principle and the universal
quantum computer, Proc.Roy.Soc. London A, 400, 96 (1985)\newline
30. (a) M.Mosca and C.Zalka, Exact quantum Fourier transforms and discrete
logarithm algorithms, http://arxiv.org/abs/quant-ph/0301093 (2003)

(b) J.Proos and C.Zalka, Shor$^{\prime }$s discrete logarithm quantum
algorithm for elliptic curves, http://arxiv.org/abs/quant-ph/0301141 (2003)%
\newline
31. A.Y.Kitaev, Quantum measurements and the Abelian stabilizer problem,
http://arxiv.org/abs/quant-ph/9511026 (1995) \newline
32. A.Barenco, C.H.Bennett, R.Cleve, D.DiVincenzo, N.Margolus, P.Shor,
T.Sleator, J.Smolin, and H.Weinfurter, Elementary gates for quantum
computation, Phys.Rev. A 52, 3457 (1995)\newline
33. D.Coppersmith, An approximate Fourier transform useful in quantum
factoring, IBM research report RC 19642 (1994);

see also: http://arxiv.org/abs/quant-ph/0201067 (2002) \newline
34. (a) R.Cleve, A note on computing quantum Fourier transforms by quantum
programs, http://www.cpsc.ucalgary.ca (1994);

(b) R.Cleve and J.Watrous, Fast parallel circuits for the quantum Fourier
transform, http://arxiv.org/abs/quant-ph/0006004 (2000) \newline
35. L.Hales and S.Hallgren, An improved quantum Fourier transform algorithm
and applications, Proc. 41st Annual Symposium on Foundations of Computer
Science, 515 (2000) \newline
36. R.Cleve, A.Ekert, C.Macchiavello, and M.Mosca, Quantum algorithms
revisited, Proc.R.Soc.Lond. A 454, 339 (1998)\newline
37. P.Benioff, The computer as a physical system: A microscopic quantum
mechanical Hamiltonian model of computers as represented by Turing machines,
J.Statist.Phys., 22, 563 (1980) \newline
38. D.Deutsch, Quantum computational networks, Proc.Roy.Soc. London A, 425,
73 (1989)\newline
39. A.Yao, Quantum circuit complexity, Proc. 34th Annual Symposium on
Foundations of Computer Science, IEEE Computer Society Press, Los Alamitos,
CA, pp. 352\newline
40. E.Bernstein and U.Vazirani, Quantum computation complexity, SIAM
J.Comput. 26, 1411 (1997)\newline
41. (a) J.M.Myers, Can a universal quantum computer be fully quantum?,
Phys.Rev.Lett. 78, 1823 (1997);

(b) M.Ozawa, Quantum Turing machines: local transition, preparation,
measurement, and halting, http://arxiv.org/abs/quant-ph/9809038 (1998);

(c) N.Linden and S.Popescu, The halting problem for quantum computers,
http://arxiv.org/abs/quant-ph/9806054 (1998);

(d) Y.Shi, Remarks on universal quantum computer, http://arxiv.org/abs
/quant-ph/9908074 (1999) \newline
42. X. Miao, Multiple-quantum operator algebra spaces and description for
the unitary time evolution of multilevel spin systems, Molec.Phys. 98, 625
(2000)\newline
43. R.R.Ernst, G.Bodenhausen, and A.Wokaun, \textit{Principles of nuclear
magnetic resonance in one and two dimensions}, Oxford university press,
Oxford, 1987\newline
44. (a) H.F.Trotter, On the product of semigroups of operators,

Proc.Am.Math.Soc. 10, 545 (1959)

(b) M.Suzuki, Decomposition formulas of exponential operators and Lie
exponentials with some applications to quantum mechanics and statistical
physics, J.Math.Phys. 26, 601 (1985) \newline
\newline

\end{document}